\documentclass[iop,revtex4-1]{emulateapj}
\usepackage{aas_macros}
\usepackage[dvipsnames,svgnames]{xcolor}
\usepackage{color,soul}
\usepackage{natbib}
\usepackage{hyperref}
\usepackage{amsmath}
\usepackage{xspace}

\usepackage{graphicx}
\usepackage{enumitem}
\usepackage{amssymb}
\usepackage{xifthen}
\usepackage{hyperref}

\hypersetup{    
  colorlinks      = {true},
  linkcolor       = {blue},
  citecolor       = {blue},
  urlcolor        = {blue},
}

\newcommand{\code}[1]{\texttt{#1}\xspace}

\newcommand{\unit}[1]{\ensuremath{\mathrm{\,#1}}\xspace}

\newcommand{\feh}         {{\rm [Fe/H]}}

\newcommand{\pmra}         {\mbox{$\mu_{\alpha} \cos{\delta}$}}
\newcommand{\pmdec}         {\mbox{$\mu_{\delta} $}}

\newcommand{\km}{\unit{km}}

\newcommand{\kpc}{\unit{kpc}}

\newcommand{\second}{\unit{s}}
\newcommand{\kms}         {\ensuremath{\km\,\second^{-1}}\xspace}

\newcommand{\milliarcsec}{\unit{mas}}
\newcommand{\yr}{\unit{yr}}
\newcommand{\pmunit}{\ensuremath{\milliarcsec\,\yr^{-1}}\xspace}

\newcommand{\Msun}{M_\odot}

\newcommand{\gaia}{\textit{Gaia}\xspace}

\shorttitle{Proper Motions of DES Satellites}
\shortauthors{Pace \& Li}

\begin{document}

\title{Proper motions of  Milky Way Ultra-Faint satellites with \gaia DR2 $\times$ DES DR1}


\author{Andrew~B.~Pace\altaffilmark{1,4}}
\author{Ting~S.~Li\altaffilmark{2,3}}

\affil{$^{1}$ George P. and Cynthia Woods Mitchell Institute for Fundamental Physics and Astronomy, and Department of Physics and Astronomy, Texas A\&M University, College Station, TX 77843, USA}
\affil{$^{2}$ Fermi National Accelerator Laboratory, P.O.\ Box 500, Batavia, IL 60510, USA}
\affil{$^{3}$ Kavli Institute for Cosmological Physics, University of Chicago, Chicago, IL 60637, USA}
\altaffiliation{$^{4}$ Mitchell Astronomy Fellow}

\email{apace@tamu.edu}
\email{tingli@fnal.gov}

\begin{abstract}

We present a new, probabilistic  method for determining the systemic proper motions of  Milky Way (MW) ultra-faint satellites in the Dark Energy Survey (DES).  We utilize the superb photometry from the first public data release (DR1) of DES to select candidate members, and cross-match them with the proper motions from $Gaia$ DR2. We model the candidate members with a mixture model (satellite and MW) in spatial and proper motion space.
This method does not require prior knowledge of satellite membership, and can successfully determine the tangential motion of thirteen DES satellites. 
With our method we present measurements of the following satellites: Columba~I, Eridanus~III, Grus~II, Phoenix~II, Pictor~I, Reticulum~III, and Tucana~IV; this is the first systemic proper motion measurement for several and the majority lack extensive spectroscopic follow-up studies. 
We compare these to the predictions of Large Magellanic Cloud satellites and to the vast polar structure. With the high precision DES photometry we conclude that most of the newly identified member stars are very metal-poor ([Fe/H] $\lesssim -2$) similar to other ultra-faint dwarf galaxies, while Reticulum III is likely more metal-rich. We also find potential members in the following satellites that might indicate their overall proper motion: Cetus~II,  Kim~2, and Horologium~II; however, due to the small number of members in each satellite, spectroscopic follow-up observations  are necessary to determine  the systemic proper motion in these satellites.

\end{abstract}


\section{INTRODUCTION}
\label{intro}

The Milky Way (MW) satellites galaxies are a diverse set of systems with sizes ranging from tens to thousands of parsecs, and luminosities between 300  to $10^9 L_{\odot}$ \citep{McConnachie2012AJ....144....4M}.
Measuring the tangential motion of a satellite was until recently only available for the largest and brightest systems with Hubble Space Telescope astrometry and long baselines \citep[e.g.][]{Piatek2002AJ....124.3198P, Kallivayalil2013ApJ...764..161K, Sohn2017ApJ...849...93S}. 
With the release of the \gaia DR2 \citep{Gaia_Brown2018A&A...616A...1G} studying the tangential motion of many more  MW satellites is now possible \citep{Gaia_Helmi2018A&A...616A..12G}.  

Learning the tangential motion of the MW satellites provides many new opportunities for further understanding their nature and origin.  
First, detailed knowledge of their orbital properties can be derived and the extent of the MW tidal influence known.
The accretion or infall time of a satellite can test  satellite star formation  quenching models (i.e. reionization versus ram pressure striping) \citep[e.g.][]{Ricotti2005ApJ...629..259R, Rocha2012MNRAS.425..231R, Fillingham2015MNRAS.454.2039F}. 
Second, we can test whether there are structures in the satellite distribution, including the hypothesis of pairs of satellites \citep[e.g. Crater-Leo, Pegasus III-Piscess II][]{Torrealba2016MNRAS.459.2370T, Kim2015ApJ...804L..44K}, the vast polar structure~\citep{Pawlowski2013MNRAS.435.2116P}, and  satellites of Large and Small Magellanic Clouds \citep{Jethwa2016MNRAS.461.2212J, Sales2017MNRAS.465.1879S}. 
Moreover, the  distribution of satellites in phase space can determine the MW mass \citep[e.g.][]{Sohn2013ApJ...768..139S, Patel2018ApJ...857...78P}. 
Many of these topics have been addressed in the first proper motion analysis of \gaia DR2 satellite papers \citep{Gaia_Helmi2018A&A...616A..12G, Simon2018ApJ...863...89S, Fritz2018A&A...619A.103F, Kallivayalil2018ApJ...867...19K}.

There have been a plethora of new candidate satellites in recent years, especially in the southern sky \citep[e.g][]{Laevens2015ApJ...813...44L, Martin2015ApJ...804L...5M,Torrealba2016MNRAS.459.2370T,Drlica-Wagner2016ApJ...833L...5D, Torrealba2018MNRAS.475.5085T, Homma2018PASJ...70S..18H}.
Many have been found in the footprint of the Dark Energy Survey (DES), a 5-year, 5000 deg$^2$ survey  \citep{Bechtol2015ApJ...807...50B, Koposov2015ApJ...805..130K, Drlica-Wagner2015ApJ...809L...4D, Kim2015ApJ...808L..39K, Luque2016MNRAS.458..603L, Luque2017MNRAS.468...97L, Luque2018MNRAS.478.2006L}.
Many of these objects remain candidates and deeper photometry \citep[e.g.][]{Carlin2017AJ....154..267C} and/or spectroscopy \citep[e.g.][]{Simon2015ApJ...808...95S, Li2018ApJ...857..145L} is required to verify the stellar overdensity and to uncover their nature as a star clusters or dwarf galaxies \citep{Willman2012AJ....144...76W}.

New systemic proper motions with \gaia have been measured for many ultra-faint  satellites \citep{Gaia_Helmi2018A&A...616A..12G, Simon2018ApJ...863...89S, Fritz2018A&A...619A.103F, Kallivayalil2018ApJ...867...19K, Massari2018A&A...620A.155M}.
Each study has utilized different methods to determine a satellite's systemic proper motion.
For example,  \citet{Gaia_Helmi2018A&A...616A..12G, Massari2018A&A...620A.155M} had a selection based on spatial positions and \gaia color-magnitude diagrams and used an iterative sigma clipping routine to measure the proper motion.
For satellites with spectroscopic follow-up, the systemic proper motion can be determined from `bright' spectroscopically confirmed members \citep{Simon2018ApJ...863...89S, Fritz2018A&A...619A.103F}.
\citet{Kallivayalil2018ApJ...867...19K} utilized a clustering algorithm to identify additional members in satellites with spectroscopically confirmed members.

In this contribution, we will introduce an independent method  to measure the systemic proper motions of satellites that do not yet have spectroscopic follow-up.
Throughout this paper we will refer to the DES candidates as satellites.  
While several objects have been confirmed via spectroscopy to be ultra-faint dwarf spheroidal galaxies \citep[Eridanus~II, Horologium~I, Reticulum~II, Tucana~II, ][]{Simon2015ApJ...808...95S,Walker2015ApJ...808..108W,Koposov2015ApJ...811...62K, Walker2016ApJ...819...53W, Li2017ApJ...838....8L} others remain ambiguous \citep[Grus~I \& Tucana~III,][]{Walker2016ApJ...819...53W, Simon2017ApJ...838...11S, Li2018ApJ...866...22L, MutluPakdil2018ApJ...863...25M}.
In addition, several of the satellites are thought to be star clusters \citep[Eridanus III, Kim 2][]{Luque2018MNRAS.478.2006L, Conn2018ApJ...852...68C, Kim2015ApJ...803...63K}. 
Several of the candidates (Tucana~V, Cetus~II) have been argued to be false positives from deeper data \citep{Conn2018ApJ...852...68C, Conn2018ApJ...857...70C}.

In~\S\ref{sec:data_methods}, we discuss the \gaia $\times$ DES DR1 data, cuts to produce a pure sample, and our methodology for determining the systemic proper motions of a satellite.  In~\S\ref{sec:results}, we validate our method by comparing our results to satellites with spectroscopic follow-up and present the initial results for our sample.
In~\S\ref{sec:discussion}, we compare the new systemic proper motions to  kinematic/dynamical predictions, discuss the metallicity from color-color diagrams, discuss individual satellites, and conclude. 

\section{Data \& Methods}
\label{sec:data_methods}
\subsection{Data}
\label{sec:data}

\begin{figure*}[th!]
\includegraphics[scale=.4]{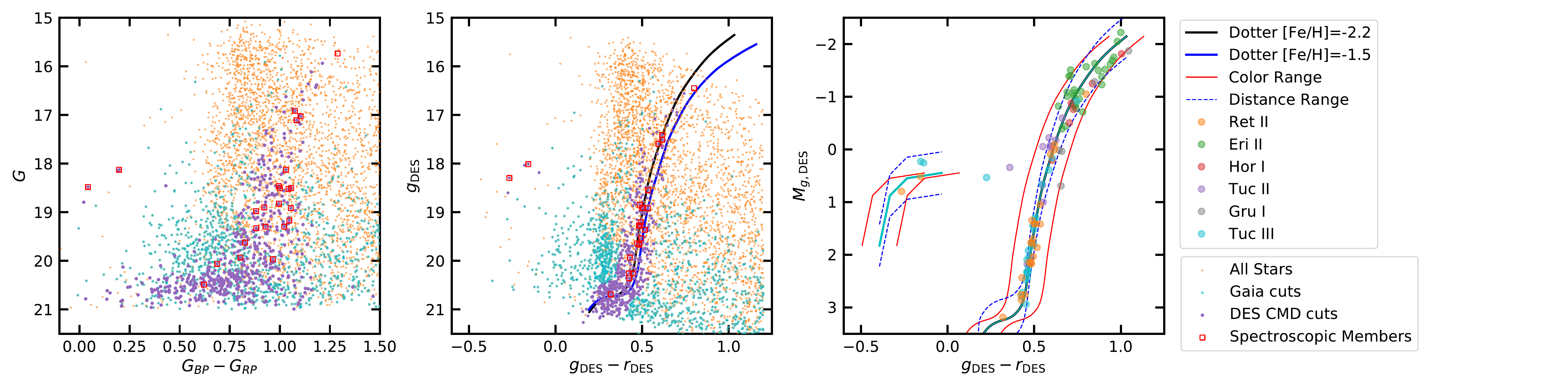}
\caption{{\bf Left \& Middle:} Example of the selection of candidate members for Reticulum II, with left (middle) showing the color-magnitude diagram (CMD) using \gaia (DES) photometry. Points are colored by: all stars (cyan), after \gaia cuts (orange; the \gaia cuts include parallax, astrometric fit, and escape velocity), after DES color-magnitude selection (purple).  Overlaid in red open squares are the  spectroscopically confirmed  members \citep{Simon2015ApJ...808...95S}.   In the middle panel we show a Dotter isochrone with age = 12.5 Gyr and  [Fe/H] = -2.2  (black) and a more metal-rich ([Fe/H] = -1.5, age= 10 Gyr; blue).  {\bf Right:} CMD with  spectroscopic members in Eridanus II, Grus I, Horologium I, Reticulum II, Tucana II, Tucana III (references in Table~\ref{table:spectra}) that are included in the DES DR1$\times$\gaia catalog. The lines shows our color-magnitude selection which includes a range based on color (red) and distance modulus (blue-dotted). }
\label{fig:cmd}
\end{figure*}

Our main objective is to determine the proper motions of all satellites found in DES \citep{Bechtol2015ApJ...807...50B, Koposov2015ApJ...805..130K, Kim2015ApJ...808L..39K, Drlica-Wagner2015ApJ...809L...4D, Luque2016MNRAS.458..603L, Luque2017MNRAS.468...97L}. 
In this section, we describe the procedures for preparing the candidate stars in each satellite, which are then used in the mixture model method  described in \S\ref{sec:method}. We list the properties of each satellite we adopt in this paper and the corresponding references in Table~\ref{table:property}.

We first perform an astrometric cross-match of DES DR1 \citep{DES2018arXiv180103181A} and \gaia DR2 \citep{Gaia_Brown2018A&A...616A...1G}   within a region of 1\degr~in radius for each satellite with a cross-match radius of 0\farcs5. 
As the astrometric precision of DES DR1 against \gaia DR1 is about 150 mas \citep{DES2018arXiv180103181A}, the cross-match radius of 0\farcs5 selects most of the stars in the magnitude range of $16 < r < 21$, where the bright end is due to  saturation in DES and the faint end is due to the limiting magnitude of \gaia. We note that the astrometry of \gaia DR2 is referenced to J2015.5 Epoch, while DES DR1 astrometry is referenced to J2000 Epoch. We did not perform any parallax or proper motion correction before the cross-match, and therefore we may miss some high-proper motion or nearby stars with this cross-match radius. As we are interested in targets that are relatively distant ($>10$~\kpc) with relatively small proper motions (a few mas\,yr$^{-1}$), the cross-match should not affect the candidate members in each satellite.

We then perform a series of astrometric cuts using \gaia DR2. 
We remove nearby stars with a parallax cut: $\varpi - 3\sigma_{\varpi}>0$ \citep{GaiaLindegren2018A&A...616A...2L}.  We remove sources with bad astrometric fits; defining $u \equiv (\mbox{\tt astrometric\_chi2\_al} / (\mbox{\tt astrometric\_n\_good\_obs\_al} - 5) )^{1/2}$. We remove stars with: $u > 1.2 \times \max{(1, \exp{(-0.2 (G - 19.5) ))}}$ \citep{GaiaLindegren2018A&A...616A...2L}.
Lastly, we perform a cut based on the MW escape velocity ($v_{\rm esc}$).
$v_{\rm esc}$ is computed with the potential \code{MWPotential2014} (with a slightly increased halo mass, $M_{\rm vir}=1.6\times 10^{12} \Msun$) from \code{galpy} \citep{Bovy2015ApJS..216...29B}.
We compute the tangential velocity ($v_{\rm tan}$) of each star by converting the proper motions into Galactic coordinates in the Galactic Standard of Rest (GSR) frame after accounting for the Sun's reflex motion, assuming $({\rm U_\odot,\, V_\odot, \, W_\odot}) = (11.1, 12.24, 7.25)\kms$, a circular velocity of $220\kms$ \citep{Schonrich2010MNRAS.403.1829S}, and each star is at the satellite's heliocentric distance.  We remove stars with the cut\footnote{We note that in principle the satellite may not be bound to the MW and the escape velocity cut would remove all members.  We manually check that there are no high proper motion stars clustered near each satellite. }: $v_{\rm tan} - 3 \sigma_{v_{\rm tan}} > v_{\rm esc}$.  
The main goal of this cut is to remove large, precise proper motions  that would increase the inferred MW dispersion parameters and pull the net MW motion towards the outliers. To be conservative we applied a relatively lose cut with a more massive MW.  
In Figure~\ref{fig:cmd}, we show the color-magnitude diagram (CMD) of candidate stars before and after the astrometric cuts, using \gaia DR2 (left panel) and DES DR1 (middle panel) photometry of Reticulum II\footnote{For examples in this paper, we select  Reticulum II as it is nearby and has the most expected number of members.  In addition, it contains a large number of stars that have been confirmed to be satellite members based on spectroscopic observations \citep{Simon2015ApJ...808...95S}.} as an example.

After the astrometric cuts, we performed additional selection criteria on the CMD using DES DR1 photometry. 
Our CMD selection is derived from the spectroscopically confirmed members in the six satellites with follow-up (see Table~\ref{table:spectra} for the satellites and the associated references).
As shown in the right panel of Figure~\ref{fig:cmd}, most of spectroscopically confirmed members on the red giant branch (RGB) in the DES satellites  lie on a Dotter isochrone \citep{Dotter2008ApJS..178...89D} with an old and metal-poor population (age $= 12.5$ Gyr, {\bf $\feh = -2.2$}). We therefore constrain our candidate members (RGB stars and main sequence turnoff stars) to be close to this isochrone. In addition, we selected blue horizontal branch (BHB) stars using an empirical isochrone of M92 from \citet{Bernard2014MNRAS.442.2999B} after transforming to DES photometric system.
Specifically, we select the targets to be either $\pm 0.1$ mag in $\Delta (g-r)$ or $\pm 0.4$ mag in $\Delta g$ to either isochrones, as illustrated by red solid and blue dashed lines in the right panel of Figure~\ref{fig:cmd}. In the left and middle panels of Figure~\ref{fig:cmd}, we show the candidate members of Reticulum II after the CMD selection along with the spectroscopically confirmed members, using \gaia DR2 and DES DR1 photometry. The spread with \gaia DR2 photometry is much larger at the faint end. Therefore, selection of a narrow isochrone window with DES DR1 photometry will largely decrease the background contamination from the Milky Way disk and halo stars.
For reference we additionally include a more metal-rich isochrone ([Fe/H]=-1.5; age=10 Gyr)  in the middle panel of Figure~\ref{fig:cmd} as several satellites have larger photometric metallicities. 

We note that our photometric selection aims for a mostly pure sample of candidate members, rather than a complete sample to include $every$ possible member star. For example, we exclude any members on the red horizontal branch (RHB) in the range of $0 < g-r < 0.4$. As these ultra-faint dwarf galaxies are old and less massive, we expect minimal RHB members in each satellite, except for some RR Lyraes in this color range. Indeed, the spectroscopic RHB member in the tidal tail of Tucana III \citep{Li2018ApJ...866...22L} turns out to be a non-member from its proper motion; however, the RHB member in Tucana II is a proper motion member \citep{Walker2016ApJ...819...53W}. Furthermore, we may miss members that are farther away from the isochrone, either due to  larger photometric uncertainties at  fainter magnitudes, or due to an intrinsic metallicity spread (e.g. see Eridanus II in the right panel of Figure~\ref{fig:cmd}). If the satellite is more metal-rich (and therefore more likely to be a star cluster rather than a dwarf galaxy), the color of its members will also deviate from the default isochrone, which may result in a null measurement.  This is further discussed  in \S\ref{sec:new} for the satellites with null results.

We note that all  DES photometry referred in this paper are dereddened photometry from DES DR1, using the $E(B-V)$ values from the reddening map of \citet{Schlegel1998ApJ...500..525S} and extinction coefficients reported in \citet{DES2018arXiv180103181A}, which were derived using the \citet{Fitzpatrick1999PASP..111...63F} reddening law and the~\citet{Schlafly2011ApJ...737..103S} adjusted reddening normalization parameter. For \gaia photometry, we refer to the observed photometry from \gaia DR2 without any reddening correction, and we note that the \gaia photometry is only used for plotting and not used for any computation.

\subsection{Method}\label{sec:method}

\begin{deluxetable*}{lrrrrrlrl}
\tablecaption{Properties of the DES Satellites}
\tablewidth{0pt}
\tablehead{
\colhead{Satellite} & \colhead{$\alpha$} & \colhead{$\delta$} & \colhead{$a_{h}$} & \colhead{$\epsilon$ } & \colhead{$\theta$} & \colhead{$D_{\odot}$ } & \colhead{$M_V$} &   \colhead{ References\tablenotemark{a}} \\
 & \colhead{deg} & \colhead{deg} & \colhead{arcmin} & & & \colhead{kpc} & 
}
\startdata
Cetus II & $19.4700$ & $-17.4200$ & 1.9 &  - & -  & 28.8 & 0.0 & 1\\
Columba I & $82.8570$ & $-28.0425$ & 2.2 & 0.30 & 24 & 183 & -4.2 & 2\\
Eridanus II & $56.0838$ & $-43.5338$ & 2.31 & 0.48 & 73 & 366 & -7.1 & 3\\
Eridanus III & $35.6888$ & $-52.2847$ & 0.315 & 0.44 & 109 & 91 & -2.07 & 4\\
Grus I & $344.1765$ & $-50.1633$ & 2.23 & 0.41 & 4 & 120 & -3.4 & 5\\
Grus II & $331.0200$ & $-46.4400$ & 6.0 &  - & -  & 53 & -3.9 & 1\\
Horologium I & $43.8820$ & $-54.1188$ & 1.41 &  - & -  & 79 & -3.4 & 5\\
Horologium II & $49.1338$ & $-50.0181$ & 2.09 & 0.52 & 127 & 78 & -2.6 & 6\\
Kim 2\tablenotemark{b} & $317.2080$ & $-51.1635$ & 0.42 & 0.12 & 35 & 104.7 & -1.5 & 7\\
Indus II & $309.7200$ & $-46.1600$ & 2.9 &  - & -  & 214 & -4.3 & 1\\
Phoenix II & $354.9975$ & $-54.4060$ & 1.38 & 0.47 & 164 & 84.3 & -2.8 & 8\\
Pictor I & $70.9475$ & $-50.2830$ & 1.18 & 0.47 & 78 & 114 & -3.1 & 5\\
Reticulum II & $53.9493$ & $-54.0466$ & 6.3 & 0.60 & 68 & 31.5 & -3.1 & 8\\
Reticulum III & $56.3600$ & $-60.4500$ & 2.4 &  - & -  & 92 & -3.3 & 1\\
Tucana II & $342.9796$ & $-58.5689$ & 12.89 & 0.39 & 107 & 57 & -3.8 & 5\\
Tucana III & $359.1500$ & $-59.6000$ & 6.0 &  - & -  & 25 & -2.4 & 1,8\\
Tucana IV & $0.7300$ & $-60.8500$ & 11.8 & 0.40 & 11 & 48 & -3.5 & 1\\
Tucana V & $354.3500$ & $-63.2700$ & 1.8 & 0.70 & 30 & 55 & -1.6 & 1\\
DES 1 & $8.4992$ & $-49.0386$ & 0.245 & 0.41 & 112 & 76 & -1.42 & 4\\
DES J0225+0304 & $36.4267$ & $3.0695$ & 2.68 & 0.61 & 31 & 23.8 & -1.1 & 9
\enddata

\tablenotetext{a}{
References:
(1) \citet{Drlica-Wagner2015ApJ...813..109D}
(2) \citet{Carlin2017AJ....154..267C}
(3) \citet{Crnojevic2016ApJ...824L..14C}
(4) \citet{Conn2018ApJ...852...68C}
(5) \citet{Koposov2015ApJ...805..130K}
(6) \citet{Kim2015ApJ...808L..39K}
(7) \citet{Kim2015ApJ...803...63K}
(8) \citet{MutluPakdil2018ApJ...863...25M}
(9) \citet{Luque2017MNRAS.468...97L}
}
\tablenotetext{b}{Also referred to as Indus~I.
}
\label{table:property}
\end{deluxetable*}

We model the candidate stars as a mixture model containing a satellite and a MW foreground.  The total likelihood ($\mathcal{L}$) is:

\begin{equation}
\mathcal{L} = (1 - f_{\rm MW}) \mathcal{L}_{\rm satellite} + f_{\rm MW} \mathcal{L}_{\rm MW} \, , \\ 
\end{equation}

\noindent where $\mathcal{L}_{\rm satellite}$ and $\mathcal{L}_{\rm MW}$ correspond to the satellite (dwarf galaxy or star cluster) and MW components respectively.  $f_{\rm MW}$ is the fraction of stars in the MW component.   Each likelihood term is  decomposed into spatial proper motion parts:

\begin{equation}
\mathcal{L}_{\rm satellite/MW} = \mathcal{L}_{\rm spatial} \mathcal{L}_{\rm PM} \, , \\ 
\end{equation}

\noindent where $\mathcal{L}_{\rm spatial}$ and $\mathcal{L}_{\rm PM}$ are terms for the spatial and proper motion distributions respectively.   
The proper motion term is modeled as a multi-variate Gaussian:

\begin{equation}
\ln \mathcal{L}_{\rm PM} = -\frac{1}{2} (\chi - \overline{\chi})^\top C^{-1} (\chi - \overline{\chi}) - \frac{1}{2}\ln{\left(4 \pi^2 \det C \right)}
\end{equation}

\noindent where $\chi = ( \mu_\alpha \cos{\delta}, \mu_\delta)$ is the data vector and $\overline{\chi} = ( \overline{\mu_\alpha \cos{\delta} }, \overline{\mu_\delta})$ is the vector containing the systemic proper motion of the satellite or MW foreground.  
The covariance matrix, $C$, includes the correlation between the proper motion errors and a term for the intrinsic proper motion dispersion.  The covariance matrix is:

\begin{equation}
C=
\begin{bmatrix}
\epsilon_{\mu_\alpha \cos{\delta}}^2 +  \sigma_{\mu_\alpha \cos{\delta}}^2 & \epsilon_{\mu_\alpha \cos{\delta} \times \mu_\delta}^2 \\
\epsilon_{\mu_\alpha \cos{\delta} \times \mu_\delta}^2 & \epsilon_{\mu_\delta}^2 +  \sigma_{\mu_\delta}^2  
\end{bmatrix} 
\, , \\ 
\end{equation}

\noindent where $\epsilon$ represents the proper motion errors and $\sigma$ the intrinsic dispersions.  
We do not include intrinsic dispersion terms for the satellite component as it is expected to be significantly smaller than the proper motion uncertainties\footnote{For example, a star with $G\sim18$ mag at 80 kpc has errors on the order of $80\kms$ while the expected intrinsic dispersion is $\sim 3-6 \kms$. }.

For the satellite spatial term, we assume a projected Plummer stellar distribution \citep[][]{Plummer1911MNRAS..71..460P}:

\begin{equation}
\Sigma (R_e) = \frac{1}{\pi a_h^2 (1-\epsilon)}(1 + R_e^2/a_h^2)^{-2} \,, 
\end{equation}

\noindent where $R_e^2 = x^2 + y^2/(1-\epsilon)^2$, is the elliptical radius, $a_h$ is the semi-major half-light radius, and $\epsilon$ is the ellipticity.  Here $x$ and $y$ are  the coordinates along the major and minor axis respectively and the on-sky coordinates ($\delta \alpha$, $\delta \delta$) have been rotated by the position angle, $\theta$, measured North to East to this frame.  The spatial scale for the MW satellites is the half-light radius, $r_h = a_h \sqrt{1-\epsilon}$, and we use the azimuthally  averaged quantity here.  The parameters for $a_h$, $\epsilon$, and $\theta$ are taken from the literature and summarized in Table~\ref{table:property}.
The satellite's probability distribution of projected ellipticity radii is given by $p_{R_e}(R_e) = {\rm d}/{\rm d} R_e \left[ \int_0^{R_e} \Sigma(R_e) R_e {\rm d}R_e  / \int_0^{R_{\rm max}} \Sigma(R_e) R_e {\rm d}R_e \right] $.  For the Plummer profile this is \citep{Walker2011ApJ...742...20W}:

\begin{equation}
\mathcal{L}_{\rm spatial} = p_{R_e}(R_e) = \frac{2 R_e/a_h^2}{(1 + R^2/a_h^2)^2}.\\
\end{equation}

We assume the MW foreground is constant the over the region probed.
We pre-compute the spatial probabilities\footnote{ We have explored  spatial parameters (i.e. $a_h$, $\epsilon$) as free parameters with Gaussian priors based on literature values and find that it does not affect our results.  We have also explored utilizing different literature structural parameter measurements for several satellites and find that it does not affect the proper motion measurements. } and the relative normalization between the two spatial components is determined with the fraction parameter ($f_{\rm MW}$).  
The spatial term in effect acts as a weight term: stars near the satellite's center are more likely to be satellite members.  
The stars at large radii will determine the MW proper motion  and assist in identifying  MW interlopers near the satellite's center. 

Overall, our model contains 7 free parameters: 2 parameters for the systemic proper motion of the satellite  ($\overline{\mu_\alpha \cos{\delta}}$, $\overline{\mu_\delta }$); 4 to describe the MW foreground model, 2 systemic proper motion  ($\overline{\mu^{{\rm MW}}_{\alpha} \cos{\delta}}$, $\overline{\mu_{\delta}^{\rm MW}}$), and 2 dispersion parameters ($\sigma_{\mu_{\alpha}\cos{\delta}}^{{\rm MW}}$, $\sigma_{\mu_{\delta}}^{{\rm MW}}$); and 1 for the normalization between the two components ($f_{\rm MW}$).
In appendix~\ref{appendix:mwmodel} we explore a two component MW foreground model. 
For priors, we assume linear priors except for the dispersion parameters where we use Jeffreys priors\footnote{ For a scaled parameter (such as the dispersion), a Jeffreys prior will be a non-informative objective prior.  For additional discussion of this prior compared to a uniform prior see Section 8 of \citet{Kim2016ApJ...833...16K}.}.  The priors ranges: are $-10 < \overline{\mu} < 10~{\rm mas \, yr^{-1}}$ for the proper motions, $-3 < \log_{10}{\sigma_{\mu}} < 1$ for the MW dispersions, and $0 < f_{\rm MW} < 1$ for the fraction parameter.
  To determine the posterior distribution we use the \code{MultiNest} algorithm \citep{Feroz2008MNRAS.384..449F, Feroz2009MNRAS.398.1601F}.

To determine a star's satellite membership, $p_i$, we take the ratio of satellite likelihood to total likelihood: $p = \mathcal{L}_{\rm satellite}/((1- f_{\rm MW}) \mathcal{L}_{\rm satellite} + f_{\rm MW}\mathcal{L}_{\rm MW})$ \citep{Martinez2011ApJ...738...55M}.  This is computed for each star at each point in the posterior.  We utilize the median value as the star's membership (which we refer to as $p_i$, for the ith star).  
$p_i$ represents the  probability for the star to be a member of the satellite population only considering its proper motion and spatial location. 
We will refers to stars as `members' if they have $p_i>0.5$.

\section{Results}\label{sec:results}

\begin{figure*}[th!]
\includegraphics[scale=.34]{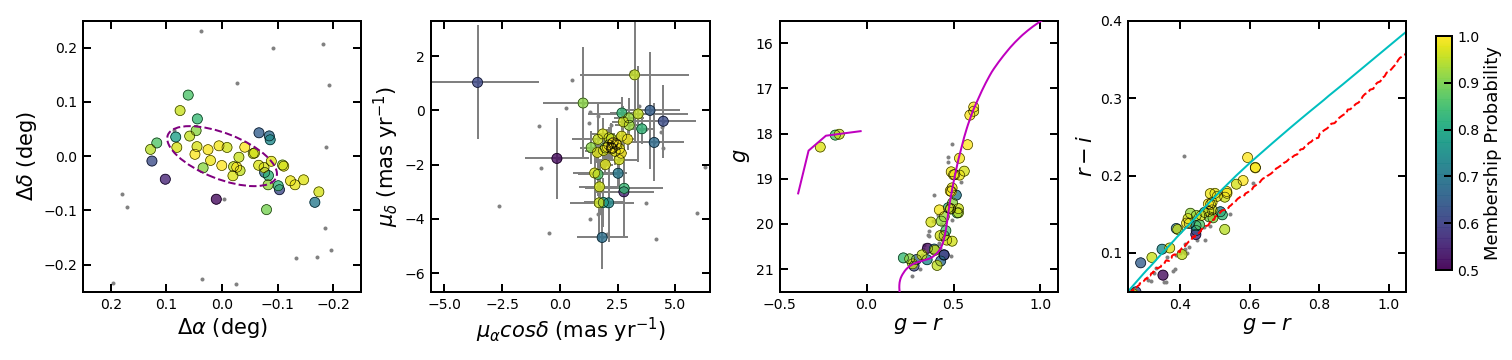}
\caption{Diagnostic plots for Reticulum II. Panels from left to right show the  spatial distribution, proper motion distribution, CMD, and color-color diagram for stars with membership probability $p_i>0.5$.  Grey dots are non-member stars ($p_i<0.1$) within $3\times$ of the half-light radius $r_h$.  The dashed purple ellipse in the first panel indicates the half-light radius of the satellite (used for computing the spatial probability); the solid purple curve in the third panel shows an Dotter isochrone with age = 12.5 Gyr and $\feh=-2.2$ and an empirical M92 isochrone for BHB stars at the distance of the satellite; the dashed red line in the fourth panel shows the empirical stellar locus of dereddened DES photometry, where stars above this lie are likely to be more metal-poor than those below this line;  the same Dotter isochrone () plotted as a cyan line  in color-color diagram (see more details regarding the 4th panel in \S\ref{sec:color}.)}
\label{fig:ret2}
\end{figure*}

\begin{figure*}[h!]
\plotone{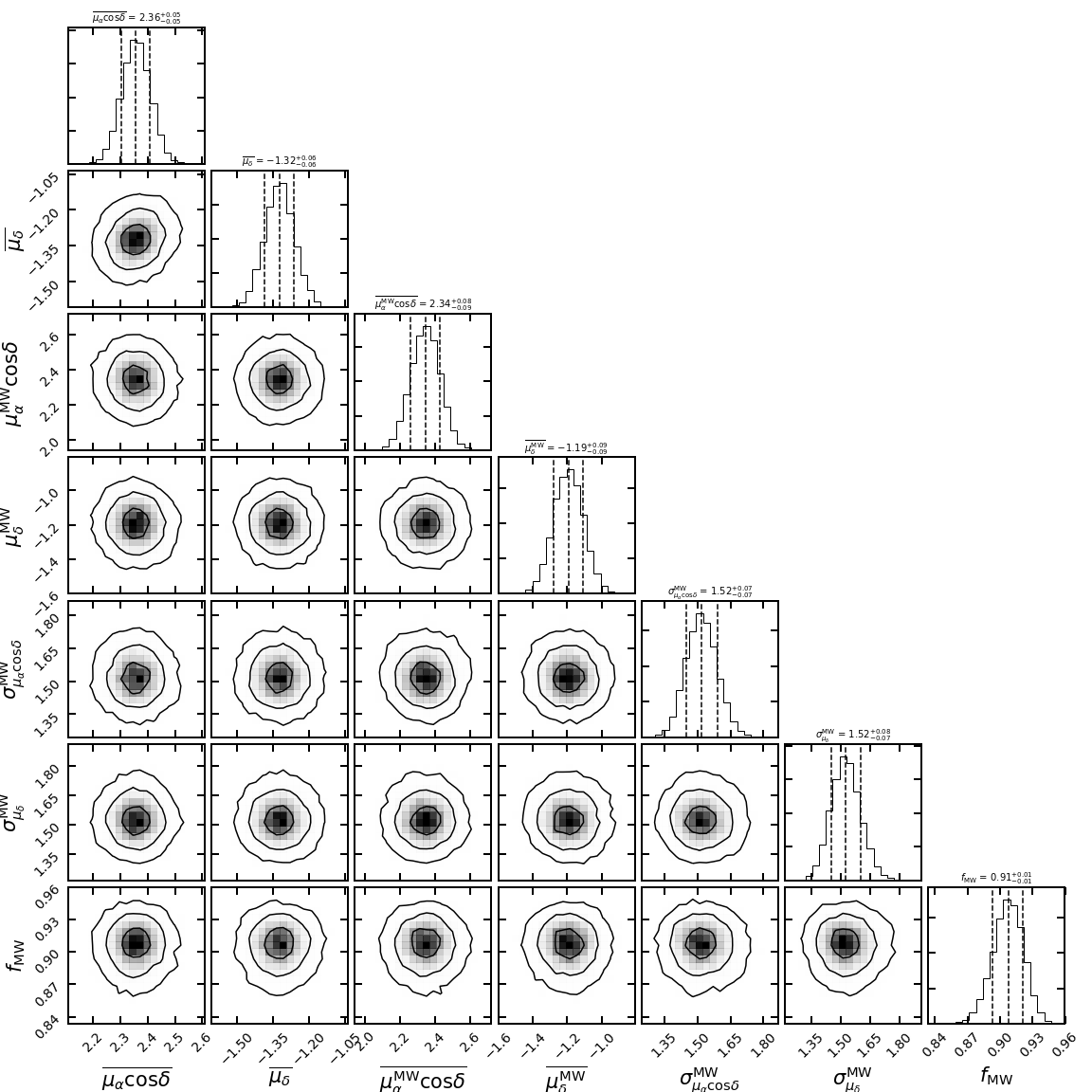}
\caption{Posterior distribution of Reticulum II.  From left to right the parameters are: $\overline{\pmra}$, $\overline{\pmdec}$ (systemic satellite proper motions), $\overline{\mu_{\alpha}^{\rm MW} \cos{\delta}}$, $\overline{\mu_{\delta}^{\rm MW}}$, $\sigma_{\mu_{\alpha}\cos{\delta}}^{\rm MW}$, $\sigma_{\mu_{\delta}}^{\rm MW}$ (MW systemic proper motions and intrinsic proper motion dispersions), and $f_{\rm MW}$. 
The contours enclose 39.4, 85.5, 98.9\% of the posterior distribution corresponding to $1,2,3-\sigma$ confidence intervals.  The dotted lines are the 16,50, 84\% confidence intervals and the numerical values are quoted above.  
}
\label{fig:ret2_corner}
\end{figure*}

Before moving to general results,  we first examine the result of one satellite, Reticulum II, in more detail.
Figure~\ref{fig:ret2} is an example of a diagnostic plot, showing the spatial distribution, proper motions, location on a CMD, and location on a color-color  ($g-r$ vs $r-i$) diagram for stars with $p_i > 0.5$.   
In Figure~\ref{fig:ret2_corner}, we show the posterior distribution of Reticulum II.  
The large number of members ($\sum p_i\approx 48$), the clear clustering of stars in proper motion space (Figure~\ref{fig:ret2}, middle left panel), and the well constrained satellite parameters (Figure~\ref{fig:ret2_corner}) all show that we have identified the systemic proper motion of Reticulum II.
Diagnostic and corner plots for other satellites are in Appendix~\ref{appendix:figures}.

\subsection{Validation}\label{sec:valid}

\begin{deluxetable}{l c cc l ll l ll}
\tablecaption{Summary of Validation with Spectroscopic Members}
\tablewidth{0pt}
\tablehead{
\colhead{Satellite} & \colhead{ ${\rm N_{ spec}}$} & \colhead{${\rm N_{\gaia}}$} & \colhead{${\rm N_{ recover}}$} & \colhead{${\rm N_{ new}}$} &  \colhead{ References\tablenotemark{a}} 
}
\startdata
Eridanus II & 26 & 12  & 11 & 8 & 1\\
Grus I & 7 & 5  & 4 & 4 & 2\\
Horologium I & 6 & 6  & 6 & 11 & 3,4\\
Reticulum II & 25 & 23  & 22 & 25 & 5,6,3\\
Tucana II & 12 & 12  & 11 & 19 & 2,7\\
Tucana III & 48 & 22\tablenotemark{b}  & 15 & 23 & 8,9
\enddata

\tablecomments{
Columns: Satellite name, number of spectroscopic members (${\rm N_{ spec}}$), number of spectroscopic members cross matched to \gaia (${\rm N_{\gaia}}$), number of spectroscopic members recovered ($p_i>0.5$) with our method (${\rm N_{ recover}}$), number of new members ($p_i>0.5$) with our method (${\rm N_{ new}})$.
}
\tablenotetext{a}{
References:
(1) \citet{Li2017ApJ...838....8L}
(2) \citet{Walker2016ApJ...819...53W}
(3) \citet{Koposov2015ApJ...811...62K}
(4) \citet{Nagasawa2018ApJ...852...99N}
(5) \citet{Simon2015ApJ...808...95S}
(6) \citet{Walker2015ApJ...808..108W}
(7) \citet{Chiti2018ApJ...857...74C}
(8) \citet{Simon2017ApJ...838...11S}
(9) \citet{Li2018ApJ...866...22L}
}
\tablenotetext{b}{This only includes the tidal tail stars within 1 degree of the center.
}
\label{table:spectra}
\end{deluxetable}

As a validation of our method we next compare our results to the 6 satellites with  spectroscopic members: Eridanus~II \citep{Li2017ApJ...838....8L}, Grus~I \citep{Walker2016ApJ...819...53W}, Horologium~I \citep{Koposov2015ApJ...811...62K, Nagasawa2018ApJ...852...99N}, Reticulum~II \citep{Simon2015ApJ...808...95S, Walker2015ApJ...808..108W, Koposov2015ApJ...811...62K}, Tucana~II \citep{Walker2016ApJ...819...53W,  Chiti2018ApJ...857...74C}, and Tucana~III \citep{Simon2017ApJ...838...11S, Li2018ApJ...866...22L}.  
We will refer to stars in a satellite that have been previously confirmed to be members with spectroscopic observations as ``spectroscopic members'' or ``spectroscopically confirmed members.''  

\begin{figure*}[th!]
\plotone{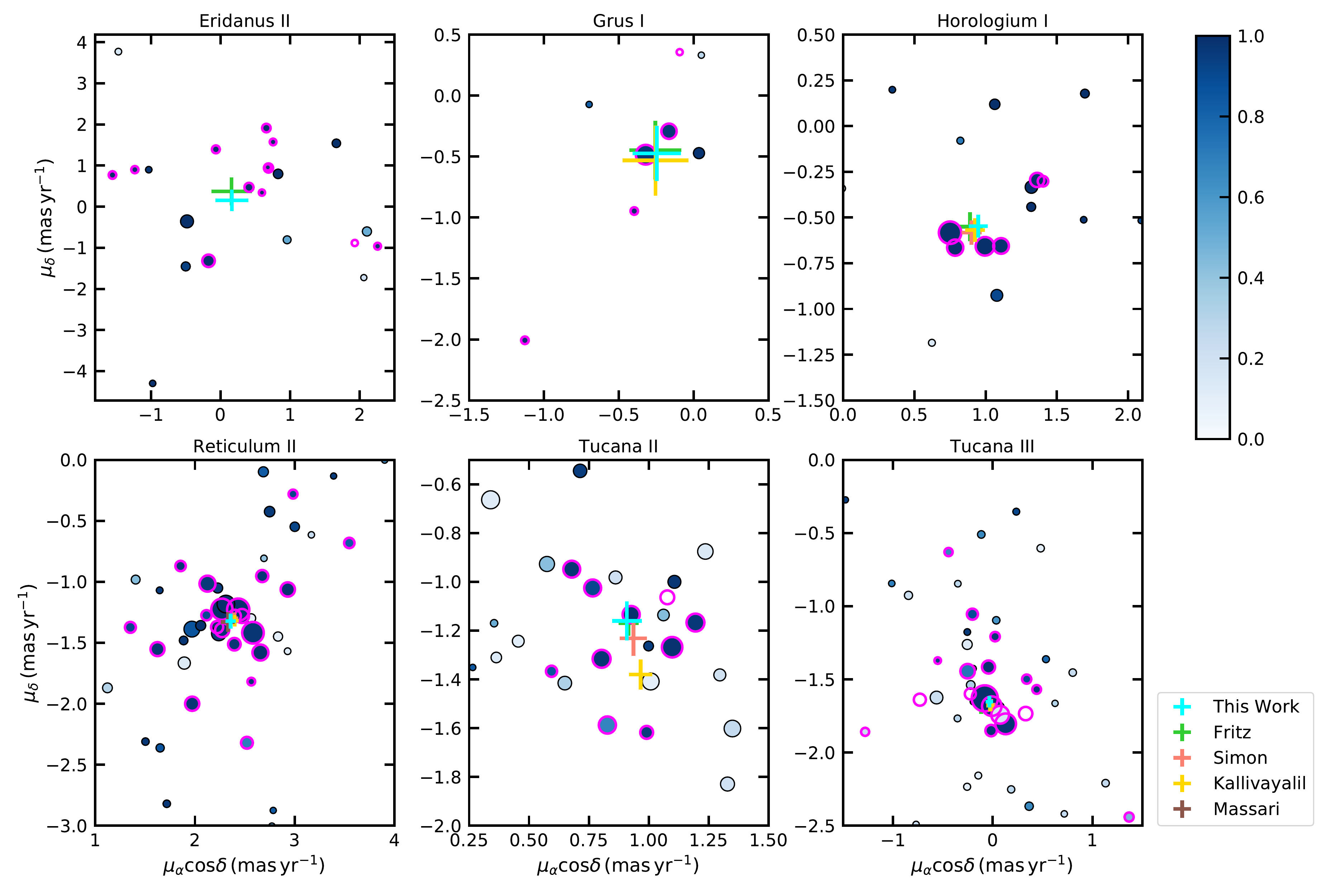}
\caption{Distribution of stars in proper motion space for the six satellites with  spectroscopy (references in Table~\ref{table:spectra}).  The size of the points are inversely related to the G-band magnitude (and therefore proper motion error bars).  The relative point size scale varies between each sub panel.  The minimum and maximum point sizes are based on the difference between the brightest and faintest members.
The points are shaded according to their membership as  assigned by our mixture model.  Spectroscopic members are circled in magenta.  Most members are recovered (see text for details on missed stars).  The points with error bars show our measurement along with  literature measurements \citep{Simon2018ApJ...863...89S, Fritz2018A&A...619A.103F, Kallivayalil2018ApJ...867...19K, Massari2018A&A...620A.155M}.
Note that many of the literature measurements overlap. 
}
\label{fig:validation}
\end{figure*}

Our results in this section are summarized in Figure~\ref{fig:validation} and Table~\ref{table:spectra}.  Figure~\ref{fig:validation} displays all potential members ($p_i>0.1$) in proper motion space.  Each panel displays a different satellite and spectroscopically confirmed members are circled in magenta. 
Overlaid are other proper motion measurements from the literature \citep{Simon2018ApJ...863...89S, Fritz2018A&A...619A.103F, Kallivayalil2018ApJ...867...19K, Massari2018A&A...620A.155M}.  We list the number of spectroscopic members in \gaia and those recovered from our method in Table~\ref{table:spectra}. 
Overall, our model recovers most known members and we find excellent agreement between our results and the literature.

We now discuss the missed spectroscopic  members of each satellite in detail.
Recall here we consider stars members if $p_i >0.5$. 
As shown in Figure~\ref{fig:validation}, we miss a single spectroscopic member in Eridanus II as it falls outside of our of color-magnitude selection and fails the astrometric quality cut. 
As this galaxy has the largest stellar mass of the satellites considered, a wider color-magnitude diagram may be indicative of extended star formation or a larger spread in metallicity.
In Grus I we miss a spectroscopic member on the red side of our color-magnitude selection. 
From medium resolution spectroscopy \citet{Walker2016ApJ...819...53W} estimate the mean metallicity to be $\overline{[{\rm Fe/H}]}\approx -1.4$, significantly more metal rich than our isochrone. 
We miss two spectroscopic members in Reticulum II.  The first is the brightest spectroscopic member and is saturated in the DES DR1.  The second is faint ($G\sim20$) and $\sim3\sigma$ off in the $\mu_\alpha \cos{\delta}$ direction.

There are 48 spectroscopic members in the core and tidal tails of Tucana~III.  Of these spectroscopic members, only 22 are within 1 degree of the center and bright enough for \gaia catalog. We miss the most stars in this satellite, with only 15 of the spectroscopic members having $p_i>0.5$.  
This is mainly due to our choice of spatial model for Tucana~III; we did not include a model to account for the tidal tails of Tucana III. Specifically, four spectroscopic members have low membership ($p_i < 0.2$) due to their large radii; with $r> 45'$,  they are the members of the tidal tails. The other three disagree with the Tucana~III mean proper motion by 1-2$\sigma$ and  are assigned lower membership.

In Tucana~II, the RHB spectroscopic member is missing as RHB were not included in our CMD selection.  
Of the six satellites we validate our measurement with, all results agree with the literature except for Tucana~II measurement in \citet{Kallivayalil2018ApJ...867...19K}.  Tucana~II is $\sim3-\sigma$ off in the $\mu_{\delta}$ direction.  We note that our Tucana~II measurement is  consistent with other Tucana~II measurements \citep{Simon2018ApJ...863...89S, Fritz2018A&A...619A.103F}.  More spectroscopic data of Tucana~II is required to validate membership and  understand this discrepancy.

We note that while we could modify our selection for an individual satellite to increase the recovery rate of spectroscopic members, we cannot do this for satellites without spectroscopic members.
For consistency, we  explore identical setups for each satellite.
As discussed in Sec.~\ref{sec:data}, our color-magnitude selection is catered towards finding a pure sample of stars in a metal-poor ultra-faint satellite.
As our model includes a MW foreground model, any interlopers will be down-weighted and  our overall results are robust to interlopers. 
To further improve satellite and MW separation we could include radial velocities in our mixture model but our goal is to apply this method to satellites without spectroscopic follow-up.

In addition to recovering the spectroscopic members, we also find additional members in these objects. We list the number of new members with $p_i > 0.5$ in Table~\ref{table:spectra}. We suggest these stars should be prioritized in future  spectroscopic observations.

\subsection{New Measurements}\label{sec:new}

\begin{deluxetable*}{lr ccc rr r}
\tablecaption{Summary of Proper Motions}
\tablewidth{0pt}
\tablehead{
\colhead{Satellite} & \colhead{$\sum p_i$} & \colhead{${\rm N}$($p_i > 0.1$)} &  \colhead{${\rm N}_{\rm expected}$} & \colhead{${\rm N}(r<3\times r_h)$}  & \colhead{$\overline{\mu_{\alpha }\cos{\delta}}$} & \colhead{$\overline{\mu_{\delta}}$ } & \colhead{$C_{\overline{\pmra} \times \overline{\pmdec}}$ } 
}
\startdata
Eridanus II & 18.44 & 21 & $20 \pm 5$ & 16 & $0.16_{-0.24}^{+0.24}$  & $0.15_{-0.26}^{+0.26}$  & -0.27\\ 
Grus I & 8.23 & 9 & $8 \pm 3$ & 8 & $-0.25_{-0.16}^{+0.16}$  & $-0.47_{-0.23}^{+0.23}$  & 0.35\\ 
Horologium I & 17.06 & 20 & $16 \pm 4$ & 15 & $0.95_{-0.07}^{+0.07}$  & $-0.55_{-0.06}^{+0.06}$  & 0.29\\ 
Reticulum II & 48.08 & 67 & $43 \pm 6$ & 64 & $2.36_{-0.05}^{+0.05}$  & $-1.32_{-0.06}^{+0.06}$  & 0.18\\ 
Tucana II & 33.53 & 65 & $32 \pm 5$ & 105 & $0.91_{-0.06}^{+0.06}$  & $-1.16_{-0.08}^{+0.08}$  & -0.42\\ 
Tucana III & 44.83 & 74 & $42 \pm 6$ & 115 & $-0.03_{-0.04}^{+0.04}$  & $-1.65_{-0.04}^{+0.04}$  & -0.38
\\\hline\\[-0.6em]
Columba I & 7.19 & 11 & $6 \pm 3$ & 8 & $-0.02_{-0.27}^{+0.24}$  & $-0.04_{-0.30}^{+0.30}$  & -0.22\\ 
Eridanus III & 5.14 & 6 & $4 \pm 2$ & 4 & $1.06_{-0.24}^{+0.24}$  & $-0.48_{-0.24}^{+0.24}$  & -0.12\\ 
Grus II & 31.49 & 58 & $39 \pm 7$ & 68 & $0.43_{-0.08}^{+0.09}$  & $-1.45_{-0.15}^{+0.11}$  & 0.24\\ 
Phoenix II & 8.78 & 10 & $9 \pm 3$ & 9 & $0.49_{-0.10}^{+0.11}$  & $-1.03_{-0.12}^{+0.12}$  & -0.48\\ 
Pictor I & 7.05 & 7 & $8 \pm 3$ & 7 & $0.01_{-0.19}^{+0.19}$  & $0.20_{-0.25}^{+0.26}$  & -0.20\\ 
Reticulum III & 5.78 & 7 & $12 \pm 4$ & 8 & $-1.02_{-0.30}^{+0.32}$  & $-1.23_{-0.36}^{+0.40}$  & 0.39\\ 
Tucana IV & 16.42 & 32 & $29 \pm 5$ & 100 & $0.63_{-0.22}^{+0.25}$  & $-1.71_{-0.24}^{+0.20}$  & -0.30
\\\hline\\[-0.6em]
Horologium II & 4.13 & 5 & $8 \pm 3$ & 5 & $0.82_{-0.40}^{+0.45}$  & $-0.04_{-0.71}^{+0.62}$  & 0.07\\ 
Cetus II & 3.19 & 4 & $3 \pm 2$ & 3&  - & - & -\\ 
Indus I & 2.54 & 3 & $2 \pm 1$ & 2&  - & - & -\\ 
Tucana V & 0.01 & 0 & $4 \pm 2$ & 2&  - & - & -
\\\hline\\[-0.6em]
Indus II & 0.01 & 0 & $4 \pm 2$ & 4&  - & - & -\\ 
DES 1 & 1.00 & 1 & $3 \pm 2$ & 1&  - & - & -\\ 
DES J0225+0304 & 0.01 & 0 & $16 \pm 4$ & 2&  - & - & -
\enddata
\tablecomments{ Columns: satellite name, sum of membership probability ($\sum p_i$), number of stars with $p_i>0.1$, number of expected members  (${\rm N}_{\rm expected}$; see text), number of stars within three times the half-light radius (${\rm N}(r<3\times r_h)$), systemic proper motion in $\alpha \cos{\delta}$ direction, systemic proper motion in $\delta$ direction, and correlation between proper motion coordinates.
The satellites are order by: satellites with spectroscopic follow-up, new proper motions,  potential proper motions, and null results.
We note that we have not included the systematic error of 0.035 ${\rm mas \, yr^{-1}}$ \citep{Gaia_Helmi2018A&A...616A..12G}.\\
(a) For Columba I, we note that two probable members are likely to be more metal-rich stars. We therefore also calculated the proper motions with 5 members which give $\mu_\alpha \cos \delta =  0.08 \pm 0.21$ mas yr$^{-1}$, $\mu_\delta = -0.11 \pm 0.28$ mas yr$^{-1}$. See details in \S\ref{sec:color} and \S\ref{sec:review}.
}
\label{table:pm}
\end{deluxetable*}

We apply the same method to 14 DES satellites that have no spectroscopic information reported in the literature. We assess detection by verifying a cluster of members in proper motion space in the diagnostic plots (e.g. Figure ~\ref{fig:ret2}; see plots for other satellites in Appendix~\ref{appendix:figures}), the total number of members in the satellite ($\sum p_i$), and by examining the satellite posteriors (e.g. Figure ~\ref{fig:ret2_corner}; see posteriors for other satellites in Appendix~\ref{appendix:figures}). Our measurements of the systemic proper motions are summarized in Table~\ref{table:pm}.

The sum of the membership probability ($\sum p_i$) in each satellite,  is essentially the number of member stars in each satellite from our mixture model (listed in Table~\ref{table:pm}). As a comparison, we  calculated the expected number of member  stars (${\rm N}_{\rm expected}$) using the luminosity and distance of each satellite in Table~\ref{table:property}.
We assume the satellite has a \citet{Chabrier2001ApJ...554.1274C} initial mass function with an age of 12.5 Gyr and metallicity of $\feh = -2.2$, then we estimate the expected number of member stars with $G<20.9$\footnote{The magnitude cut is close to the faintest potential members in our sample.   We note that that \gaia DR2 is not complete to this magnitude and that the limiting magnitude may change between satellites.  } from 100 realizations of stellar populations randomly sampled using \code{ugali}\footnote{\url{https://github.com/DarkEnergySurvey/ugali}}. 
The expected number of members are given in Table~\ref{table:pm} in the ${\rm N}_{\rm expected}$ column. 
For most satellites, ${\rm N}_{\rm expected}$ and $\sum p_i$ are in excellent agreement (i.e. the mixture model agrees with the stellar population sampling) however Grus~II, Reticulum~III, Tucana~IV, and DES~J0225+0304 are all lower than expected. 
The number of stars within three times the half-light radius that pass all of our cuts (${\rm N}(r<3\times r_h)$) is given in Table~\ref{table:pm}.  For a stellar distribution that follows a Plummer density profile, $3\times r_h$ includes 90\% of the stars.  Comparing this column to $\sum p_i$ shows that in some satellites our cuts are enough to identify most members (e.g., Grus~I, Pictor~I) whereas in other satellites the mixture model is necessary to remove MW interlopers (e.g. Grus~II, Tucana~II).

We are able to measure the systemic proper motion of the following seven satellites which lack extensive spectrosopic follow-up and were not used to validate our method: Columba~I, Eridanus~III, Grus~II, Phoenix~II, Pictor~I, Reticulum~III, and Tucana~IV. 
The satellites Eridanus~III and Horologium~II lie at the border of what we consider a measurement of the systemic proper motion; in both cases the `detection' would be dependent on a single red giant branch star and two blue horizontal branch stars.  While interlopers are rare at blue colors we only claim a signal in Eridanus III as the stars are brighter and more tightly clustered in proper motion space and the posteriors of Horologuim~II are not well constrained; we leave Horologium~II as a plausible measurement. We review the seven satellites with new measurements individually in~\S\ref{sec:review}.

In additional to Horologium~II, we are not able to conclusively determine the proper motions of Cetus~II, or  Kim~2\footnote{We note that this satellite was later independently discovered by two groups \citep{Koposov2015ApJ...805..130K, Bechtol2015ApJ...807...50B} working with DES data and both referred to as Indus~I. As it was first discovered by \citet{Kim2015ApJ...803...63K} and they were the first to consider it a globular cluster we have denoted it as Kim~2 throughout the paper.}.  
The satellite posteriors are not well constrained; this is due to the low number of inferred members (generally a `bright' star plus a few very faint stars).
As the number of expected members was low in these satellites, it is not surprising to see the ambiguous result.
We further discuss these satellites in  Appendix~\ref{appendix:nodetect} and we do not claim  measurements of the systemic proper motion for these satellites. 
However, the potential members are excellent targets for future radial velocity measurements.

We are not able to determine the systemic proper motion of DES~1, DES~J0225+0304,  Indus~II, or Tucana~V. 
The likelihood fit returns zero signal (i.e. $\sum p_i\approx0$). There are two possibilities for this non-detection. First, the satellites could be more metal-rich and therefore our CMD cut removes some  members. To test this, we explored a `metal-rich' isochrone ($\overline{[Fe/H]}\approx -1.5$, age=10 Gyr) but were still not able to locate any members.  Second, due to the low luminosity of these satellites, the number of expected members brighter than the \gaia magnitude limit is minimal. 
This is true for for DES~1 and Indus~II.  As Poisson noise is expected for `bright' members, it is not  unexpected to  see the no detection for a couple of faint satellites.  However, given that ${\rm N}_{\rm expected}=16$ for DES~J0225+0304, it is surprising to have no detection for the satellite. 
As DES~J0225+0304 is located within the Sagittarius stream it is possible that the overdensity is driven by the stellar stream.  
Within the central region of DES~J0225+0304 ($2\times r_h$), only  six stars are left after applying much looser isochrone cuts.
As noted in \citet{Luque2017MNRAS.468...97L} the age, metallicity, and distance of the stellar candidate and the Sagittarius stream overlap and this candidate may be a false positive \citep[see][]{Conn2018ApJ...852...68C, Conn2018ApJ...857...70C}. 
Deeper imaging is necessary to further explore the nature of DES~J0225+0304. 
We note that as described in \S\ref{sec:data}, we applied an escape velocity cut to remove nearby stars with large proper motions, as the MW hypervelocity stars will slightly affect the inferred MW parameters. In principle the satellite may not be bound to the MW and the escape velocity cut  will remove all members. To test this we examined all stars at small radius without the  escape velocity cut to check for members. No  satellites have an  overdensity of `hypervelocity'  stars and we conclude that this cut did not lead us to miss the signal from any of the satellites.

\section{Discussion and Conclusion}
\label{sec:discussion}
\subsection{Comparison to Predicted Dynamics and Kinematics}\label{sec:lmc}

\begin{figure*}[th!]
\plotone{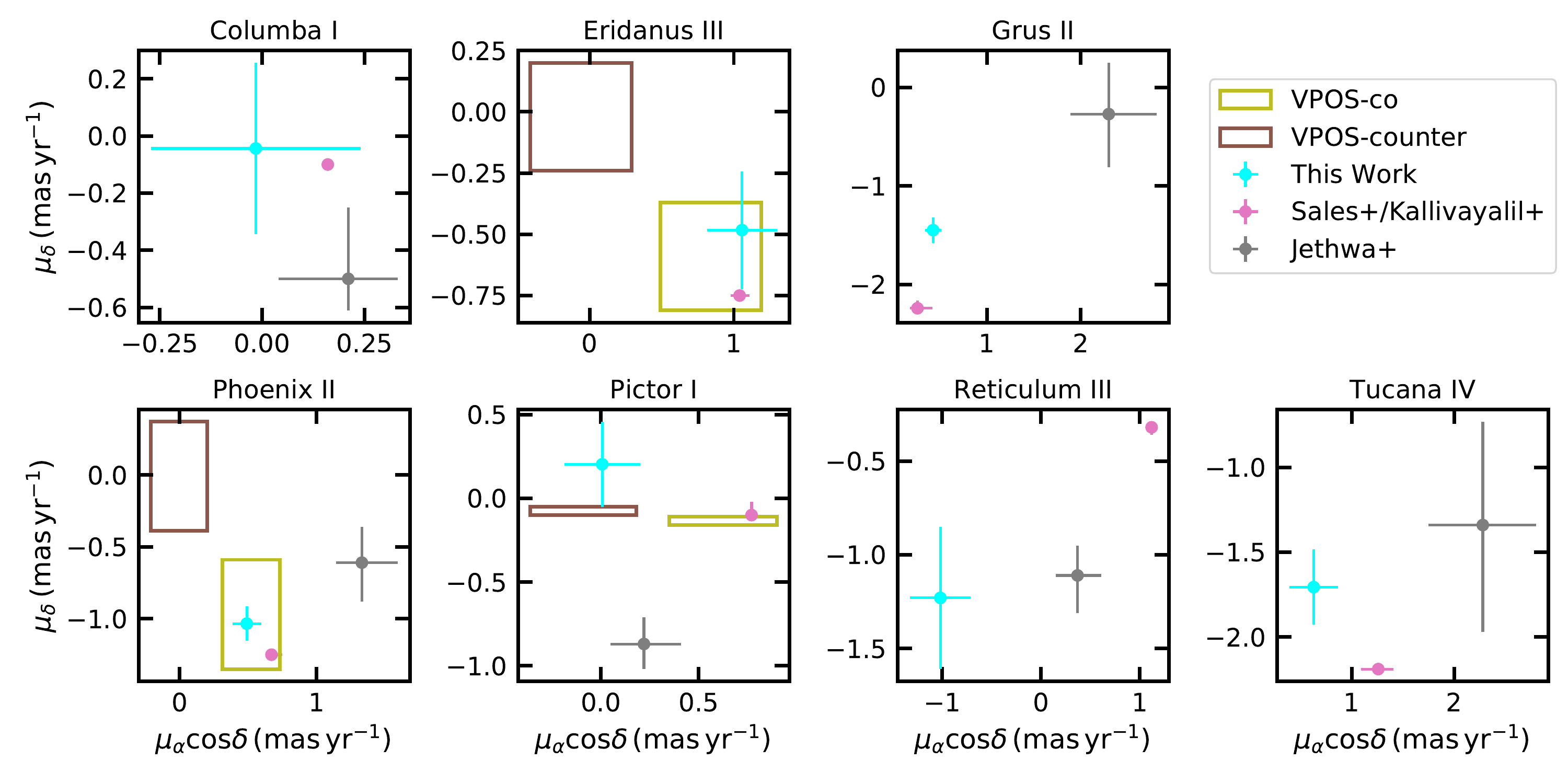}
\caption{Comparison of our proper motion results to predictions of LMC satellites infall models \citep{Jethwa2016MNRAS.461.2212J, Sales2017MNRAS.465.1879S, Kallivayalil2018ApJ...867...19K} and to predictions of satellites co- and counter-orbiting  the vast polar structure (VPOS) of the MW \citep{Pawlowski2015MNRAS.453.1047P}.  }
\label{fig:model}
\end{figure*}

In this section, we compare our measurements of the  systemic proper motion with  predictions of LMC satellites dynamics and kinematic membership in the vast polar structure (VPOS). The summary of the model comparison is shown in Figure~\ref{fig:model}.

Due to their proximity to the LMC, it was immediately suggested that some of the new DES satellites are (or were) associated with the LMC  \citep{Bechtol2015ApJ...807...50B, Koposov2015ApJ...805..130K, Drlica-Wagner2015ApJ...813..109D}.
The LMC is predicted to have many of its own satellites and there is even a dearth of LMC satellites with higher stellar masses ($\approx10^4 M_{\odot}$) \citep{Dooley2017MNRAS.472.1060D}. 
More detailed analytic modeling and cosmological N-body simulations suggest that  several of the new DES satellites were accreted by the MW with the LMC \citep{Deason2015MNRAS.453.3568D, Yozin2015MNRAS.453.2302Y, Jethwa2016MNRAS.461.2212J, Sales2017MNRAS.465.1879S}.
In order to confirm the association of a satellite with the LMC, full phase space knowledge and orbit modeling is required.
Our results provide part of the kinematic input to test these predictions.

There are two independent analyses that have made predictions for the radial velocities and proper motions for DES satellites assuming an LMC association\footnote{One of the analysis is spread over two  papers   \citep{Sales2017MNRAS.465.1879S, Kallivayalil2018ApJ...867...19K}.  \citet{Sales2017MNRAS.465.1879S} provides the theoretical framework and discuss the N-Body simulation  while \citet{Kallivayalil2018ApJ...867...19K} provides the observational counterpart and proper motion predictions. } \citep{Jethwa2016MNRAS.461.2212J, Sales2017MNRAS.465.1879S, Kallivayalil2018ApJ...867...19K}. 
\citet{Jethwa2016MNRAS.461.2212J} consider the distribution of LMC satellites after simulating the accretion of a LMC analog into a MW halo.  They perform multiple simulations, varying the masses of both the LMC and MW. 
\citet{Sales2017MNRAS.465.1879S, Kallivayalil2018ApJ...867...19K} examine the accretion of a LMC analog in a cosmological simulation.
In Figure~\ref{fig:model}, systemic proper motion predictions from both models are compared to our results.
Based on the \citet{Kallivayalil2018ApJ...867...19K} models, our proper motion results suggest that  Eridanus III, and Phoenix II are (or were) associated with the LMC.  
In fact, \citet{Kallivayalil2018ApJ...867...19K} find an overdensity of stars in Phoenix II  with proper motions consistent with their prediction and suggest that these stars are Phoenix II members.  
None of the predictions from \citet{Jethwa2016MNRAS.461.2212J} models agree with our measurements.
To further confirm any association with the LMC for these satellites, 
the radial velocity is required in addition to orbit modeling with a LMC potential.
We note that the two models have quite different predictions for the proper motions of the seven satellites.  This may be due to the setup of the simulation (cosmological versus isolated) or the choices in mass for each LMC, SMC, and MW components \citep[only][included the SMC]{Jethwa2016MNRAS.461.2212J}.  Moreover, there may be subtle differences based on how the simulations were transformed into the observed frame and local standard of rest \citep{Fritz2018arXiv180507350F}.

Another peculiarity in the distribution of the MW satellites is the so-called vast polar structure (VPOS).  The VPOS is a planar structure of satellite galaxies and distant globular clusters that is roughly perpendicular ($\theta\sim82^{\circ}$) to the MW disk \citep{Pawlowski2018MPLA...3330004P}. 
New proper motions from \gaia have already confirmed some satellites are consistent with  membership in the VPOS \citep{Simon2018ApJ...863...89S,Fritz2018A&A...619A.103F}.
The systemic proper motion predictions \citep{Pawlowski2015MNRAS.453.1047P} of satellites co-orbiting (VPOS-co) and counter-orbiting (VPOS-counter) are additionally included in Figure~\ref{fig:model}.
These predictions are represented as boxes, if the systemic proper motion is anywhere within the boxes the satellite is consistent with VPOS membership. We find Eridanus III and Phoenix II are consistent with co-orbiting, while Pictor I is consistent at $\sim1\sigma$ with counter-orbiting.  The other 4 satellites do not have published predictions.
We note that Eridanus III and Pictor I are spatially consistent with membership while Phoenix II is $1\sigma$ away from the plane.  
Of ultra-faints with new proper motions, 16 have been consistent with VPOS membership while 6 are not \citep{Fritz2018A&A...619A.103F}.  
Similar to the LMC model comparisons, further confirmation of VPOS  membership   requires radial velocities.

\subsection{Metallicity with Color-Color Diagram}\label{sec:color}

\begin{figure*}[th!]
\plotone{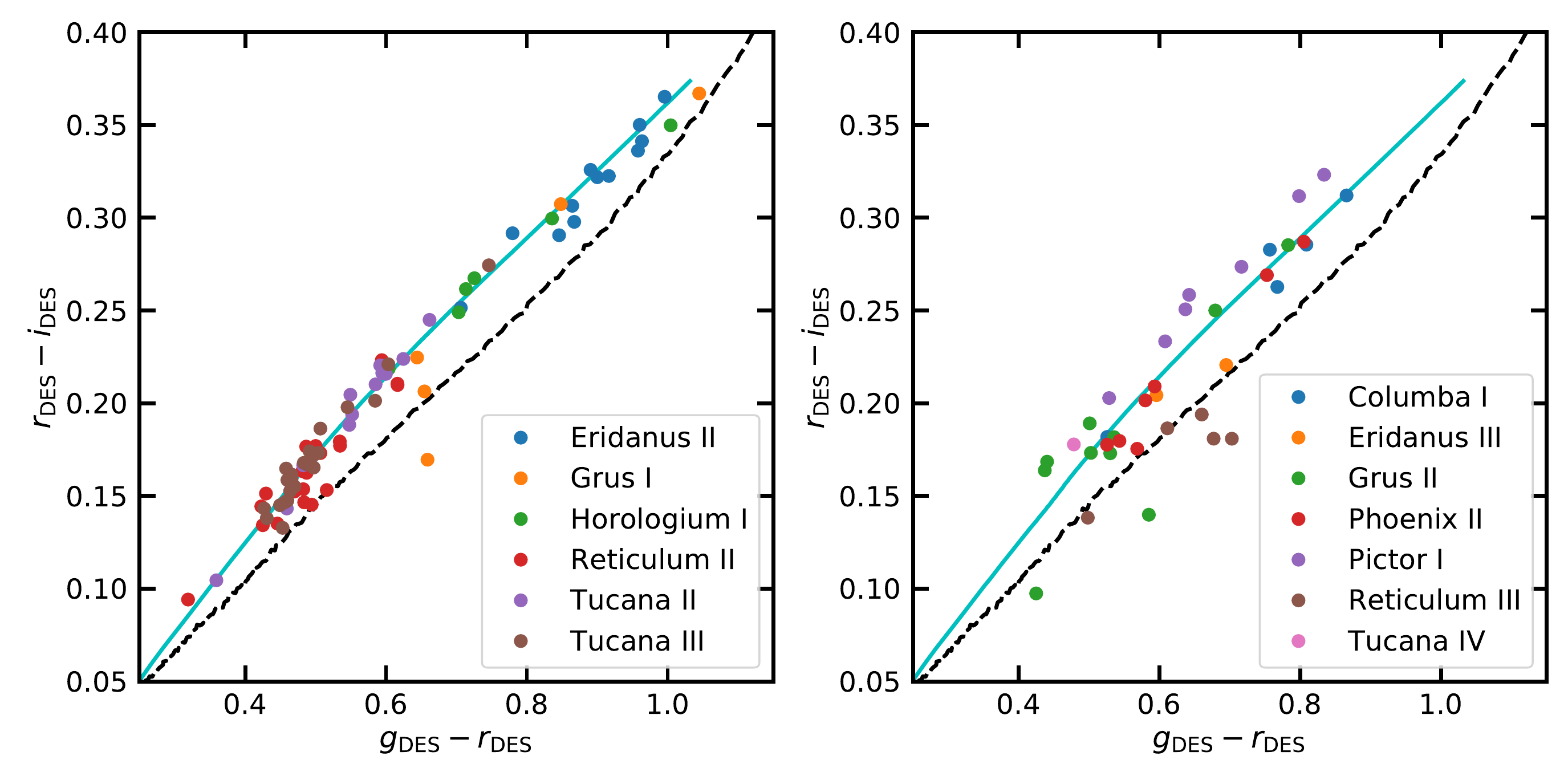}
\caption{$g-r$ vs $r-i$ color-color diagrams for stars in the DES satellites. {\bf Left:} Spectroscopic members in the DES DR1 $\times$ \gaia catalog of the six DES satellites with spectroscopic follow-up (see Table~\ref{table:spectra} for citations). Excluding Grus I all satellites where found to be extremely metal-poor ($\overline{[{\rm Fe/H}]}\lesssim -2.0$) from the spectroscopic follow-up. Correspondingly, the spectroscopic members in these satellites (excluding Grus I) all lie above the DES empirical stellar locus (dashed black line) and can be well described by a Dotter isochrone with a metal-poor population ($\feh = -2.2$; cyan line) . {\bf Right:} High probability members ($p_i > 0.8$) of the seven satellites with new proper motion measurements: Columba I, Eridanus III, Grus II, Phoenix II, Pictor I, Reticulum III, and Tucana IV.  Most of these proper motion members in new satellites are predicted to be metal-poor with the exception of Reticulum III, where all of the members are below the stellar locus.  }
\label{fig:color_color}
\end{figure*}

The DES photometry is precise enough to identify metal-poor RGB stars in a $g-r$ vs $r-i$ color-color diagram with color range of around $0.4 \lesssim g-r \lesssim 0.8$, first shown in the spectroscopic follow-up study of the Tucana~III stream in \citet{Li2018ApJ...866...22L}.
In the color range $0.4 \lesssim g-r \lesssim 0.8$, at a given $r-i$ color metal poor stars are bluer  relative to metal rich stars in  $g-r$ color.  
This is a similar effect to the traditional ultraviolet excess or line-blanketing effect \citep[ e.g.,][]{Wildey1962ApJ...135...94W, Sandage1969ApJ...158.1115S} as seen in the SDSS $u-g$ vs $g-r$ diagram~\citep{Ivezic2008ApJ...684..287I}.


To further verify that  metal-poor RGB stars can be identified, we examined all spectroscopically confirmed members in the DES satellites (see Table~\ref{table:spectra} for references), as shown in the left hand panel of Figure~\ref{fig:color_color}.  At $g-r \gtrsim 0.35$, all spectroscopic members are found above the empirical stellar locus\footnote{The empirical stellar locus is constructed as the median $r-i$ color at every $g-r$ color bin using dereddened DES photometry with $16 \lesssim r \lesssim 21$ (sampled over the full survey footprint). See more details in~\citet{Li2018ApJ...866...22L}.} except for Grus I which is more metal rich \citep{Walker2016ApJ...819...53W}. 
This further confirms the correlation between the metallicity and  DES stellar colors.
In the right hand panel we show a similar diagram for stars with high membership probability in the mixture model ($p_i>0.8$) for the seven satellites with new proper motion measurements. 
The candidate selection is not biased to only include stars above the empirical stellar locus and our mixture model does not consider color in the fit. 
Interestingly, the members are preferentially located above the empirical stellar locus,  suggesting the metal-poor nature of these satellites.  
The only exception is Reticulum~III, where all RGB members lie below the empirical stellar locus line.

Though the colors of  stars are not used in the mixture model,  the locations of the stars in the color-color diagram will help us assess their membership  when we review  individual satellites in \S\ref{sec:review}.

\subsection{Review of Individual Galaxies}\label{sec:review}

We discuss the seven satellites with systemic proper motion measurements that we did not sure to validate of measurements in this section.  The seven satellites are: Columba~I, Eridanus~III, Grus~II, Phoenix~II, Pictor~I, Reticulum~III, and Tucana~IV.  
This is the first proper motion measurement of Eridanus~III and Pictor~I.
We discuss star-by-star comparisons with  \citet{Fritz2018arXiv180507350F} that identified members with VLT/FLAMES/GIRAFE spectroscopy in three satellites (Columba~I, Reticulum~III, Phoenix~II).
The diagnostic and posterior plots for each satellite are available in Appendix~\ref{appendix:figures}.

We divide the potential members into three categories: high probability ($p_i > 0.8$), medium probability ($0.3 < p_i < 0.8$), and low probability ($0.1 < p_i < 0.3$). The full list of potential members ($p_i>0.1$) is in Table~\ref{tab:membership}\footnote{ Full membership files can be found at \url{https://github.com/apace7/gaia_cross_des_proper_motions}.} and includes  astrometry, photometry, proper motions and membership probabilities.

{\bf Columba~I} has 5  members with high probability and 2 with medium probability. However, both medium probability members are below the stellar locus and are therefore likely non-members. If we calculate the proper motion with the 5 high probability members, we get $\mu_\alpha \cos \delta =  0.08 \pm 0.21$ mas yr$^{-1}$, $\mu_\delta = -0.11 \pm 0.28$ mas yr$^{-1}$, which is still consistent to our results with the mixture model in Table~\ref{table:pm}.

\citet{Fritz2018arXiv180507350F} presents spectroscopic and astrometric data of Columba~I using \gaia DR2 and VLT/FLAMES spectroscopy.  They find 8 potential members, considering 4 as confident members and 4 as potential members.
Of these 8 stars, 2 confident and 3 possible members are in the \gaia DR2 catalog. 
They find $(\mu_{\alpha}\cos{\delta}, \mu_\delta) = (0.33 \pm 0.28, -0.38\pm 0.38) \pmunit$ which is consistent at the  $1.25-\sigma, 0.9\sigma$ level with our measurement.  The two stars they consider members we also consider members  ($p_i=0.98, 0.97$).  
Of the three stars they consider possible members, one has medium membership ($p_i=0.70$), while the other two are at the non-member/low-membership boarder  ($p_i=0.09, .011$).  
We note they prefer their confident membership proper motion measurement.
As our mixture model method considers all \gaia proper motion data their identified members are a subset of ours.

{\bf Eridanus III} has the lowest significance with what we consider a `detection'.  It has 4  members with high probability (including 2 BHBs) and 2 with medium probability. The confidence of our detection is mainly based on the agreement in proper motion space for the brightest RGB and 2 BHB stars and the constrained satellite parameters in the posterior distribution. Interestingly, the proper motion overlaps with the  \citet{Kallivayalil2018ApJ...867...19K} LMC satellite accretion models and as a satellite co-orbiting in the VPOS \citep{Pawlowski2015MNRAS.453.1047P}. 
Of the satellites with new measurements it is the only satellite considered to be a star cluster \citep{Luque2018MNRAS.478.2006L, Conn2018ApJ...852...68C}.

{\bf Grus II} is relative bright and nearby,  therefore it has the most potential members among the satellites with new proper motion measurements.
It has 11 RGB and 3 BHB members with high probability. 
For Grus II, \citet{Massari2018A&A...620A.155M} find $(\mu_{\alpha}\cos{\delta}, \mu_\delta) = (0.37 \pm 0.07, -1.33\pm 0.08) \pmunit$ which is consistent with our result at the $1-\sigma$ level.
We find the total membership to be $\sum p_i = 32$, which is smaller than expected from  the stellar population simulations (${\rm N_{expected}}=39\pm7$) whereas
 \citet{Massari2018A&A...620A.155M} find 45 members. 
We suspect our results are driven by the partial overlap in  proper motion space of Grus II and the MW foreground.
Almost  all stars at larger radii ($r>1.5\times r_h$) are considered MW members.  
However, we still find that the stars follow the Plummer distribution; for stars with $p_i>0.5$, 16 are within the half-light radius and 12 are outside of it.
Due to the large number of members near the center of the satellite we can still successfully recover the systemic proper motion of the satellite.
To further disentangle Grus II and the MW foreground, radial velocities are required.

{\bf Phoenix II} has 9 members with high probability, 2 of which are BHBs, and the remaining 7 RGB members are all above the empirical stellar locus and are therefore likely to be metal-poor members. 
The systemic proper motion of Phoenix II agrees with the accretion models of \citet{Sales2017MNRAS.465.1879S, Kallivayalil2018ApJ...867...19K} and is candidate for a LMC satellite.
\citet{Kallivayalil2018ApJ...867...19K} starting from their model predictions, used a clustering algorithm to identify 4 stars near Phoenix II with a similar systemic proper motion of $(\mu_{\alpha}\cos{\delta}, \mu_\delta) = (0.54 \pm 0.10, -1.17\pm 0.12) \pmunit$.  Our model identifies additional members but it is consistent with their result.

\citet{Fritz2018arXiv180507350F} find six members from their combination of  \gaia DR2 and VLT/FLAMES spectroscopy (one is considered a possible member).  They find $(\mu_{\alpha}\cos{\delta}, \mu_\delta) = (0.50 \pm 0.12, -1.16\pm 0.14) \pmunit$ which is consistent with our measurement. 
Two of these stars  fall outside of  our color-magnitude selection (one of these stars is the possible member), but they are likely members based on their proper motions and velocities.  One star is located on the red horizontal branch (a potential RR Lyrae) and we specifically excluded this region from our analysis. 
The other star is much redder than the isochrone ($\Delta g-r \approx 0.14$); its red color may be due to odd abundances \citep[e.g. Carbon enhanced star][]{Koposov2018MNRAS.479.5343K}.
The four other stars are all considered members in our analysis ($p_i=0.99, 0.92, 0.82, 1.00$).

{\bf Pictor I\footnote{We note that in both discovery papers, \citet{Bechtol2015ApJ...807...50B} and \citet{Koposov2015ApJ...805..130K}, this satellite (DES J0443.8-5017) was incorrectly referred to as Pictoris I.  This mistake was due to latin case.  Pictoris is the latin genitive of Pictor, and is used to refer to stars in  constellations. }} has 7 RGB members with high probability. The 5 brighter stars form a tight cluster in proper motion space, indicating that this is the likely signal of Pictor I. All stars are found above the stellar locus suggesting that this is a metal-poor satellite.

{\bf Reticulum III} has 5 members with high probability and 1 member with medium probability (which is also a BHB star). Interestingly, all 5 high probability members lie below the stellar locus. As these 5 stars are clumped in  proper motion space, it is unambiguous that these stars are  Reticulum III members. The location in color-color space indicates that it might have a relatively more more-metal rich population.
Based on its size $r_{1/2}\approx 60 \, {\rm pc}$ and luminosity it is not expected to be a star cluster \citep{Drlica-Wagner2015ApJ...809L...4D}. 
From the stellar mass-metallicity relation \citep{Kirby2013ApJ...779..102K}, a dwarf galaxy of it's luminosity is expected to have $\overline{\feh}\approx-2.5$. 
It is possible that it is a remnant of a much more massive satellite, similar to what has been suggested for Segue 2 \citep{Kirby2013ApJ...770...16K}.
Alternatively, these stars may still be metal-poor but have some interesting chemical composition  \citep[e.g. carbon enhancement][]{Koposov2018MNRAS.479.5343K}. 
Spectroscopic follow-up is needed to conclusively provide information on their metallicity and chemical abundance. 
As mentioned in \S\ref{sec:new}, we find fewer members in Reticulum III from our mixture model ($\sum p_i = 5.78$) than expected from the luminosity estimation ($N_{\rm expected} = 12 \pm 4$; Table~\ref{table:pm}). If Reticulum III is more metal-rich then the metal-poor isochrone, the CMD selection may have missed some members.

\citet{Fritz2018arXiv180507350F} find three members from their combination of  \gaia DR2 and VLT/FLAMES spectroscopy.  They find $(\mu_{\alpha}\cos{\delta}, \mu_\delta) = (-0.39 \pm 0.53, -0.32\pm 0.63) \pmunit$ which is $1.2-\sigma, 1.4\sigma$ away from our measurement.  Their faintest member we consider a non-member as it does not satisfy our color-magnitude selection while the other two we consider members  ($p_i=0.99, 0.98$).  As our method is able to identify all members in the  \gaia DR2 data their measurement is a subset of the total \gaia DR2 sample.  
We do note that they find all three stars to be very metal poor ([Fe/H]$<-2.5$) from calcium triplet measurements  in contrast to our findings with the g-r vs r-i color-color diagram.

{\bf Tucana~IV} has few members with high probability but many with medium probability. 
Similar to Grus~II, we find that the MW foreground overlaps with the Tucana~IV proper motion and makes recovery of the satellite systemic proper motion difficult.  Because of its low surface brightness and large size  the mixture model has difficulty separating nearby members from the MW.  
As mentioned in \S\ref{sec:new}, the number of members from the likelihood fit ($\sum p_i = 16$) is much smaller than the expected number of members based on its luminosity and distance ($N_{\rm expected} = 29 \pm 5$; Table~\ref{table:pm}).
In fact, our method has trouble identifying any high probability members outside the half-light radius; only 3 stars outside the half-light radius have $p_i>0.5$.
Increasing the complexity of foreground model does not increase the number of satellite  stars found (see Appendix~\ref{appendix:mwmodel} for a full description and discussion of the two component MW modeling). 
\citet{Massari2018A&A...620A.155M} find $(\mu_{\alpha}\cos{\delta}, \mu_\delta) = (0.75 \pm 0.06, -1.70\pm 0.08) \pmunit$ while we find a similar result with much larger errors $(\mu_{\alpha}\cos{\delta}, \mu_\delta) = (0.63 \pm 0.24, -1.71 \pm 0.22) \pmunit$.  We also note that the posterior distribution for the satellite parameters contain non-Gaussian tails.  If this satellite was less luminous we likely would not have been able to identify it.  



\subsection{Conclusions}\label{sec:conclusion}

We have presented a method for determining the proper motion of ultra-faint satellites utilizing \gaia DR2 proper motions and DES DR1 photometry. 
Our mixture model successfully recovered the systemic proper motion of the six DES satellites with spectroscopic members as a validation of our method.
We were  able to measure the systemic proper motion of seven additional satellites: Columba~I, Eridanus~III, Grus~II, Phoenix~II, Pictor~I, Reticulum~III, and Tucana~IV, five of which are new measurements.
We found that Eridanus~III and Phoenix~II are consistent with the dynamics of LMC satellites but additional verification with the satellite's systemic radial velocity and orbit modeling is required.
Of the three satellites with vast polar structure proper motion predictions, all three are consistent with membership. Eridanus~III and Phoenix~II are co-orbiting while is Pictor~I counter-orbiting.  
With DES photometry most of the new satellites are predicted to be extremely metal-poor ($\feh \lesssim -2$); the exception is Reticulum~III, which we predict to be more metal rich ($\overline{\feh} \sim -1.5$) than most ultra-faint satellite. 

Although the main motivation of this work was to measure the systemic proper motion of the satellites in the Milky Way, as a byproduct, our study also provides a list of satellite members based on their photometry and proper motion. In Table~\ref{tab:membership}, we list all stars with membership probability $p_i > 0.1$ in 17 satellites. For the 6 satellites with spectroscopic follow-up (i.e. Horologium I, Reticulum II, Eridanus II, Grus I, Tucana II, Tucana III), we find additional  members in each satellite (see \S\ref{sec:valid} and Table~\ref{table:spectra}); these  members are relatively bright and are excellent targets to increase the spectroscopic sample sizes to improve the dynamical mass measurements. For the 7 satellites that without any or extensive spectroscopic follow-up but with systemic proper motion measurements (i.e. Columba~I, Eridanus~III, Grus~II, Phoenix~II, Pictor~I, Reticulum~III, and Tucana~IV), the list of members can aid target selection in future spectroscopic follow-up. We are not able to conclusively determine the proper motions of the remaining 3 satellites (i.e. Cetus~II, Kim~2,  and Horologium~II); however, we suggest that spectroscopic follow-up/and membership confirmation  of these potential members could determine the systemic proper motion of the satellites. We note that due to our selection in ~\S\ref{sec:data}, we may miss some stars in each satellite. Therefore, the list we provide is not meant to be complete, but should be considered with higher priority for  target selection of spectroscopic follow-up observations.

\acknowledgements{

We thank Josh Simon, Jen Marshall, Louie Strigari, Alex Drlica-Wagner, and Keith Bechtol for helpful discussions and comments.
We thank Alex Drlica-Wagner and Sergey Kosopov for their help on accessing DES DR1 and \gaia DR2.
We thank the referee for helpful comments and suggestions that improved the paper.
A.B.P. acknowledges generous support from the George P. and Cynthia Woods Institute for Fundamental Physics and Astronomy at Texas A\&M University. T.S.L in particular thanks the 2018 KITP program ``The Small-Scale Structure of Cold(?) Dark Matter'' that helped frame the perspective of this paper. This research was supported in part by the National Science Foundation under Grant No. NSF PHY-1748958.

This work has made use of data from the European Space Agency (ESA)
mission {\it Gaia} (\url{https://www.cosmos.esa.int/gaia}), processed by the {\it Gaia} Data Processing and Analysis Consortium (DPAC,
\url{https://www.cosmos.esa.int/web/gaia/dpac/consortium}). Funding
for the DPAC has been provided by national institutions, in particular
the institutions participating in the {\it Gaia} Multilateral Agreement.

This research has made use of NASA's Astrophysics Data System Bibliographic Services.

This project used public archival data from the Dark Energy Survey (DES). Funding for the DES Projects has been provided by the U.S. Department of Energy, the U.S. National Science Foundation, the Ministry of Science and Education of Spain, the Science and Technology FacilitiesCouncil of the United Kingdom, the Higher Education Funding Council for England, the National Center for Supercomputing Applications at the University of Illinois at Urbana-Champaign, the Kavli Institute of Cosmological Physics at the University of Chicago, the Center for Cosmology and Astro-Particle Physics at the Ohio State University, the Mitchell Institute for Fundamental Physics and Astronomy at Texas A\&M University, Financiadora de Estudos e Projetos, Funda{\c c}{\~a}o Carlos Chagas Filho de Amparo {\`a} Pesquisa do Estado do Rio de Janeiro, Conselho Nacional de Desenvolvimento Cient{\'i}fico e Tecnol{\'o}gico and the Minist{\'e}rio da Ci{\^e}ncia, Tecnologia e Inova{\c c}{\~a}o, the Deutsche Forschungsgemeinschaft, and the Collaborating Institutions in the Dark Energy Survey.
The Collaborating Institutions are Argonne National Laboratory, the University of California at Santa Cruz, the University of Cambridge, Centro de Investigaciones Energ{\'e}ticas, Medioambientales y Tecnol{\'o}gicas-Madrid, the University of Chicago, University College London, the DES-Brazil Consortium, the University of Edinburgh, the Eidgen{\"o}ssische Technische Hochschule (ETH) Z{\"u}rich,  Fermi National Accelerator Laboratory, the University of Illinois at Urbana-Champaign, the Institut de Ci{\`e}ncies de l'Espai (IEEC/CSIC), the Institut de F{\'i}sica d'Altes Energies, Lawrence Berkeley National Laboratory, the Ludwig-Maximilians Universit{\"a}t M{\"u}nchen and the associated Excellence Cluster Universe, the University of Michigan, the National Optical Astronomy Observatory, the University of Nottingham, The Ohio State University, the OzDES Membership Consortium, the University of Pennsylvania, the University of Portsmouth, SLAC National Accelerator Laboratory, Stanford University, the University of Sussex, and Texas A\&M University.
Based in part on observations at Cerro Tololo Inter-American Observatory, National Optical Astronomy Observatory, which is operated by the Association of Universities for Research in Astronomy (AURA) under a cooperative agreement with the National Science Foundation.
}

{\it Facilities:} 
 \facility{Gaia}
 \facility{DES}
 
{\it Software:} 
\code{astropy} \citep{Astropy2013}, 
\code{corner.py} \citep{corner}, 
\code{matplotlib} \citep{matplotlib}, 
\code{numpy} \citep{numpy}, 
\code{PyGaia}\footnote{\url{https://github.com/agabrown/PyGaia}},
\code{scipy} \citep{scipy}, 
\code{ugali} \citep{Bechtol2015ApJ...807...50B}\footnote{\url{https://github.com/DarkEnergySurvey/ugali}}.
\code{galpy} \citep{Bovy2015ApJS..216...29B}\footnote{\url{https://github.com/jobovy/galpy}}, 

\bibliographystyle{apj}
\bibliography{temp}{}

\begin{thebibliography}{}
\expandafter\ifx\csname natexlab\endcsname\relax\def\natexlab#1{#1}\fi

\bibitem[{{Astropy Collaboration} {et~al.}(2013){Astropy Collaboration},
  {Robitaille}, {Tollerud}, {Greenfield}, {Droettboom}, {Bray}, {Aldcroft},
  {Davis}, {Ginsburg}, {Price-Whelan}, {Kerzendorf}, {Conley}, {Crighton},
  {Barbary}, {Muna}, {Ferguson}, {Grollier}, {Parikh}, {Nair}, {Unther},
  {Deil}, {Woillez}, {Conseil}, {Kramer}, {Turner}, {Singer}, {Fox}, {Weaver},
  {Zabalza}, {Edwards}, {Azalee Bostroem}, {Burke}, {Casey}, {Crawford},
  {Dencheva}, {Ely}, {Jenness}, {Labrie}, {Lim}, {Pierfederici}, {Pontzen},
  {Ptak}, {Refsdal}, {Servillat}, \& {Streicher}}]{Astropy2013}
{Astropy Collaboration}, {Robitaille}, T.~P., {Tollerud}, E.~J., {et~al.} 2013,
  \aap, 558, A33

\bibitem[{{Bechtol} {et~al.}(2015){Bechtol}, {Drlica-Wagner}, {Balbinot},
  {Pieres}, {Simon}, {Yanny}, {Santiago}, {Wechsler}, {Frieman}, {Walker},
  {Williams}, {Rozo}, {Rykoff}, {Queiroz}, {Luque}, {Benoit-L{\'e}vy},
  {Tucker}, {Sevilla}, {Gruendl}, {da Costa}, {Fausti Neto}, {Maia}, {Abbott},
  {Allam}, {Armstrong}, {Bauer}, {Bernstein}, {Bernstein}, {Bertin}, {Brooks},
  {Buckley-Geer}, {Burke}, {Carnero Rosell}, {Castander}, {Covarrubias},
  {DAndrea}, {DePoy}, {Desai}, {Diehl}, {Eifler}, {Estrada}, {Evrard},
  {Fernandez}, {Finley}, {Flaugher}, {Gaztanaga}, {Gerdes}, {Girardi},
  {Gladders}, {Gruen}, {Gutierrez}, {Hao}, {Honscheid}, {Jain}, {James},
  {Kent}, {Kron}, {Kuehn}, {Kuropatkin}, {Lahav}, {Li}, {Lin}, {Makler},
  {March}, {Marshall}, {Martini}, {Merritt}, {Miller}, {Miquel}, {Mohr},
  {Neilsen}, {Nichol}, {Nord}, {Ogando}, {Peoples}, {Petravick}, {Plazas},
  {Romer}, {Roodman}, {Sako}, {Sanchez}, {Scarpine}, {Schubnell}, {Smith},
  {Soares-Santos}, {Sobreira}, {Suchyta}, {Swanson}, {Tarle}, {Thaler},
  {Thomas}, {Wester}, {Zuntz}, \& {The DES
  Collaboration}}]{Bechtol2015ApJ...807...50B}
{Bechtol}, K., {Drlica-Wagner}, A., {Balbinot}, E., {et~al.} 2015, \apj, 807,
  50

\bibitem[{{Bernard} {et~al.}(2014){Bernard}, {Ferguson}, {Schlafly}, {Platais},
  {Bell}, {Martin}, {Rix}, {Slater}, {Burgett}, {Chambers}, {Draper}, {Hodapp},
  {Kaiser}, {Kudritzki}, {Magnier}, {Metcalfe}, {Tonry}, {Wainscoat}, \&
  {Waters}}]{Bernard2014MNRAS.442.2999B}
{Bernard}, E.~J., {Ferguson}, A.~M.~N., {Schlafly}, E.~F., {et~al.} 2014,
  \mnras, 442, 2999

\bibitem[{{Bovy}(2015)}]{Bovy2015ApJS..216...29B}
{Bovy}, J. 2015, \apjs, 216, 29

\bibitem[{{Carlin} {et~al.}(2017){Carlin}, {Sand}, {Mu{\~n}oz}, {Spekkens},
  {Willman}, {Crnojevi{\'c}}, {Forbes}, {Hargis}, {Kirby}, {Peter},
  {Romanowsky}, \& {Strader}}]{Carlin2017AJ....154..267C}
{Carlin}, J.~L., {Sand}, D.~J., {Mu{\~n}oz}, R.~R., {et~al.} 2017, \aj, 154,
  267

\bibitem[{{Chabrier}(2001)}]{Chabrier2001ApJ...554.1274C}
{Chabrier}, G. 2001, \apj, 554, 1274

\bibitem[{{Chiti} {et~al.}(2018){Chiti}, {Frebel}, {Ji}, {Jerjen}, {Kim}, \&
  {Norris}}]{Chiti2018ApJ...857...74C}
{Chiti}, A., {Frebel}, A., {Ji}, A.~P., {et~al.} 2018, \apj, 857, 74

\bibitem[{{Conn} {et~al.}(2018{\natexlab{a}}){Conn}, {Jerjen}, {Kim}, \&
  {Schirmer}}]{Conn2018ApJ...852...68C}
{Conn}, B.~C., {Jerjen}, H., {Kim}, D., \& {Schirmer}, M. 2018{\natexlab{a}},
  \apj, 852, 68

\bibitem[{{Conn} {et~al.}(2018{\natexlab{b}}){Conn}, {Jerjen}, {Kim}, \&
  {Schirmer}}]{Conn2018ApJ...857...70C}
---. 2018{\natexlab{b}}, \apj, 857, 70

\bibitem[{{Crnojevi{\'c}} {et~al.}(2016){Crnojevi{\'c}}, {Sand}, {Zaritsky},
  {Spekkens}, {Willman}, \& {Hargis}}]{Crnojevic2016ApJ...824L..14C}
{Crnojevi{\'c}}, D., {Sand}, D.~J., {Zaritsky}, D., {et~al.} 2016, \apjl, 824,
  L14

\bibitem[{{Deason} {et~al.}(2015){Deason}, {Wetzel}, {Garrison-Kimmel}, \&
  {Belokurov}}]{Deason2015MNRAS.453.3568D}
{Deason}, A.~J., {Wetzel}, A.~R., {Garrison-Kimmel}, S., \& {Belokurov}, V.
  2015, \mnras, 453, 3568

\bibitem[{{DES Collaboration}(2018)}]{DES2018arXiv180103181A}
{DES Collaboration}. 2018, ArXiv e-prints, arXiv:1801.03181

\bibitem[{{Dooley} {et~al.}(2017){Dooley}, {Peter}, {Carlin}, {Frebel},
  {Bechtol}, \& {Willman}}]{Dooley2017MNRAS.472.1060D}
{Dooley}, G.~A., {Peter}, A.~H.~G., {Carlin}, J.~L., {et~al.} 2017, \mnras,
  472, 1060

\bibitem[{{Dotter} {et~al.}(2008){Dotter}, {Chaboyer}, {Jevremovi{\'c}},
  {Kostov}, {Baron}, \& {Ferguson}}]{Dotter2008ApJS..178...89D}
{Dotter}, A., {Chaboyer}, B., {Jevremovi{\'c}}, D., {et~al.} 2008, \apjs, 178,
  89

\bibitem[{{Drlica-Wagner} {et~al.}(2015{\natexlab{a}}){Drlica-Wagner},
  {Bechtol}, {Rykoff}, {Luque}, {Queiroz}, {Mao}, {Wechsler}, {Simon},
  {Santiago}, {Yanny}, {Balbinot}, {Dodelson}, {Fausti Neto}, {James}, {Li},
  {Maia}, {Marshall}, {Pieres}, {Stringer}, {Walker}, {Abbott}, {Abdalla},
  {Allam}, {Benoit-L{\'e}vy}, {Bernstein}, {Bertin}, {Brooks}, {Buckley-Geer},
  {Burke}, {Carnero Rosell}, {Carrasco Kind}, {Carretero}, {Crocce}, {da
  Costa}, {Desai}, {Diehl}, {Dietrich}, {Doel}, {Eifler}, {Evrard}, {Finley},
  {Flaugher}, {Fosalba}, {Frieman}, {Gaztanaga}, {Gerdes}, {Gruen}, {Gruendl},
  {Gutierrez}, {Honscheid}, {Kuehn}, {Kuropatkin}, {Lahav}, {Martini},
  {Miquel}, {Nord}, {Ogando}, {Plazas}, {Reil}, {Roodman}, {Sako}, {Sanchez},
  {Scarpine}, {Schubnell}, {Sevilla-Noarbe}, {Smith}, {Soares-Santos},
  {Sobreira}, {Suchyta}, {Swanson}, {Tarle}, {Tucker}, {Vikram}, {Wester},
  {Zhang}, {Zuntz}, \& {DES Collaboration}}]{Drlica-Wagner2015ApJ...813..109D}
{Drlica-Wagner}, A., {Bechtol}, K., {Rykoff}, E.~S., {et~al.}
  2015{\natexlab{a}}, \apj, 813, 109

\bibitem[{{Drlica-Wagner} {et~al.}(2015{\natexlab{b}}){Drlica-Wagner},
  {Albert}, {Bechtol}, {Wood}, {Strigari}, {S{\'a}nchez-Conde}, {Baldini},
  {Essig}, {Cohen-Tanugi}, {Anderson}, {Bellazzini}, {Bloom}, {Caputo},
  {Cecchi}, {Charles}, {Chiang}, {de Angelis}, {Funk}, {Fusco}, {Gargano},
  {Giglietto}, {Giordano}, {Guiriec}, {Gustafsson}, {Kuss}, {Loparco},
  {Lubrano}, {Mirabal}, {Mizuno}, {Morselli}, {Ohsugi}, {Orlando}, {Persic},
  {Rain{\`o}}, {Sehgal}, {Spada}, {Suson}, {Zaharijas}, {Zimmer}, {Fermi-LAT
  Collaboration}, {Abbott}, {Allam}, {Balbinot}, {Bauer}, {Benoit-L{\'e}vy},
  {Bernstein}, {Bernstein}, {Bertin}, {Brooks}, {Buckley-Geer}, {Burke},
  {Carnero Rosell}, {Castander}, {Covarrubias}, {D'Andrea}, {da Costa},
  {DePoy}, {Desai}, {Diehl}, {Cunha}, {Eifler}, {Estrada}, {Evrard}, {Fausti
  Neto}, {Fernandez}, {Finley}, {Flaugher}, {Frieman}, {Gaztanaga}, {Gerdes},
  {Gruen}, {Gruendl}, {Gutierrez}, {Honscheid}, {Jain}, {James}, {Jeltema},
  {Kent}, {Kron}, {Kuehn}, {Kuropatkin}, {Lahav}, {Li}, {Luque}, {Maia},
  {Makler}, {March}, {Marshall}, {Martini}, {Merritt}, {Miller}, {Miquel},
  {Mohr}, {Neilsen}, {Nord}, {Ogando}, {Peoples}, {Petravick}, {Pieres},
  {Plazas}, {Queiroz}, {Romer}, {Roodman}, {Rykoff}, {Sako}, {Sanchez},
  {Santiago}, {Scarpine}, {Schubnell}, {Sevilla}, {Smith}, {Soares-Santos},
  {Sobreira}, {Suchyta}, {Swanson}, {Tarle}, {Thaler}, {Thomas}, {Tucker},
  {Walker}, {Wechsler}, {Wester}, {Williams}, {Yanny}, {Zuntz}, \& {DES
  Collaboration}}]{Drlica-Wagner2015ApJ...809L...4D}
{Drlica-Wagner}, A., {Albert}, A., {Bechtol}, K., {et~al.} 2015{\natexlab{b}},
  \apjl, 809, L4

\bibitem[{{Drlica-Wagner} {et~al.}(2016){Drlica-Wagner}, {Bechtol}, {Allam},
  {Tucker}, {Gruendl}, {Johnson}, {Walker}, {James}, {Nidever}, {Olsen},
  {Wechsler}, {Cioni}, {Conn}, {Kuehn}, {Li}, {Mao}, {Martin}, {Neilsen},
  {Noel}, {Pieres}, {Simon}, {Stringfellow}, {van der Marel}, \&
  {Yanny}}]{Drlica-Wagner2016ApJ...833L...5D}
{Drlica-Wagner}, A., {Bechtol}, K., {Allam}, S., {et~al.} 2016, \apjl, 833, L5

\bibitem[{{Feroz} \& {Hobson}(2008)}]{Feroz2008MNRAS.384..449F}
{Feroz}, F., \& {Hobson}, M.~P. 2008, \mnras, 384, 449

\bibitem[{{Feroz} {et~al.}(2009){Feroz}, {Hobson}, \&
  {Bridges}}]{Feroz2009MNRAS.398.1601F}
{Feroz}, F., {Hobson}, M.~P., \& {Bridges}, M. 2009, \mnras, 398, 1601

\bibitem[{{Fillingham} {et~al.}(2015){Fillingham}, {Cooper}, {Wheeler},
  {Garrison-Kimmel}, {Boylan-Kolchin}, \&
  {Bullock}}]{Fillingham2015MNRAS.454.2039F}
{Fillingham}, S.~P., {Cooper}, M.~C., {Wheeler}, C., {et~al.} 2015, \mnras,
  454, 2039

\bibitem[{{Fitzpatrick}(1999)}]{Fitzpatrick1999PASP..111...63F}
{Fitzpatrick}, E.~L. 1999, \pasp, 111, 63

\bibitem[{Foreman-Mackey(2016)}]{corner}
Foreman-Mackey, D. 2016, The Journal of Open Source Software, 24,
  doi:10.21105/joss.00024

\bibitem[{{Fritz} {et~al.}(2018{\natexlab{a}}){Fritz}, {Battaglia},
  {Pawlowski}, {Kallivayalil}, {van der Marel}, {Sohn}, {Brook}, \&
  {Besla}}]{Fritz2018A&A...619A.103F}
{Fritz}, T.~K., {Battaglia}, G., {Pawlowski}, M.~S., {et~al.}
  2018{\natexlab{a}}, \aap, 619, A103

\bibitem[{{Fritz} {et~al.}(2018{\natexlab{b}}){Fritz}, {Carrera}, \&
  {Battaglia}}]{Fritz2018arXiv180507350F}
{Fritz}, T.~K., {Carrera}, R., \& {Battaglia}, G. 2018{\natexlab{b}}, ArXiv
  e-prints, arXiv:1805.07350

\bibitem[{{Gaia Collaboration} {et~al.}(2018{\natexlab{a}}){Gaia
  Collaboration}, {Helmi}, {van Leeuwen}, {McMillan}, {Massari}, {Antoja},
  {Robin}, {Lindegren}, {Bastian}, {Arenou}, \&
  et~al.}]{Gaia_Helmi2018A&A...616A..12G}
{Gaia Collaboration}, {Helmi}, A., {van Leeuwen}, F., {et~al.}
  2018{\natexlab{a}}, \aap, 616, A12

\bibitem[{{Gaia Collaboration} {et~al.}(2018{\natexlab{b}}){Gaia
  Collaboration}, {Brown}, {Vallenari}, {Prusti}, {de Bruijne}, {Babusiaux},
  {Bailer-Jones}, {Biermann}, {Evans}, {Eyer}, \&
  et~al.}]{Gaia_Brown2018A&A...616A...1G}
{Gaia Collaboration}, {Brown}, A.~G.~A., {Vallenari}, A., {et~al.}
  2018{\natexlab{b}}, \aap, 616, A1

\bibitem[{{Homma} {et~al.}(2018){Homma}, {Chiba}, {Okamoto}, {Komiyama},
  {Tanaka}, {Tanaka}, {Ishigaki}, {Hayashi}, {Arimoto}, {Garmilla}, {Lupton},
  {Strauss}, {Miyazaki}, {Wang}, \& {Murayama}}]{Homma2018PASJ...70S..18H}
{Homma}, D., {Chiba}, M., {Okamoto}, S., {et~al.} 2018, \pasj, 70, S18

\bibitem[{Hunter(2007)}]{matplotlib}
Hunter, J.~D. 2007, Computing In Science \& Engineering, 9, 90

\bibitem[{{Ivezi{\'c}} {et~al.}(2008){Ivezi{\'c}}, {Sesar}, {Juri{\'c}},
  {Bond}, {Dalcanton}, {Rockosi}, {Yanny}, {Newberg}, {Beers}, {Allende
  Prieto}, {Wilhelm}, {Lee}, {Sivarani}, {Norris}, {Bailer-Jones}, {Re
  Fiorentin}, {Schlegel}, {Uomoto}, {Lupton}, {Knapp}, {Gunn}, {Covey}, {Allyn
  Smith}, {Miknaitis}, {Doi}, {Tanaka}, {Fukugita}, {Kent}, {Finkbeiner},
  {Munn}, {Pier}, {Quinn}, {Hawley}, {Anderson}, {Kiuchi}, {Chen}, {Bushong},
  {Sohi}, {Haggard}, {Kimball}, {Barentine}, {Brewington}, {Harvanek},
  {Kleinman}, {Krzesinski}, {Long}, {Nitta}, {Snedden}, {Lee}, {Harris},
  {Brinkmann}, {Schneider}, \& {York}}]{Ivezic2008ApJ...684..287I}
{Ivezi{\'c}}, {\v Z}., {Sesar}, B., {Juri{\'c}}, M., {et~al.} 2008, \apj, 684,
  287

\bibitem[{{Jethwa} {et~al.}(2016){Jethwa}, {Erkal}, \&
  {Belokurov}}]{Jethwa2016MNRAS.461.2212J}
{Jethwa}, P., {Erkal}, D., \& {Belokurov}, V. 2016, \mnras, 461, 2212

\bibitem[{Jones {et~al.}(2001)Jones, Oliphant, Peterson, {et~al.}}]{scipy}
Jones, E., Oliphant, T., Peterson, P., {et~al.} 2001, {SciPy}: Open source
  scientific tools for {Python}, [Online; accessed 2015-08-25]

\bibitem[{{Kallivayalil} {et~al.}(2013){Kallivayalil}, {van der Marel},
  {Besla}, {Anderson}, \& {Alcock}}]{Kallivayalil2013ApJ...764..161K}
{Kallivayalil}, N., {van der Marel}, R.~P., {Besla}, G., {Anderson}, J., \&
  {Alcock}, C. 2013, \apj, 764, 161

\bibitem[{{Kallivayalil} {et~al.}(2018){Kallivayalil}, {Sales}, {Zivick},
  {Fritz}, {Del Pino}, {Sohn}, {Besla}, {van der Marel}, {Navarro}, \&
  {Sacchi}}]{Kallivayalil2018ApJ...867...19K}
{Kallivayalil}, N., {Sales}, L.~V., {Zivick}, P., {et~al.} 2018, \apj, 867, 19

\bibitem[{{Kim} \& {Jerjen}(2015)}]{Kim2015ApJ...808L..39K}
{Kim}, D., \& {Jerjen}, H. 2015, \apjl, 808, L39

\bibitem[{{Kim} {et~al.}(2015{\natexlab{a}}){Kim}, {Jerjen}, {Mackey}, {Da
  Costa}, \& {Milone}}]{Kim2015ApJ...804L..44K}
{Kim}, D., {Jerjen}, H., {Mackey}, D., {Da Costa}, G.~S., \& {Milone}, A.~P.
  2015{\natexlab{a}}, \apjl, 804, L44

\bibitem[{{Kim} {et~al.}(2015{\natexlab{b}}){Kim}, {Jerjen}, {Milone},
  {Mackey}, \& {Da Costa}}]{Kim2015ApJ...803...63K}
{Kim}, D., {Jerjen}, H., {Milone}, A.~P., {Mackey}, D., \& {Da Costa}, G.~S.
  2015{\natexlab{b}}, \apj, 803, 63

\bibitem[{{Kim} {et~al.}(2016){Kim}, {Jerjen}, {Geha}, {Chiti}, {Milone}, {Da
  Costa}, {Mackey}, {Frebel}, \& {Conn}}]{Kim2016ApJ...833...16K}
{Kim}, D., {Jerjen}, H., {Geha}, M., {et~al.} 2016, \apj, 833, 16

\bibitem[{{Kirby} {et~al.}(2013{\natexlab{a}}){Kirby}, {Boylan-Kolchin},
  {Cohen}, {Geha}, {Bullock}, \& {Kaplinghat}}]{Kirby2013ApJ...770...16K}
{Kirby}, E.~N., {Boylan-Kolchin}, M., {Cohen}, J.~G., {et~al.}
  2013{\natexlab{a}}, \apj, 770, 16

\bibitem[{{Kirby} {et~al.}(2013{\natexlab{b}}){Kirby}, {Cohen}, {Guhathakurta},
  {Cheng}, {Bullock}, \& {Gallazzi}}]{Kirby2013ApJ...779..102K}
{Kirby}, E.~N., {Cohen}, J.~G., {Guhathakurta}, P., {et~al.}
  2013{\natexlab{b}}, \apj, 779, 102

\bibitem[{{Koposov} {et~al.}(2015{\natexlab{a}}){Koposov}, {Belokurov},
  {Torrealba}, \& {Evans}}]{Koposov2015ApJ...805..130K}
{Koposov}, S.~E., {Belokurov}, V., {Torrealba}, G., \& {Evans}, N.~W.
  2015{\natexlab{a}}, \apj, 805, 130

\bibitem[{{Koposov} {et~al.}(2015{\natexlab{b}}){Koposov}, {Casey},
  {Belokurov}, {Lewis}, {Gilmore}, {Worley}, {Hourihane}, {Randich}, {Bensby},
  {Bragaglia}, {Bergemann}, {Carraro}, {Costado}, {Flaccomio}, {Francois},
  {Heiter}, {Hill}, {Jofre}, {Lando}, {Lanzafame}, {de Laverny}, {Monaco},
  {Morbidelli}, {Sbordone}, {Mikolaitis}, \&
  {Ryde}}]{Koposov2015ApJ...811...62K}
{Koposov}, S.~E., {Casey}, A.~R., {Belokurov}, V., {et~al.} 2015{\natexlab{b}},
  \apj, 811, 62

\bibitem[{{Koposov} {et~al.}(2018){Koposov}, {Walker}, {Belokurov}, {Casey},
  {Geringer-Sameth}, {Mackey}, {Da Costa}, {Erkal}, {Jethwa}, {Mateo},
  {Olszewski}, \& {Bailey}}]{Koposov2018MNRAS.479.5343K}
{Koposov}, S.~E., {Walker}, M.~G., {Belokurov}, V., {et~al.} 2018, \mnras, 479,
  5343

\bibitem[{{Laevens} {et~al.}(2015){Laevens}, {Martin}, {Bernard}, {Schlafly},
  {Sesar}, {Rix}, {Bell}, {Ferguson}, {Slater}, {Sweeney}, {Wyse}, {Huxor},
  {Burgett}, {Chambers}, {Draper}, {Hodapp}, {Kaiser}, {Magnier}, {Metcalfe},
  {Tonry}, {Wainscoat}, \& {Waters}}]{Laevens2015ApJ...813...44L}
{Laevens}, B.~P.~M., {Martin}, N.~F., {Bernard}, E.~J., {et~al.} 2015, \apj,
  813, 44

\bibitem[{{Li} {et~al.}(2017){Li}, {Simon}, {Drlica-Wagner}, {Bechtol}, {Wang},
  {Garc{\'{\i}}a-Bellido}, {Frieman}, {Marshall}, {James}, {Strigari}, {Pace},
  {Balbinot}, {Zhang}, {Abbott}, {Allam}, {Benoit-L{\'e}vy}, {Bernstein},
  {Bertin}, {Brooks}, {Burke}, {Carnero Rosell}, {Carrasco Kind}, {Carretero},
  {Cunha}, {D'Andrea}, {da Costa}, {DePoy}, {Desai}, {Diehl}, {Eifler},
  {Flaugher}, {Goldstein}, {Gruen}, {Gruendl}, {Gschwend}, {Gutierrez},
  {Krause}, {Kuehn}, {Lin}, {Maia}, {March}, {Menanteau}, {Miquel}, {Plazas},
  {Romer}, {Sanchez}, {Santiago}, {Schubnell}, {Sevilla-Noarbe}, {Smith},
  {Sobreira}, {Suchyta}, {Tarle}, {Thomas}, {Tucker}, {Walker}, {Wechsler},
  {Wester}, {Yanny}, \& {(DES Collaboration}}]{Li2017ApJ...838....8L}
{Li}, T.~S., {Simon}, J.~D., {Drlica-Wagner}, A., {et~al.} 2017, \apj, 838, 8

\bibitem[{{Li} {et~al.}(2018{\natexlab{a}}){Li}, {Simon}, {Pace}, {Torrealba},
  {Kuehn}, {Drlica-Wagner}, {Bechtol}, {Vivas}, {van der Marel}, {Wood},
  {Yanny}, {Belokurov}, {Jethwa}, {Zucker}, {Lewis}, {Kron}, {Nidever},
  {S{\'a}nchez-Conde}, {Ji}, {Conn}, {James}, {Martin}, {Martinez-Delgado},
  {No{\"e}l}, \& {MagLiteS Collaboration}}]{Li2018ApJ...857..145L}
{Li}, T.~S., {Simon}, J.~D., {Pace}, A.~B., {et~al.} 2018{\natexlab{a}}, \apj,
  857, 145

\bibitem[{{Li} {et~al.}(2018{\natexlab{b}}){Li}, {Simon}, {Kuehn}, {Pace},
  {Erkal}, {Bechtol}, {Yanny}, {Drlica-Wagner}, {Marshall}, {Lidman},
  {Balbinot}, {Carollo}, {Jenkins}, {Mart{\'{\i}}nez-V{\'a}zquez}, {Shipp},
  {Stringer}, {Vivas}, {Walker}, {Wechsler}, {Abdalla}, {Allam}, {Annis},
  {Avila}, {Bertin}, {Brooks}, {Buckley-Geer}, {Burke}, {Carnero Rosell},
  {Carrasco Kind}, {Carretero}, {Cunha}, {D'Andrea}, {da Costa}, {Davis}, {De
  Vicente}, {Doel}, {Eifler}, {Evrard}, {Flaugher}, {Frieman},
  {Garc{\'{\i}}a-Bellido}, {Gaztanaga}, {Gerdes}, {Gruen}, {Gruendl},
  {Gschwend}, {Gutierrez}, {Hartley}, {Hollowood}, {Honscheid}, {James},
  {Krause}, {Maia}, {March}, {Menanteau}, {Miquel}, {Plazas}, {Sanchez},
  {Santiago}, {Scarpine}, {Schindler}, {Schubnell}, {Sevilla-Noarbe}, {Smith},
  {Smith}, {Soares-Santos}, {Sobreira}, {Suchyta}, {Swanson}, {Tarle},
  {Tucker}, \& {DES Collaboration}}]{Li2018ApJ...866...22L}
{Li}, T.~S., {Simon}, J.~D., {Kuehn}, K., {et~al.} 2018{\natexlab{b}}, \apj,
  866, 22

\bibitem[{{Lindegren} {et~al.}(2018){Lindegren}, {Hern{\'a}ndez}, {Bombrun},
  {Klioner}, {Bastian}, {Ramos-Lerate}, {de Torres}, {Steidelm{\"u}ller},
  {Stephenson}, {Hobbs}, {Lammers}, {Biermann}, {Geyer}, {Hilger}, {Michalik},
  {Stampa}, {McMillan}, {Casta{\~n}eda}, {Clotet}, {Comoretto}, {Davidson},
  {Fabricius}, {Gracia}, {Hambly}, {Hutton}, {Mora}, {Portell}, {van Leeuwen},
  {Abbas}, {Abreu}, {Altmann}, {Andrei}, {Anglada}, {Balaguer-N{\'u}{\~n}ez},
  {Barache}, {Becciani}, {Bertone}, {Bianchi}, {Bouquillon}, {Bourda},
  {Br{\"u}semeister}, {Bucciarelli}, {Busonero}, {Buzzi}, {Cancelliere},
  {Carlucci}, {Charlot}, {Cheek}, {Crosta}, {Crowley}, {de Bruijne}, {de
  Felice}, {Drimmel}, {Esquej}, {Fienga}, {Fraile}, {Gai}, {Garralda},
  {Gonz{\'a}lez-Vidal}, {Guerra}, {Hauser}, {Hofmann}, {Holl}, {Jordan},
  {Lattanzi}, {Lenhardt}, {Liao}, {Licata}, {Lister}, {L{\"o}ffler},
  {Marchant}, {Martin-Fleitas}, {Messineo}, {Mignard}, {Morbidelli}, {Poggio},
  {Riva}, {Rowell}, {Salguero}, {Sarasso}, {Sciacca}, {Siddiqui}, {Smart},
  {Spagna}, {Steele}, {Taris}, {Torra}, {van Elteren}, {van Reeven}, \&
  {Vecchiato}}]{GaiaLindegren2018A&A...616A...2L}
{Lindegren}, L., {Hern{\'a}ndez}, J., {Bombrun}, A., {et~al.} 2018, \aap, 616,
  A2

\bibitem[{{Luque} {et~al.}(2016){Luque}, {Queiroz}, {Santiago}, {Pieres},
  {Balbinot}, {Bechtol}, {Drlica-Wagner}, {Neto}, {da Costa}, {Maia}, {Yanny},
  {Abbott}, {Allam}, {Benoit-L{\'e}vy}, {Bertin}, {Brooks}, {Buckley-Geer},
  {Burke}, {Rosell}, {Kind}, {Carretero}, {Cunha}, {Desai}, {Diehl},
  {Dietrich}, {Eifler}, {Finley}, {Flaugher}, {Fosalba}, {Frieman}, {Gerdes},
  {Gruen}, {Gutierrez}, {Honscheid}, {James}, {Kuehn}, {Kuropatkin}, {Lahav},
  {Li}, {March}, {Marshall}, {Martini}, {Miquel}, {Neilsen}, {Nichol}, {Nord},
  {Ogando}, {Plazas}, {Romer}, {Roodman}, {Sanchez}, {Scarpine}, {Schubnell},
  {Sevilla-Noarbe}, {Smith}, {Soares-Santos}, {Sobreira}, {Suchyta}, {Swanson},
  {Tarle}, {Thaler}, {Tucker}, {Walker}, \& {Zhang}}]{Luque2016MNRAS.458..603L}
{Luque}, E., {Queiroz}, A., {Santiago}, B., {et~al.} 2016, \mnras, 458, 603

\bibitem[{{Luque} {et~al.}(2017){Luque}, {Pieres}, {Santiago}, {Yanny},
  {Vivas}, {Queiroz}, {Drlica-Wagner}, {Morganson}, {Balbinot}, {Marshall},
  {Li}, {Neto}, {da Costa}, {Maia}, {Bechtol}, {Kim}, {Bernstein}, {Dodelson},
  {Whiteway}, {Diehl}, {Finley}, {Abbott}, {Abdalla}, {Allam}, {Annis},
  {Benoit-L{\'e}vy}, {Bertin}, {Brooks}, {Burke}, {Rosell}, {Kind},
  {Carretero}, {Cunha}, {D'Andrea}, {Desai}, {Doel}, {Evrard}, {Flaugher},
  {Fosalba}, {Gerdes}, {Goldstein}, {Gruen}, {Gruendl}, {Gutierrez}, {James},
  {Kuehn}, {Kuropatkin}, {Lahav}, {Martini}, {Miquel}, {Nord}, {Ogando},
  {Plazas}, {Romer}, {Sanchez}, {Scarpine}, {Schubnell}, {Sevilla-Noarbe},
  {Smith}, {Soares-Santos}, {Sobreira}, {Suchyta}, {Swanson}, {Tarle},
  {Thomas}, \& {Walker}}]{Luque2017MNRAS.468...97L}
{Luque}, E., {Pieres}, A., {Santiago}, B., {et~al.} 2017, \mnras, 468, 97

\bibitem[{{Luque} {et~al.}(2018){Luque}, {Santiago}, {Pieres}, {Marshall},
  {Pace}, {Kron}, {Drlica-Wagner}, {Queiroz}, {Balbinot}, {dal Ponte}, {Fausti
  Neto}, {da Costa}, {Maia}, {Walker}, {Abdalla}, {Allam}, {Annis}, {Bechtol},
  {Benoit-L{\'e}vy}, {Bertin}, {Brooks}, {Carnero Rosell}, {Carrasco Kind},
  {Carretero}, {Crocce}, {Davis}, {Doel}, {Eifler}, {Flaugher},
  {Garc{\'{\i}}a-Bellido}, {Gerdes}, {Gruen}, {Gruendl}, {Gutierrez},
  {Honscheid}, {James}, {Kuehn}, {Kuropatkin}, {Miquel}, {Nichol}, {Plazas},
  {Sanchez}, {Scarpine}, {Schindler}, {Sevilla-Noarbe}, {Smith},
  {Soares-Santos}, {Sobreira}, {Suchyta}, {Tarle}, \&
  {Thomas}}]{Luque2018MNRAS.478.2006L}
{Luque}, E., {Santiago}, B., {Pieres}, A., {et~al.} 2018, \mnras, 478, 2006

\bibitem[{{Martin} {et~al.}(2015){Martin}, {Nidever}, {Besla}, {Olsen},
  {Walker}, {Vivas}, {Gruendl}, {Kaleida}, {Mu{\~n}oz}, {Blum}, {Saha}, {Conn},
  {Bell}, {Chu}, {Cioni}, {de Boer}, {Gallart}, {Jin}, {Kunder}, {Majewski},
  {Martinez-Delgado}, {Monachesi}, {Monelli}, {Monteagudo}, {No{\"e}l},
  {Olszewski}, {Stringfellow}, {van der Marel}, \&
  {Zaritsky}}]{Martin2015ApJ...804L...5M}
{Martin}, N.~F., {Nidever}, D.~L., {Besla}, G., {et~al.} 2015, \apjl, 804, L5

\bibitem[{{Martinez} {et~al.}(2011){Martinez}, {Minor}, {Bullock},
  {Kaplinghat}, {Simon}, \& {Geha}}]{Martinez2011ApJ...738...55M}
{Martinez}, G.~D., {Minor}, Q.~E., {Bullock}, J., {et~al.} 2011, \apj, 738, 55

\bibitem[{{Massari} \& {Helmi}(2018)}]{Massari2018A&A...620A.155M}
{Massari}, D., \& {Helmi}, A. 2018, \aap, 620, A155

\bibitem[{{McConnachie}(2012)}]{McConnachie2012AJ....144....4M}
{McConnachie}, A.~W. 2012, \aj, 144, 4

\bibitem[{{Mutlu-Pakdil} {et~al.}(2018){Mutlu-Pakdil}, {Sand}, {Carlin},
  {Spekkens}, {Caldwell}, {Crnojevi{\'c}}, {Hughes}, {Willman}, \&
  {Zaritsky}}]{MutluPakdil2018ApJ...863...25M}
{Mutlu-Pakdil}, B., {Sand}, D.~J., {Carlin}, J.~L., {et~al.} 2018, \apj, 863,
  25

\bibitem[{{Nagasawa} {et~al.}(2018){Nagasawa}, {Marshall}, {Li}, {Hansen},
  {Simon}, {Bernstein}, {Balbinot}, {Drlica-Wagner}, {Pace}, {Strigari},
  {Pellegrino}, {DePoy}, {Suntzeff}, {Bechtol}, {Walker}, {Abbott}, {Abdalla},
  {Allam}, {Annis}, {Benoit-L{\'e}vy}, {Bertin}, {Brooks}, {Carnero Rosell},
  {Carrasco Kind}, {Carretero}, {Cunha}, {DAndrea}, {da Costa}, {Davis},
  {Desai}, {Doel}, {Eifler}, {Flaugher}, {Fosalba}, {Frieman},
  {Garc{\'{\i}}a-Bellido}, {Gaztanaga}, {Gerdes}, {Gruen}, {Gruendl},
  {Gschwend}, {Gutierrez}, {Hartley}, {Honscheid}, {James}, {Jeltema},
  {Krause}, {Kuehn}, {Kuhlmann}, {Kuropatkin}, {March}, {Miquel}, {Nord},
  {Roodman}, {Sanchez}, {Santiago}, {Scarpine}, {Schindler}, {Schubnell},
  {Sevilla-Noarbe}, {Smith}, {Smith}, {Soares-Santos}, {Sobreira}, {Suchyta},
  {Tarle}, {Thomas}, {Tucker}, {Wechsler}, {Wolf}, \&
  {Yanny}}]{Nagasawa2018ApJ...852...99N}
{Nagasawa}, D.~Q., {Marshall}, J.~L., {Li}, T.~S., {et~al.} 2018, \apj, 852, 99

\bibitem[{{Patel} {et~al.}(2018){Patel}, {Besla}, {Mandel}, \&
  {Sohn}}]{Patel2018ApJ...857...78P}
{Patel}, E., {Besla}, G., {Mandel}, K., \& {Sohn}, S.~T. 2018, \apj, 857, 78

\bibitem[{{Pawlowski}(2018)}]{Pawlowski2018MPLA...3330004P}
{Pawlowski}, M.~S. 2018, Modern Physics Letters A, 33, 1830004

\bibitem[{{Pawlowski} \& {Kroupa}(2013)}]{Pawlowski2013MNRAS.435.2116P}
{Pawlowski}, M.~S., \& {Kroupa}, P. 2013, \mnras, 435, 2116

\bibitem[{{Pawlowski} {et~al.}(2015){Pawlowski}, {McGaugh}, \&
  {Jerjen}}]{Pawlowski2015MNRAS.453.1047P}
{Pawlowski}, M.~S., {McGaugh}, S.~S., \& {Jerjen}, H. 2015, \mnras, 453, 1047

\bibitem[{{Piatek} {et~al.}(2002){Piatek}, {Pryor}, {Olszewski}, {Harris},
  {Mateo}, {Minniti}, {Monet}, {Morrison}, \&
  {Tinney}}]{Piatek2002AJ....124.3198P}
{Piatek}, S., {Pryor}, C., {Olszewski}, E.~W., {et~al.} 2002, \aj, 124, 3198

\bibitem[{{Plummer}(1911)}]{Plummer1911MNRAS..71..460P}
{Plummer}, H.~C. 1911, \mnras, 71, 460

\bibitem[{{Ricotti} \& {Gnedin}(2005)}]{Ricotti2005ApJ...629..259R}
{Ricotti}, M., \& {Gnedin}, N.~Y. 2005, \apj, 629, 259

\bibitem[{{Rocha} {et~al.}(2012){Rocha}, {Peter}, \&
  {Bullock}}]{Rocha2012MNRAS.425..231R}
{Rocha}, M., {Peter}, A.~H.~G., \& {Bullock}, J. 2012, \mnras, 425, 231

\bibitem[{{Sales} {et~al.}(2017){Sales}, {Navarro}, {Kallivayalil}, \&
  {Frenk}}]{Sales2017MNRAS.465.1879S}
{Sales}, L.~V., {Navarro}, J.~F., {Kallivayalil}, N., \& {Frenk}, C.~S. 2017,
  \mnras, 465, 1879

\bibitem[{{Sandage}(1969)}]{Sandage1969ApJ...158.1115S}
{Sandage}, A. 1969, \apj, 158, 1115

\bibitem[{{Schlafly} \& {Finkbeiner}(2011)}]{Schlafly2011ApJ...737..103S}
{Schlafly}, E.~F., \& {Finkbeiner}, D.~P. 2011, \apj, 737, 103

\bibitem[{{Schlegel} {et~al.}(1998){Schlegel}, {Finkbeiner}, \&
  {Davis}}]{Schlegel1998ApJ...500..525S}
{Schlegel}, D.~J., {Finkbeiner}, D.~P., \& {Davis}, M. 1998, \apj, 500, 525

\bibitem[{{Sch{\"o}nrich} {et~al.}(2010){Sch{\"o}nrich}, {Binney}, \&
  {Dehnen}}]{Schonrich2010MNRAS.403.1829S}
{Sch{\"o}nrich}, R., {Binney}, J., \& {Dehnen}, W. 2010, \mnras, 403, 1829

\bibitem[{{Simon}(2018)}]{Simon2018ApJ...863...89S}
{Simon}, J.~D. 2018, \apj, 863, 89

\bibitem[{{Simon} {et~al.}(2015){Simon}, {Drlica-Wagner}, {Li}, {Nord}, {Geha},
  {Bechtol}, {Balbinot}, {Buckley-Geer}, {Lin}, {Marshall}, {Santiago},
  {Strigari}, {Wang}, {Wechsler}, {Yanny}, {Abbott}, {Bauer}, {Bernstein},
  {Bertin}, {Brooks}, {Burke}, {Capozzi}, {Carnero Rosell}, {Carrasco Kind},
  {DAndrea}, {da Costa}, {DePoy}, {Desai}, {Diehl}, {Dodelson}, {Cunha},
  {Estrada}, {Evrard}, {Fausti Neto}, {Fernandez}, {Finley}, {Flaugher},
  {Frieman}, {Gaztanaga}, {Gerdes}, {Gruen}, {Gruendl}, {Honscheid}, {James},
  {Kent}, {Kuehn}, {Kuropatkin}, {Lahav}, {Maia}, {March}, {Martini}, {Miller},
  {Miquel}, {Ogando}, {Romer}, {Roodman}, {Rykoff}, {Sako}, {Sanchez},
  {Schubnell}, {Sevilla}, {Smith}, {Soares-Santos}, {Sobreira}, {Suchyta},
  {Swanson}, {Tarle}, {Thaler}, {Tucker}, {Vikram}, {Walker}, {Wester}, \& {The
  DES Collaboration}}]{Simon2015ApJ...808...95S}
{Simon}, J.~D., {Drlica-Wagner}, A., {Li}, T.~S., {et~al.} 2015, \apj, 808, 95

\bibitem[{{Simon} {et~al.}(2017){Simon}, {Li}, {Drlica-Wagner}, {Bechtol},
  {Marshall}, {James}, {Wang}, {Strigari}, {Balbinot}, {Kuehn}, {Walker},
  {Abbott}, {Allam}, {Annis}, {Benoit-L{\'e}vy}, {Brooks}, {Buckley-Geer},
  {Burke}, {Carnero Rosell}, {Carrasco Kind}, {Carretero}, {Cunha}, {D'Andrea},
  {da Costa}, {DePoy}, {Desai}, {Doel}, {Fernandez}, {Flaugher}, {Frieman},
  {Garc{\'{\i}}a-Bellido}, {Gaztanaga}, {Goldstein}, {Gruen}, {Gutierrez},
  {Kuropatkin}, {Maia}, {Martini}, {Menanteau}, {Miller}, {Miquel}, {Neilsen},
  {Nord}, {Ogando}, {Plazas}, {Romer}, {Rykoff}, {Sanchez}, {Santiago},
  {Scarpine}, {Schubnell}, {Sevilla-Noarbe}, {Smith}, {Sobreira}, {Suchyta},
  {Swanson}, {Tarle}, {Whiteway}, {Yanny}, \& {DES
  Collaboration}}]{Simon2017ApJ...838...11S}
{Simon}, J.~D., {Li}, T.~S., {Drlica-Wagner}, A., {et~al.} 2017, \apj, 838, 11

\bibitem[{{Sohn} {et~al.}(2013){Sohn}, {Besla}, {van der Marel},
  {Boylan-Kolchin}, {Majewski}, \& {Bullock}}]{Sohn2013ApJ...768..139S}
{Sohn}, S.~T., {Besla}, G., {van der Marel}, R.~P., {et~al.} 2013, \apj, 768,
  139

\bibitem[{{Sohn} {et~al.}(2017){Sohn}, {Patel}, {Besla}, {van der Marel},
  {Bullock}, {Strigari}, {van de Ven}, {Walker}, \&
  {Bellini}}]{Sohn2017ApJ...849...93S}
{Sohn}, S.~T., {Patel}, E., {Besla}, G., {et~al.} 2017, \apj, 849, 93

\bibitem[{{Torrealba} {et~al.}(2016){Torrealba}, {Koposov}, {Belokurov}, \&
  {Irwin}}]{Torrealba2016MNRAS.459.2370T}
{Torrealba}, G., {Koposov}, S.~E., {Belokurov}, V., \& {Irwin}, M. 2016,
  \mnras, 459, 2370

\bibitem[{{Torrealba} {et~al.}(2018){Torrealba}, {Belokurov}, {Koposov},
  {Bechtol}, {Drlica-Wagner}, {Olsen}, {Vivas}, {Yanny}, {Jethwa}, {Walker},
  {Li}, {Allam}, {Conn}, {Gallart}, {Gruendl}, {James}, {Johnson}, {Kuehn},
  {Kuropatkin}, {Martin}, {Martinez-Delgado}, {Nidever}, {No{\"e}l}, {Simon},
  {Stringfellow}, \& {Tucker}}]{Torrealba2018MNRAS.475.5085T}
{Torrealba}, G., {Belokurov}, V., {Koposov}, S.~E., {et~al.} 2018, \mnras, 475,
  5085

\bibitem[{{Trotta}(2008)}]{Trotta2008ConPh..49...71T}
{Trotta}, R. 2008, Contemporary Physics, 49, 71

\bibitem[{{Walker} {et~al.}(2015){Walker}, {Mateo}, {Olszewski}, {Bailey},
  {Koposov}, {Belokurov}, \& {Evans}}]{Walker2015ApJ...808..108W}
{Walker}, M.~G., {Mateo}, M., {Olszewski}, E.~W., {et~al.} 2015, \apj, 808, 108

\bibitem[{{Walker} \& {Pe{\~n}arrubia}(2011)}]{Walker2011ApJ...742...20W}
{Walker}, M.~G., \& {Pe{\~n}arrubia}, J. 2011, \apj, 742, 20

\bibitem[{{Walker} {et~al.}(2016){Walker}, {Mateo}, {Olszewski}, {Koposov},
  {Belokurov}, {Jethwa}, {Nidever}, {Bonnivard}, {Bailey}, {Bell}, \&
  {Loebman}}]{Walker2016ApJ...819...53W}
{Walker}, M.~G., {Mateo}, M., {Olszewski}, E.~W., {et~al.} 2016, \apj, 819, 53

\bibitem[{Walt {et~al.}(2011)Walt, Colbert, \& Varoquaux}]{numpy}
Walt, S. v.~d., Colbert, S.~C., \& Varoquaux, G. 2011, Computing in Science \&
  Engineering, 13, 22

\bibitem[{{Wildey} {et~al.}(1962){Wildey}, {Burbidge}, {Sandage}, \&
  {Burbidge}}]{Wildey1962ApJ...135...94W}
{Wildey}, R.~L., {Burbidge}, E.~M., {Sandage}, A.~R., \& {Burbidge}, G.~R.
  1962, \apj, 135, 94

\bibitem[{{Willman} \& {Strader}(2012)}]{Willman2012AJ....144...76W}
{Willman}, B., \& {Strader}, J. 2012, \aj, 144, 76

\bibitem[{{Yozin} \& {Bekki}(2015)}]{Yozin2015MNRAS.453.2302Y}
{Yozin}, C., \& {Bekki}, K. 2015, \mnras, 453, 2302

\end{thebibliography}

\appendix

\section{A. Two Component Milky Way Foreground Modeling}\label{appendix:mwmodel} 

Here we explore a more complex MW foreground model to verify that our results are robust to choice of foreground model.
We expand the proper motion component of the MW model from one to two Gaussian components. 
We will refer to this model as the two component model.  
Physically the two components could represent the MW halo and disk, however, note that the cuts we apply ($\varpi$ and $v_{esc}$) preferentially removes MW disk  stars from the sample.
To distinguish the components, we assume that one component has an overall larger proper motion dispersion: $\sigma_{\mu, 1}^2 > \sigma_{\mu, 2}^2$ (where $\sigma_{\mu}^2=\left(\sigma_{\mu_{\alpha}\cos{\delta}} \right)^2 + \sigma_{\mu_{\delta}}^2$).  
In the disk and halo interpretation the `disk' component will have a larger dispersion as the disk stars are closer and have larger  proper motions.
Overall, adding the second component increases the number of free parameters by five; two  mean MW proper motions, two MW proper motion dispersions, and a fraction parameter to weigh the two components.
The priors for these parameters are set to be the same as in the single foreground model case. 

\begin{figure*}[h!]
\plotone{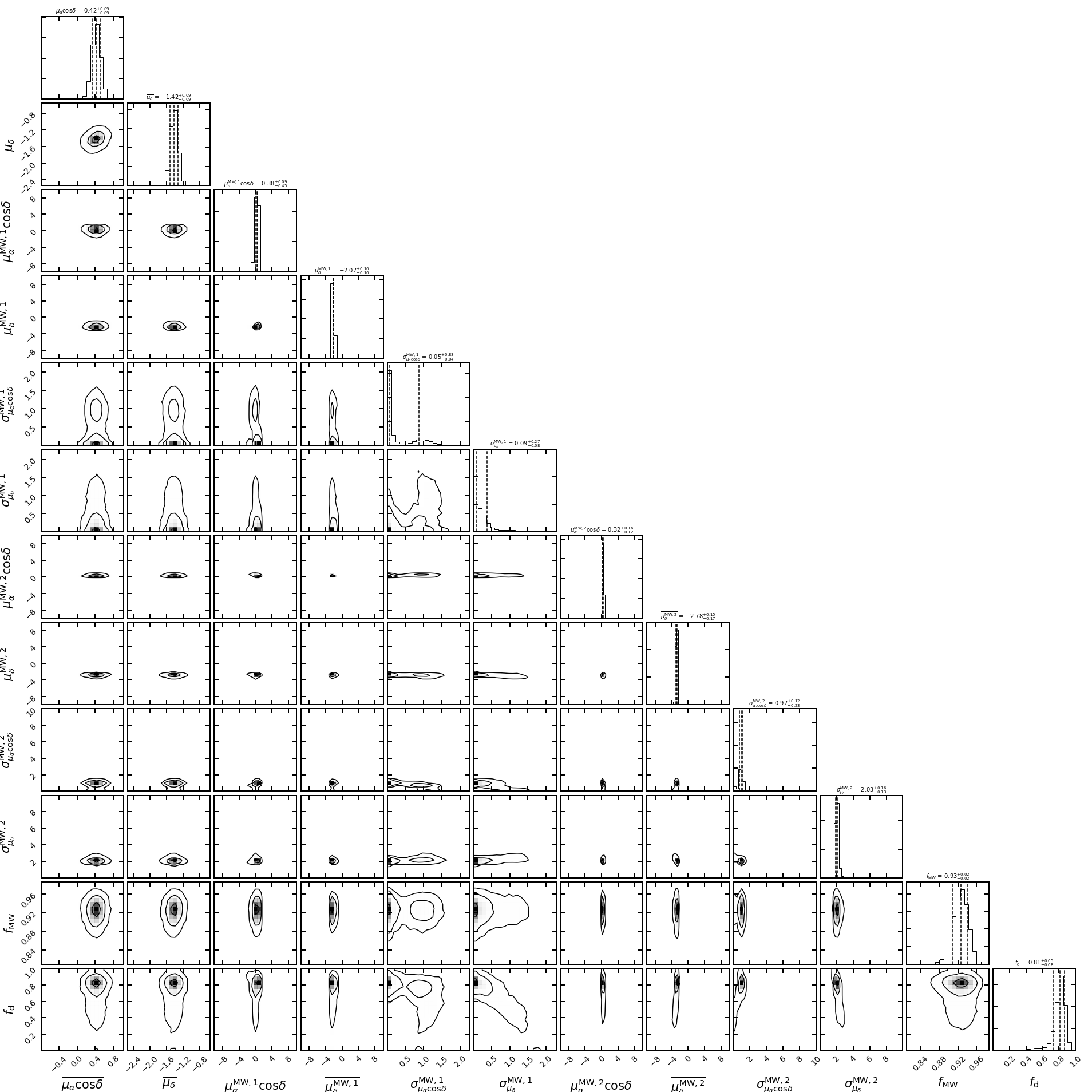}
\caption{Corner plot of Grus~II for the two component MW foreground model.  The two MW components are labeled 1,2 and $f_d$ is the fraction of stars in the 2nd MW population.}
\label{fig:corner_mw_model}
\end{figure*}

We apply this additional foreground model to the 13 satellites with a signal and Horologium~II.
In all cases our measurements in the two component model are within the errors of the single component model, with similar  precision on the systemic proper motion with the exception of Horologium~II and Tucana~IV.  
In Figure~\ref{fig:corner_mw_model}, we show the posterior distribution of Grus~II as an example of the two component model.  The posterior distribution of the satellite systemic proper motion parameters are extremely similar to the original foreground model.  
In general, the  MW foreground parameters are not well constrained.
The satellite proper motions parameters do not correlate with any of the foreground MW parameters.  The MW parameters are highly correlated among  other MW parameters (in particular between the `halo' and `disk' parameters).   The dispersions are correlated and in most cases not resolved into individual components. The number of stars in the background is not large enough to separate the foreground into multiple components. 

To quantify which of the foreground models  is a better fit we compute the logarithmic Bayes Factor ($\ln{\rm B}$), the ratio of Bayesian evidence between the two models,  a commonly utilized model comparison test  \citep{Trotta2008ConPh..49...71T}.   The ranges of $\ln{\rm B}: 0<1<2.5<5$ correspond to insignificant, weak, moderate and strong evidence in favor of one model (negative values indicate evidence in favor the other model).  We find  in all but one satellite (Eridanus~III) the two component model is highly disfavored; for 12 satellites the $\ln{\rm B}$ range is between  $-20 < \ln{\rm B} < -5.5$ (where positive values imply the two component foreground  is favored).  Eridanus~III has $\ln{\rm B} = +4.6$. Based on the model selection criteria increasing  the complexity and number of parameters of the foreground model does not improve the fit and is not justified based on the sample sizes.

We find that in most cases the overall change in membership is lower but small; for 9/13 satellites the change is  ($\sum\Delta p_i < -1$).  The overall range of membership changes is $ \sum \Delta p_i \sim -8.0 -  +0.6 $.
When restricted to only  brighter stars ($G < 19.5$) the change in membership is much smaller and the  maximum change is only $\sum \Delta p_i\sim1.7$.
In general, the stars `moved' to the MW population are faint ($G > 19.5$) with large proper motion errors.  
The satellites with the largest number of members (e.g. Grus~II, Tucana~II, and Tucana~III) had the largest decrease in membership, however, their  systemic proper motion does not change (with the exception of Tucana~IV). 
While some satellites have large changes in membership, they are almost all faint stars.

Of the satellites with a signal, Tucana~IV has the largest change in the systemic motion; it changes by $\approx0.7 \sigma/0.3 \sigma$ ($\mu_{\alpha}\cos{\delta}, \,\mu_{\delta}$) and the errors increase by about a factor of two.  
The membership decreases by $\sum \delta p_i \sim 4.4$.  The significance of the detection of the proper motion signal of Tucana~IV is decreased  relative to the original foreground model (however, this foreground model is disfavored compared to the original).
As there are not many bright stars to anchor the satellite measurement it is much harder to distinguish between MW and satellite stars for this satellite.  In addition this was the only satellite where all four MW dispersions  parameters  were constrained to be non-zero.  In most cases the MW dispersion parameters had large tails to zero-dispersion.  Tucana~IV is one of the most diffuse satellites in our sample and as there is some overlap in MW proper motion our method has trouble disentangling it from the MW.  Radial velocities and stellar chemistry will be key to improve the Tucana~IV systemic proper motion measurement.

\section{B. Discussion of Individual Satellites without a Conclusive Detection}\label{appendix:nodetect} 

While we are not able to conclusively determine the systemic proper motion of Cetus~II, Kim~2, and Horologium~II, we do find several potential members in each satellite.
We have provided diagnostic plots and corner plots for these four satellites in Appendix~\ref{appendix:figures}.  
If  stars in these systems are verified as members with spectroscopic follow-up, the systemic proper motions could be determined.
Horologium~II has 3 members with high probability (including 1 BHB) and 2 with medium probability (including 1 BHB).  Three RGB stars are all above stellar locus, indicating that they are metal-poor stars. Although these 5 stars cluster in  proper motion space, the errors are too large to claim that they are members of the same source.  In addition, the satellite posterior is not well constrained.
Spectrosopically confirming their membership in Horologium~II or improving the precision of the proper motions (i.e. later \gaia releases) are required for future studies of this object.   We note that in the two component foreground model, the significance of the Horologium~II signal deceases.

We note that there has been spectroscopic follow-up of Horologium~II with VLT/FLAMES/GIRAFFE \citep{Fritz2018arXiv180507350F}, however, the poor velocity precision of only three potential members and lack of a clear cold spike makes it unclear whether the heliocentric velocity has been measured.  
They find $(\mu_{\alpha}\cos{\delta}, \mu_\delta) = (1.52 \pm 0.25, -0.47\pm 0.39) \pmunit$ which is consistent at the $1.6-\sigma, 0.6\sigma$ with our measurement.
They find three members, the brightest is outside our color-magnitude diagram selection.  The other two have membership of 0.68 and 0.98 in our model.

In Kim~2 and Cetus~II we find a single `bright' star and 2-3 faint stars, in agreement with the expected number of stars  in these objects (${\rm N_{expected}=2-4}$).  In Tucana~V, if we assume $\epsilon=0$ we find a non-zero signal (although it is at lower significance than the other satellites discussed in this section). 
Deeper photometry of Kim~2 suggests that it is significantly more metal rich than our assumed isochrone \citep[$\overline{\feh}\approx-1$][]{Kim2015ApJ...803...63K}.  While we searched for more members with a more metal-rich isochrone ($\overline{\feh}\approx-1.5$), we were not able to find any.  Given the low luminosity of this object we did not expect to find many members (see Table~\ref{table:pm}).
The reality as stellar overdensities for Cetus~II and Tucana~V has been questioned with deep  data \citep{Conn2018ApJ...852...68C, Conn2018ApJ...857...70C}.  

\section{C. Figures}
\label{appendix:figures}

\begin{figure*}[h!]
\includegraphics[scale=.34]{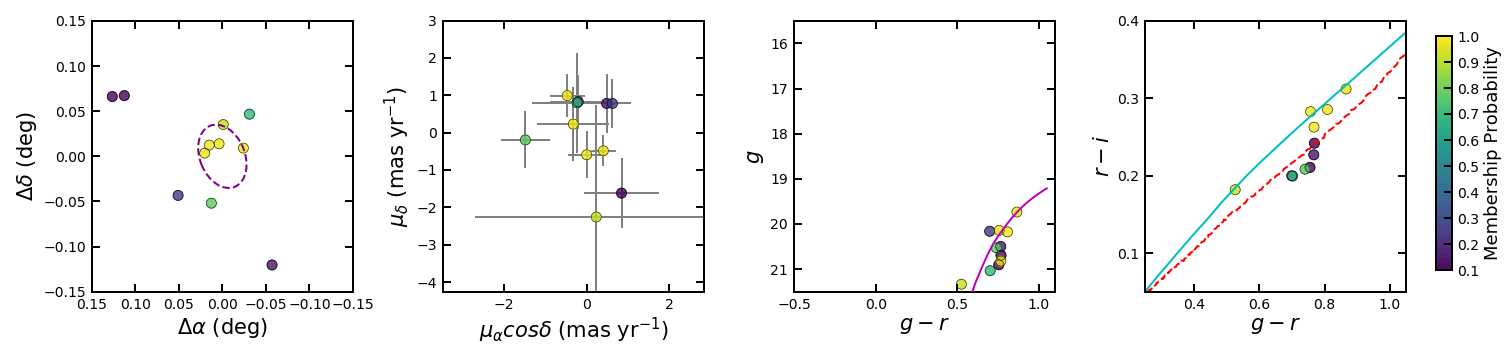}
\caption{Analogous to Figure~\ref{fig:ret2} but for Columba I.}
\label{fig:col1}
\end{figure*}

\begin{figure*}[h!]
\includegraphics[scale=.34]{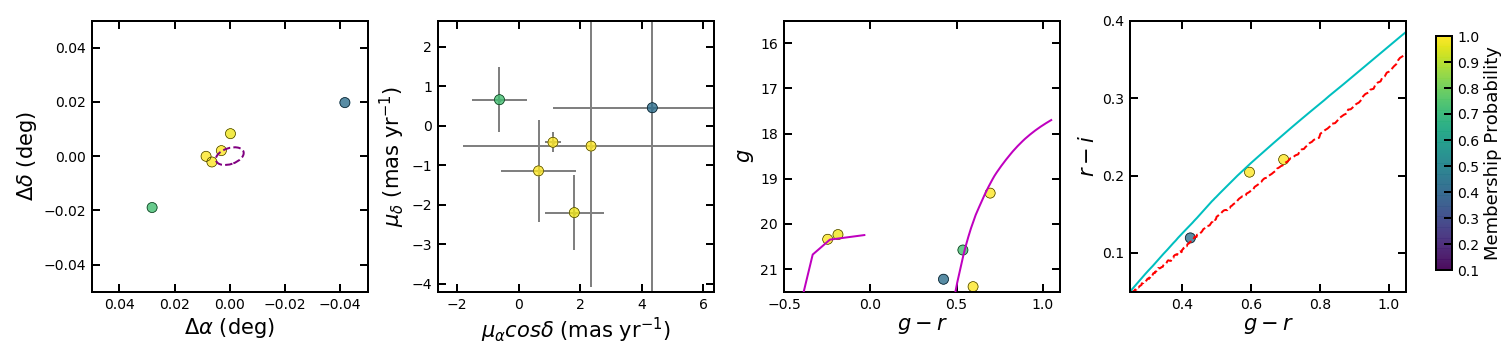}
\caption{Analogous to Figure~\ref{fig:ret2} but for Eridanus III.}
\label{fig:eri3}
\end{figure*}

\begin{figure*}[h!]
\includegraphics[scale=.34]{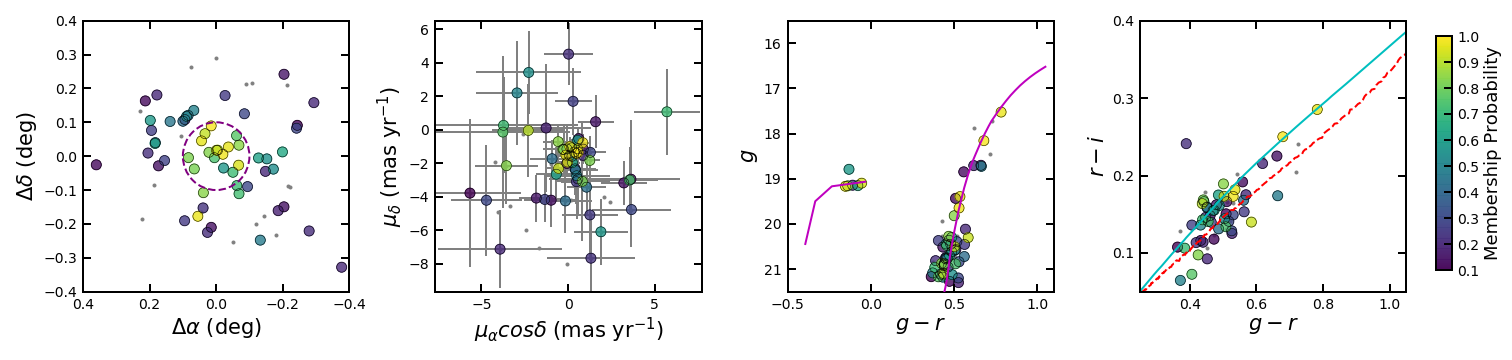}
\caption{Analogous to Figure~\ref{fig:ret2} but for Grus II.}
\label{fig:gru2}
\end{figure*}

\begin{figure*}[h!]
\includegraphics[scale=.34]{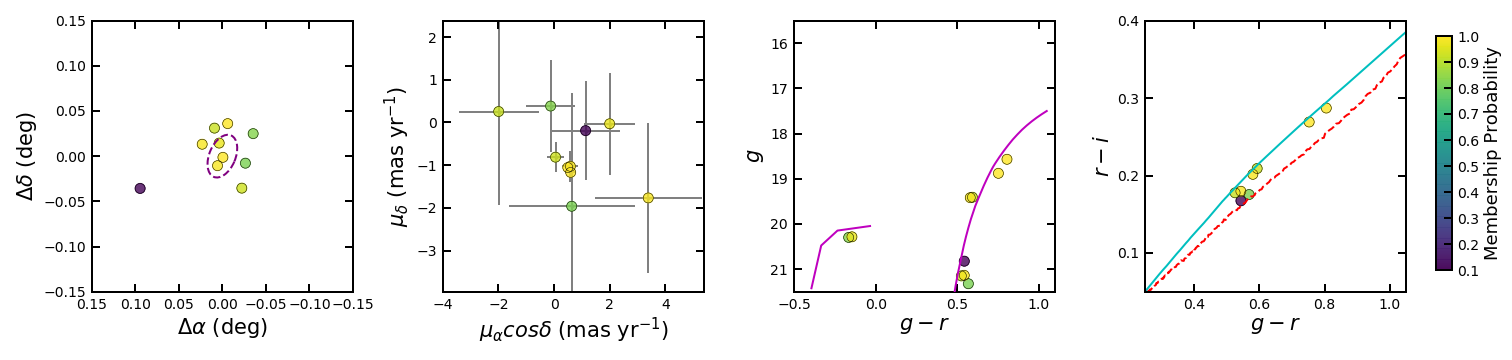}
\caption{Analogous to Figure~\ref{fig:ret2} but for Phoenix II.}
\label{fig:phe2}
\end{figure*}

\begin{figure*}[h!]
\includegraphics[scale=.34]{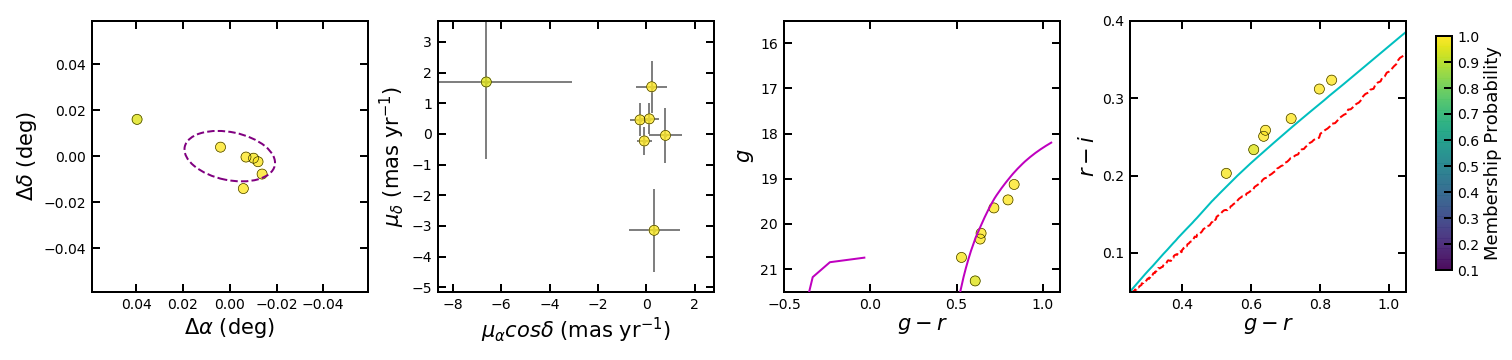}
\caption{Analogous to Figure~\ref{fig:ret2} but for Pictor I.}
\label{fig:pic1}
\end{figure*}

\begin{figure*}[h!]
\includegraphics[scale=.34]{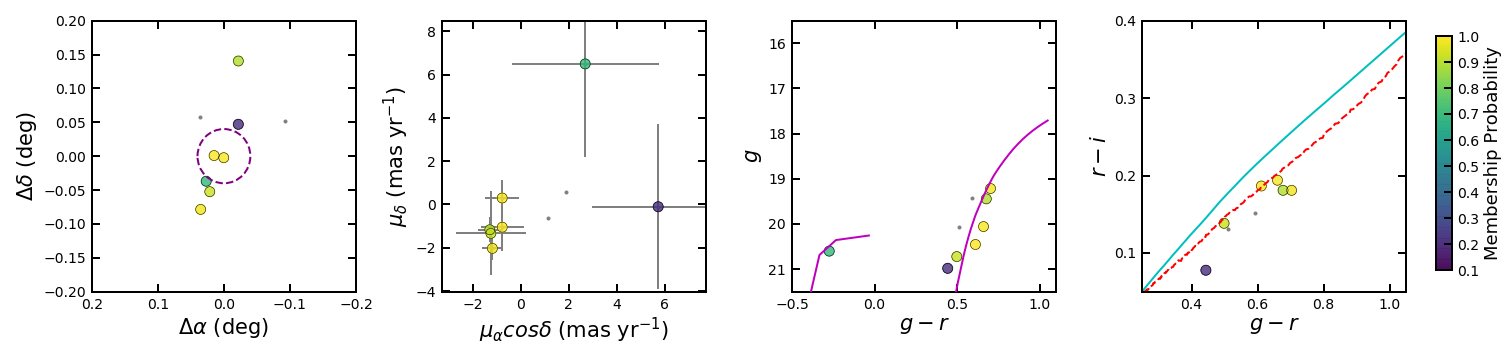}
\caption{Analogous to Figure~\ref{fig:ret2} but for Reticulum III.}
\label{fig:ret3}
\end{figure*}

\begin{figure*}[h!]
\includegraphics[scale=.34]{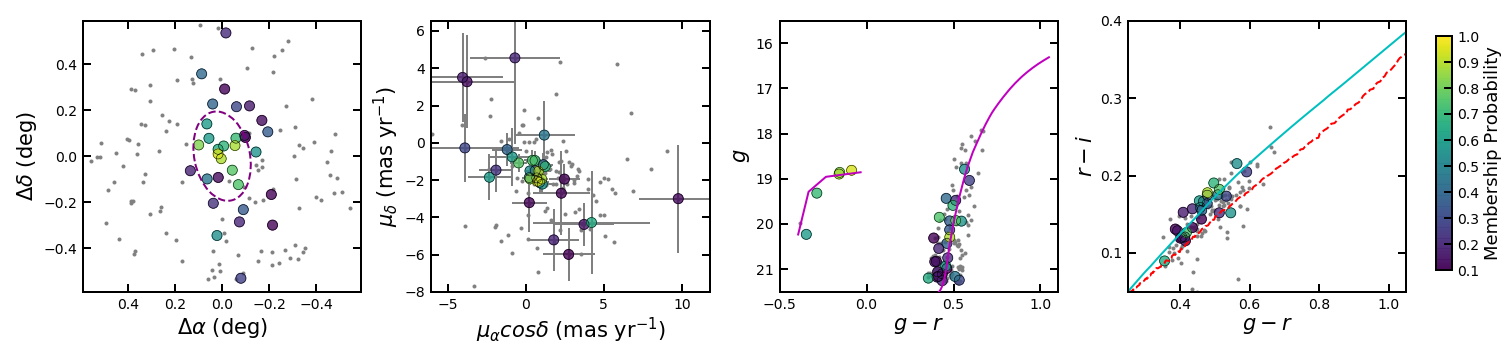}
\caption{Analogous to Figure~\ref{fig:ret2} but for Tucana IV.}
\label{fig:tuc4}
\end{figure*}

\begin{figure*}[h!]
\includegraphics[scale=.34]{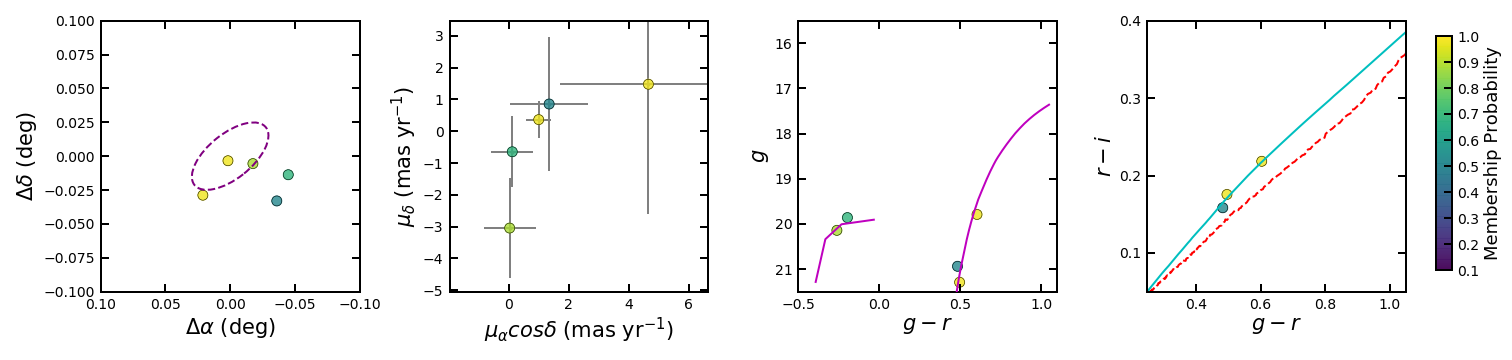}
\caption{Analogous to Figure~\ref{fig:ret2} but for Horologium II.}
\label{fig:hor2}
\end{figure*}

\begin{figure*}[h!]
\includegraphics[scale=.34]{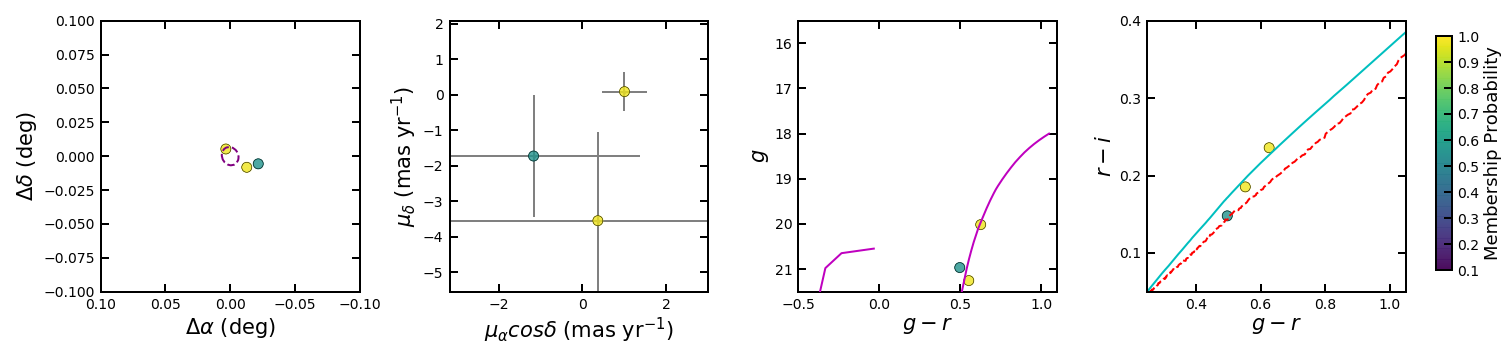}
\caption{Analogous to Figure~\ref{fig:ret2} but for Kim 2.}
\label{fig:ind1}
\end{figure*}

\begin{figure*}[h!]
\includegraphics[scale=.34]{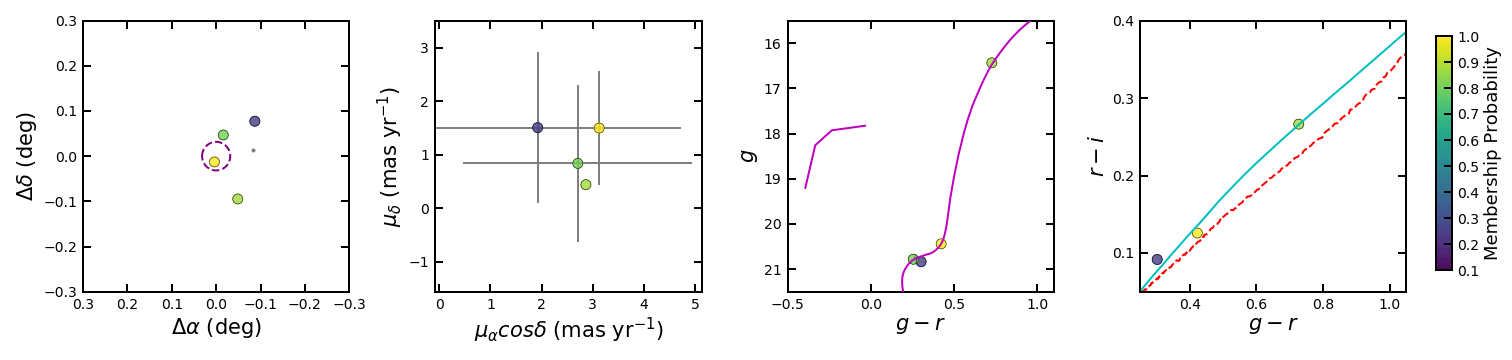}
\caption{Analogous to Figure~\ref{fig:ret2} but for Cetus II.}
\label{fig:cet2}
\end{figure*}

\begin{figure*}[h!]
\plotone{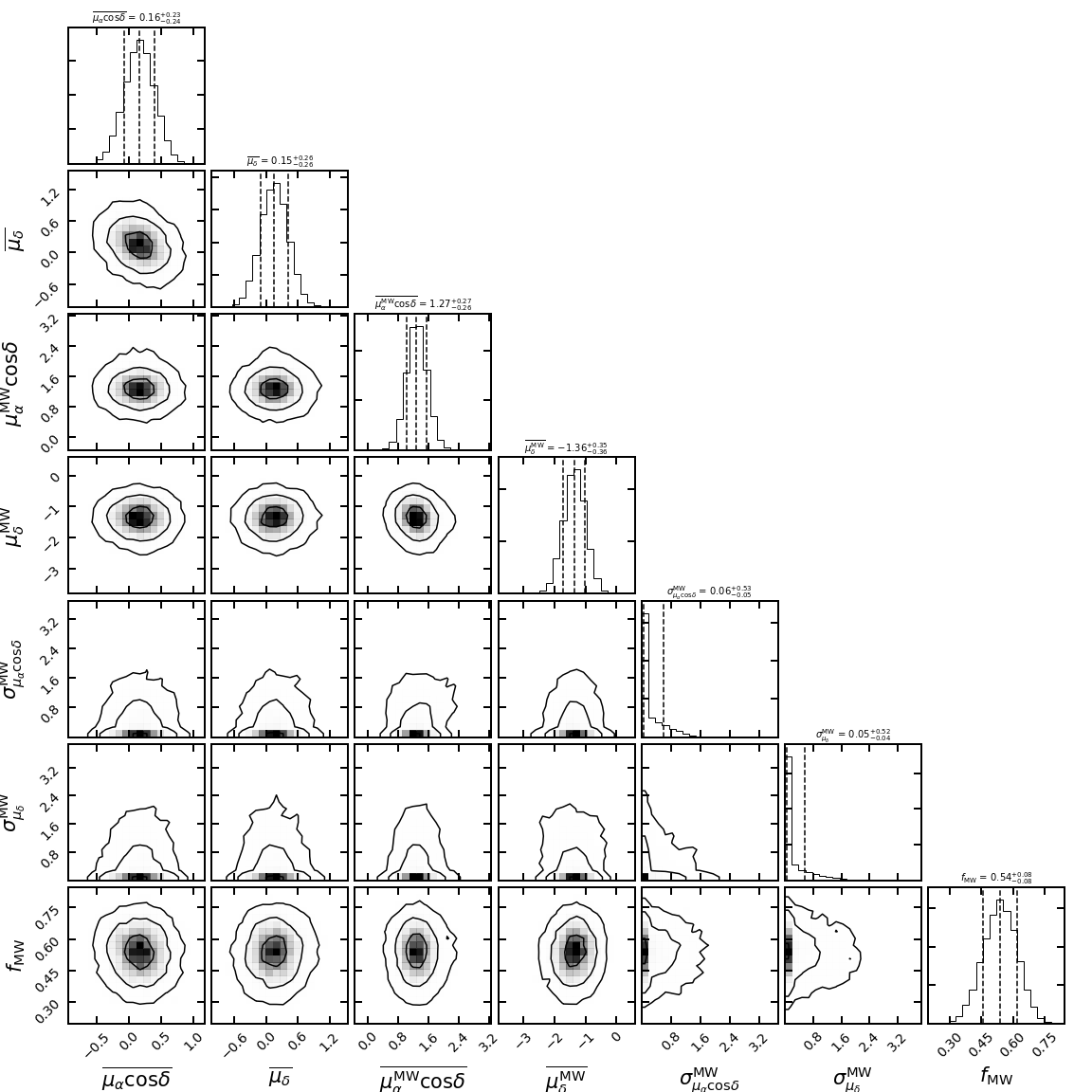}
\caption{Analogous to Figure~\ref{fig:ret2_corner} but for Eridanus II.}
\label{fig:eri2_corner}
\end{figure*}

\begin{figure*}[h!]
\plotone{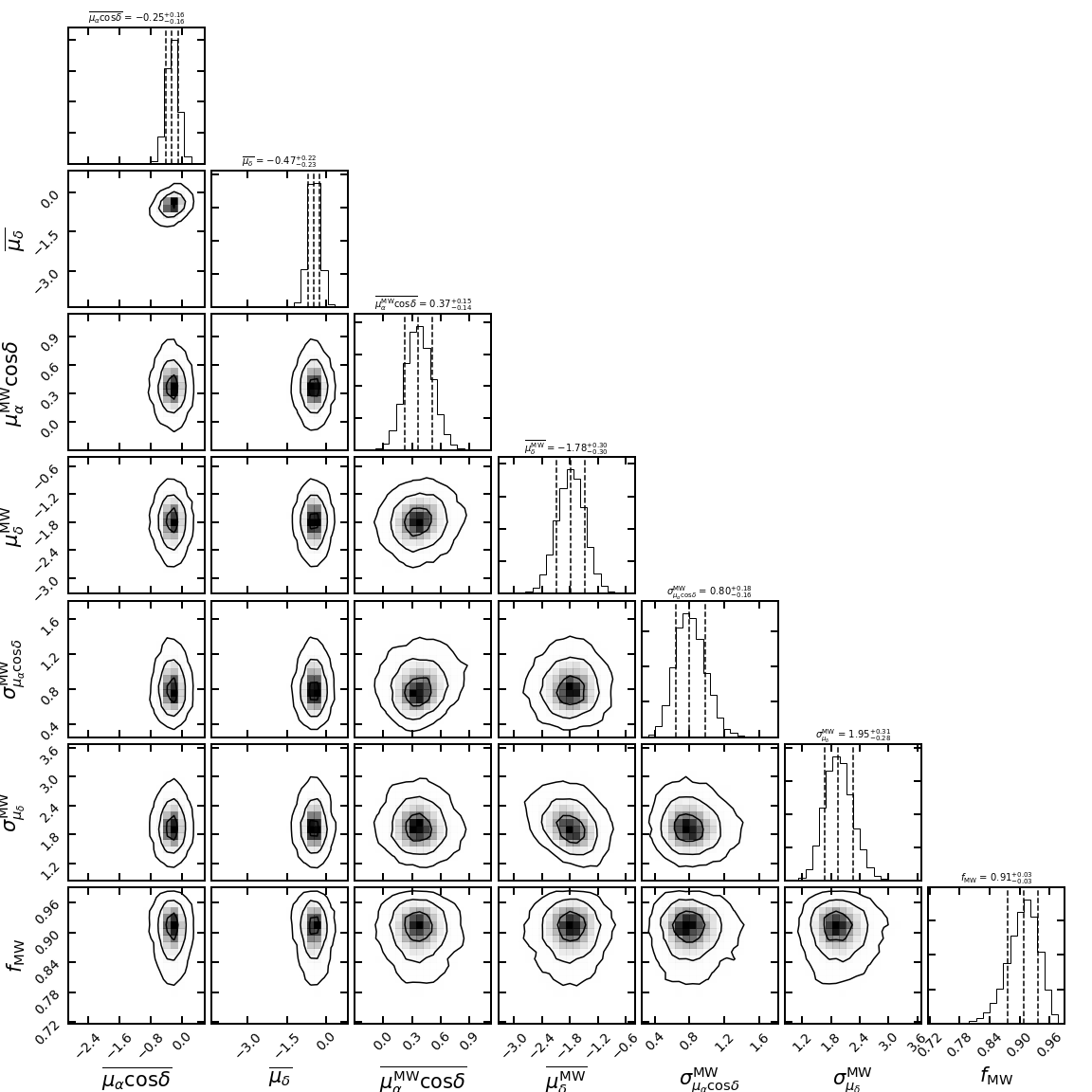}
\caption{Analogous to Figure~\ref{fig:ret2_corner} but for Grus I.}
\label{fig:gru1_corner}
\end{figure*}

\begin{figure*}[h!]
\plotone{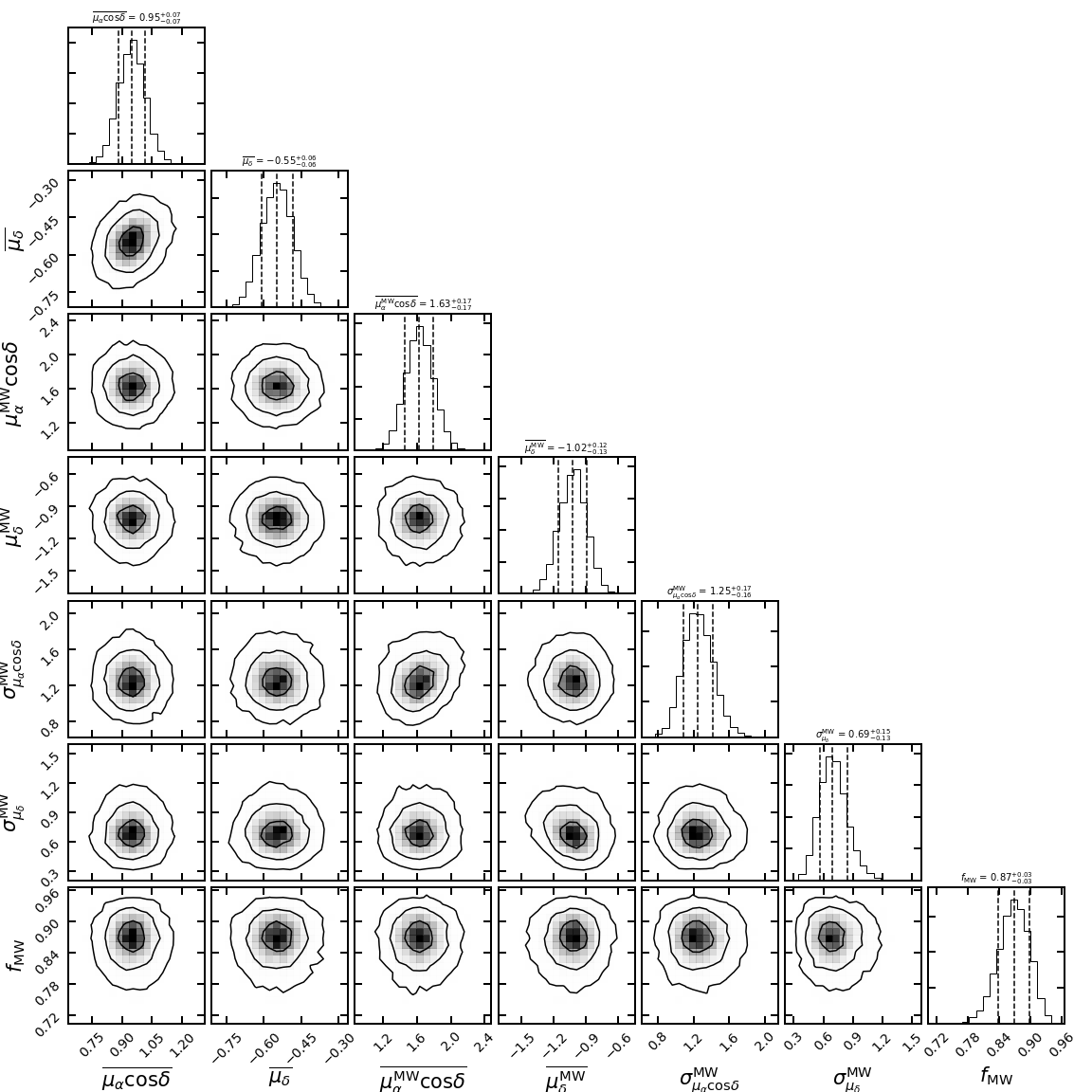}
\caption{Analogous to Figure~\ref{fig:ret2_corner} but for Horologium I.}
\label{fig:hor1_corner}
\end{figure*}

\begin{figure*}[h!]
\plotone{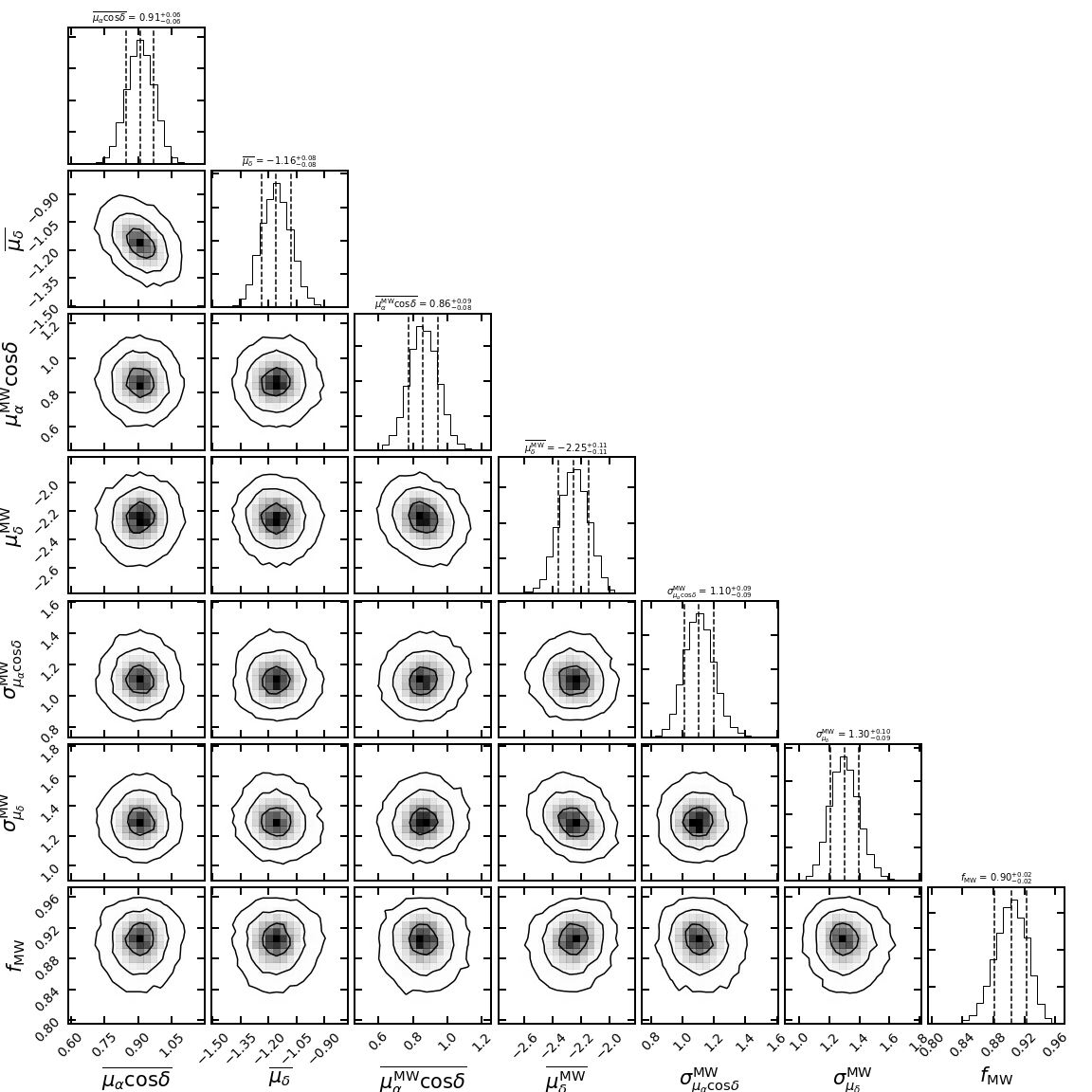}
\caption{Analogous to Figure~\ref{fig:ret2_corner} but for Tucana II.}
\label{fig:tuc2_corner}
\end{figure*}

\begin{figure*}[h!]
\plotone{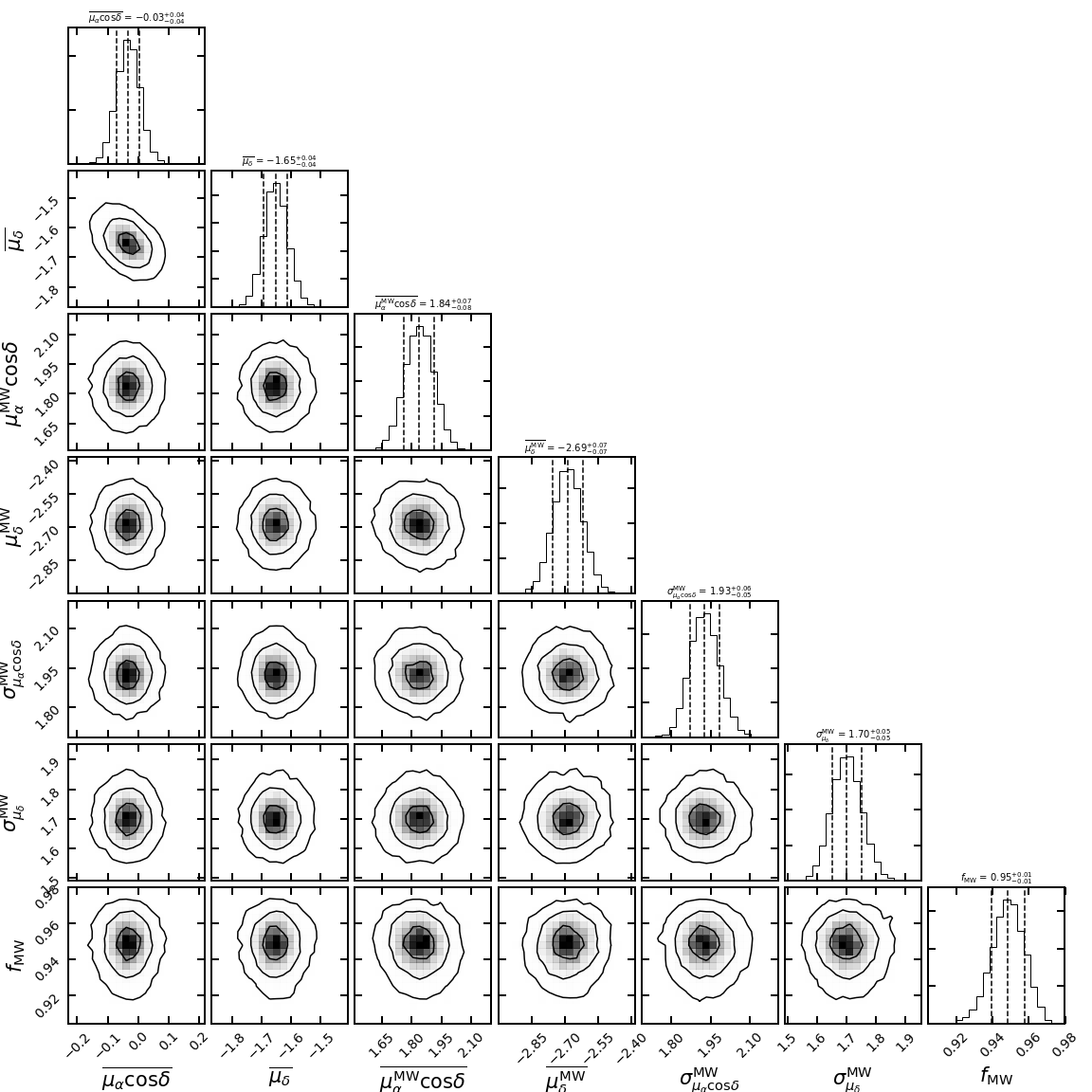}
\caption{Analogous to Figure~\ref{fig:ret2_corner} but for Tucana III.}
\label{fig:tuc3_corner}
\end{figure*}

\begin{figure*}[h!]
\plotone{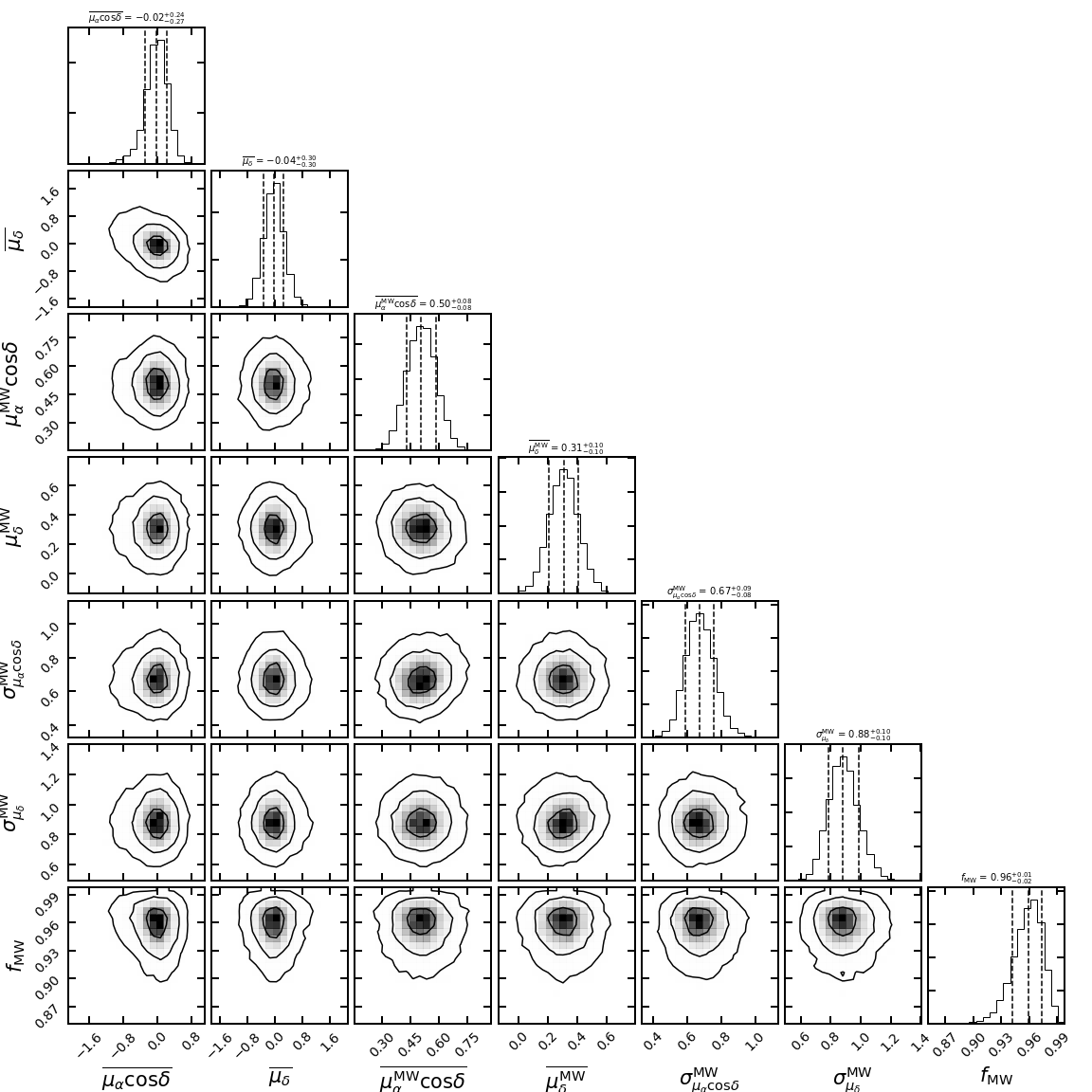}
\caption{Analogous to Figure~\ref{fig:ret2_corner} but for Columba I.}
\label{fig:col1_corner}
\end{figure*}

\begin{figure*}[h!]
\plotone{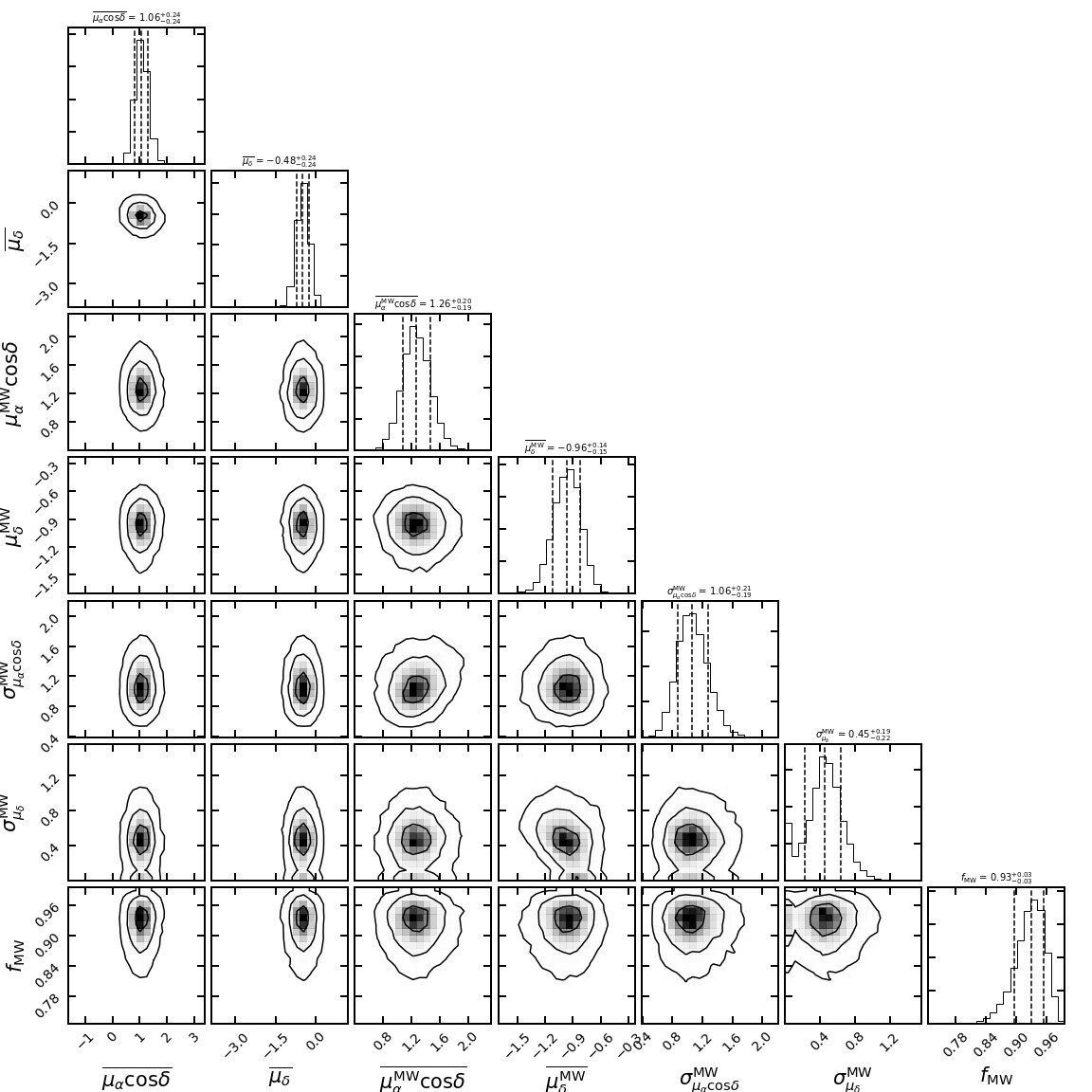}
\caption{Analogous to Figure~\ref{fig:ret2_corner} but for Eridanus III.}
\label{fig:eri3_corner}
\end{figure*}

\begin{figure*}[h!]
\plotone{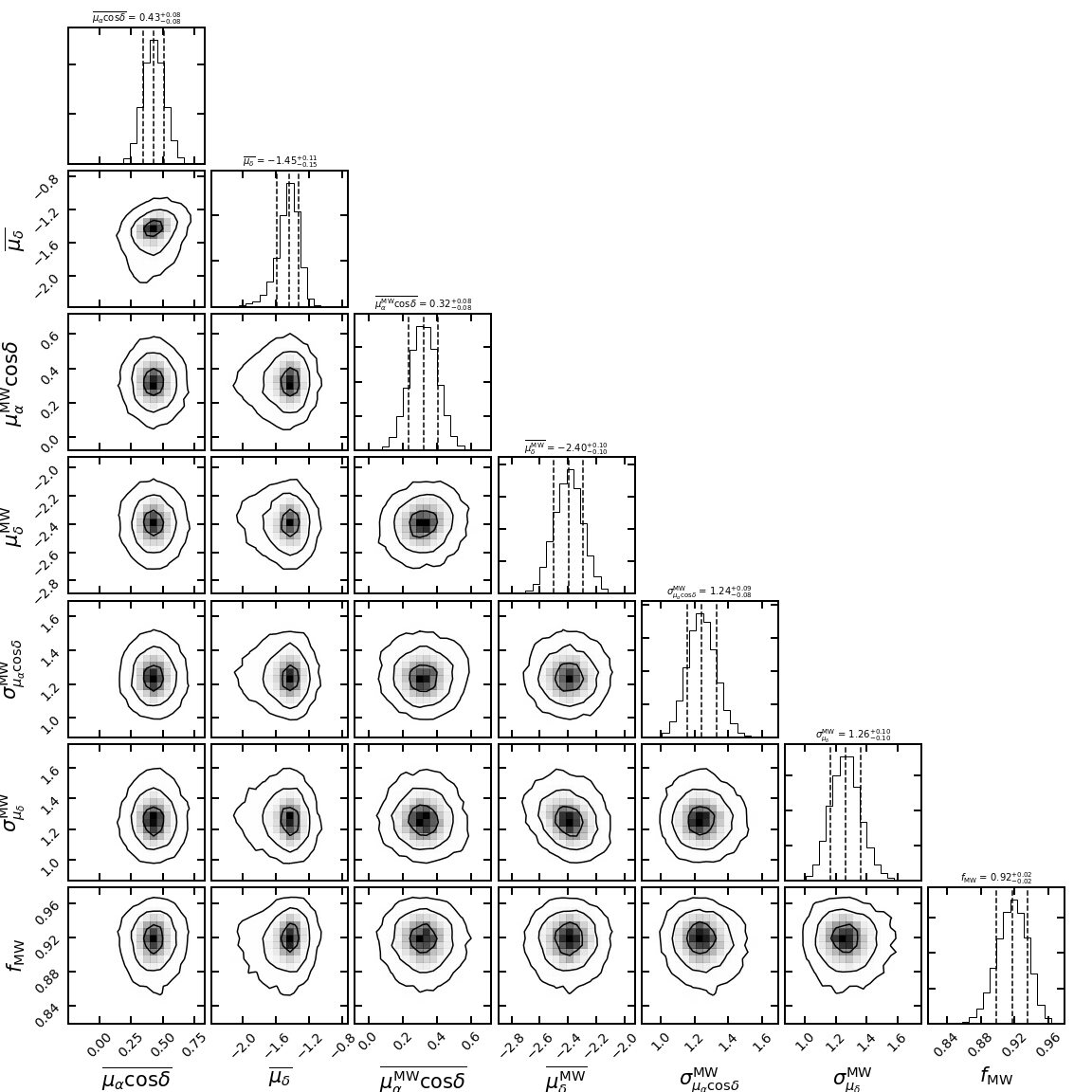}
\caption{Analogous to Figure~\ref{fig:ret2_corner} but for Grus II.}
\label{fig:gru2_corner}
\end{figure*}

\begin{figure*}[h!]
\plotone{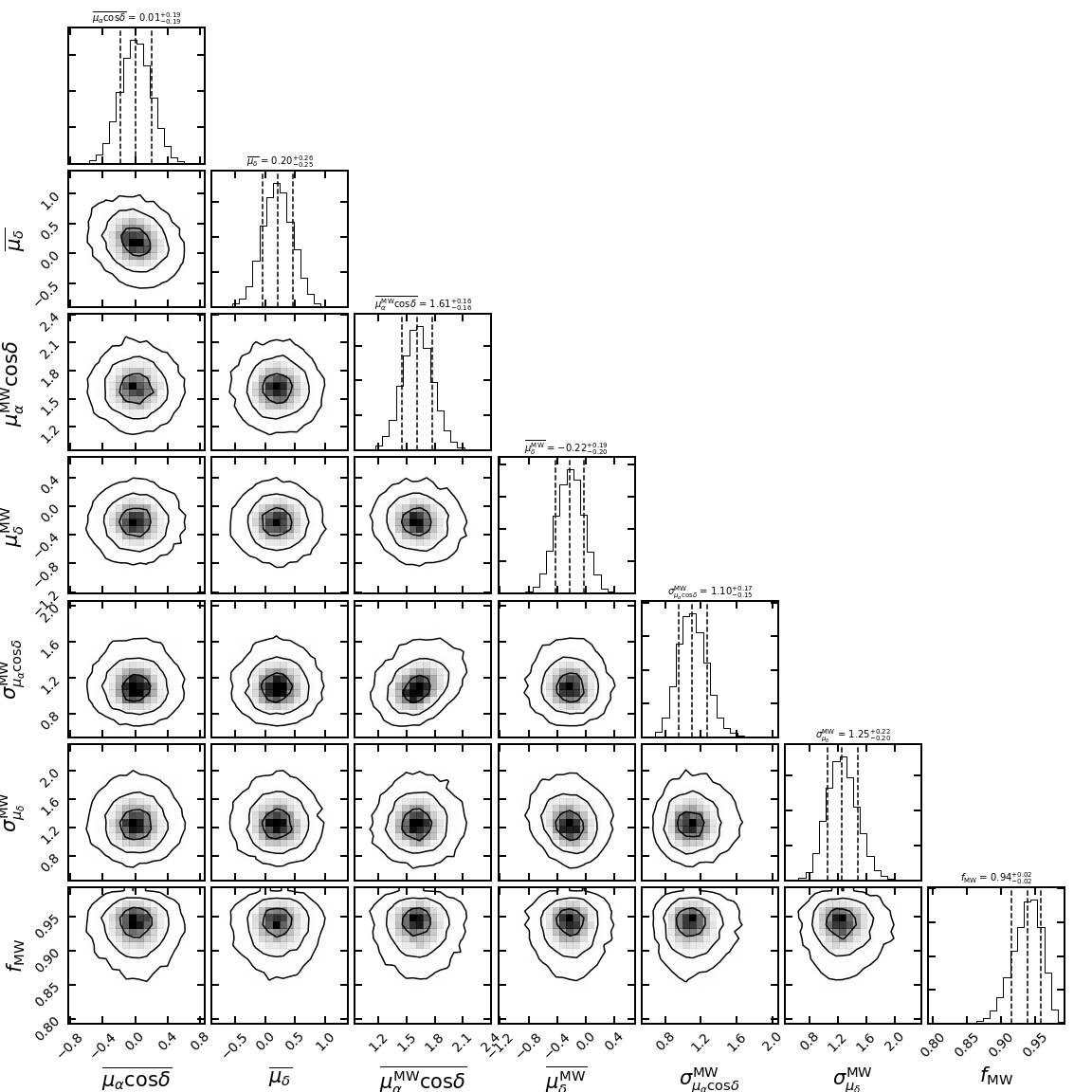}
\caption{Analogous to Figure~\ref{fig:ret2_corner} but for Pictor I.}
\label{fig:pic1_corner}
\end{figure*}

\begin{figure*}[h!]
\plotone{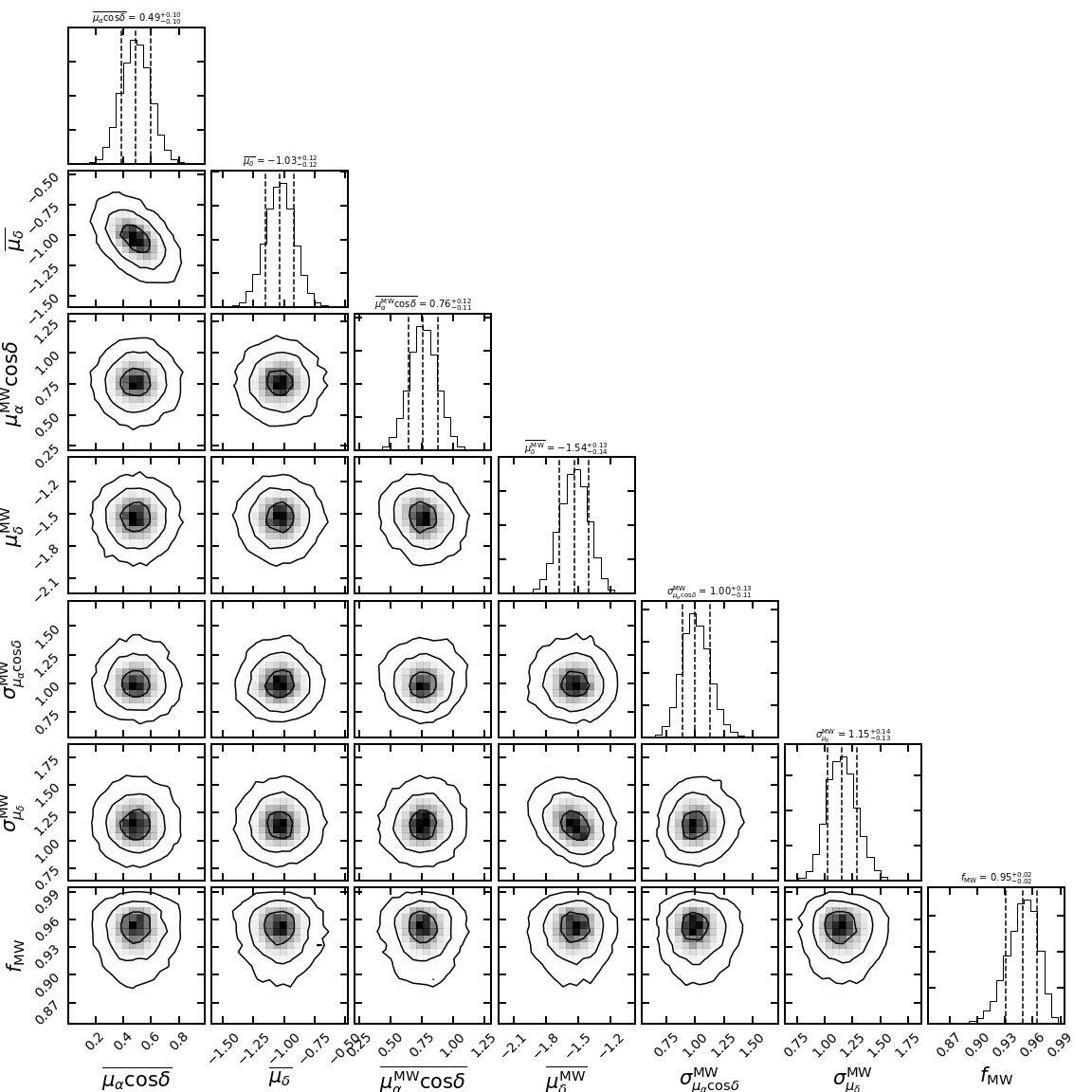}
\caption{Analogous to Figure~\ref{fig:ret2_corner} but for Phoenix II.}
\label{fig:phe2_corner}
\end{figure*}

\begin{figure*}[h!]
\plotone{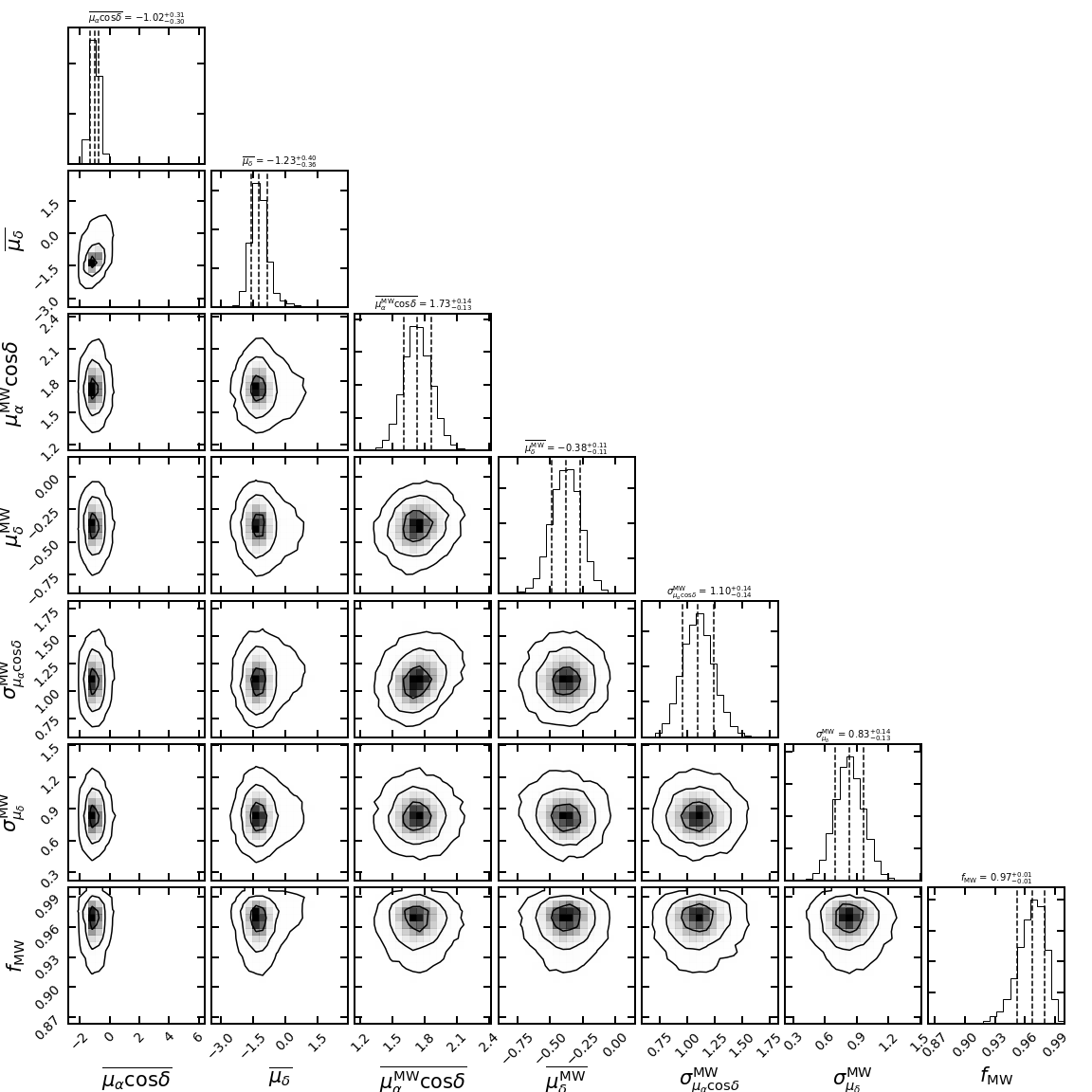}
\caption{Analogous to Figure~\ref{fig:ret2_corner} but for Reticulum III.}
\label{fig:ret3_corner}
\end{figure*}

\begin{figure*}[h!]
\plotone{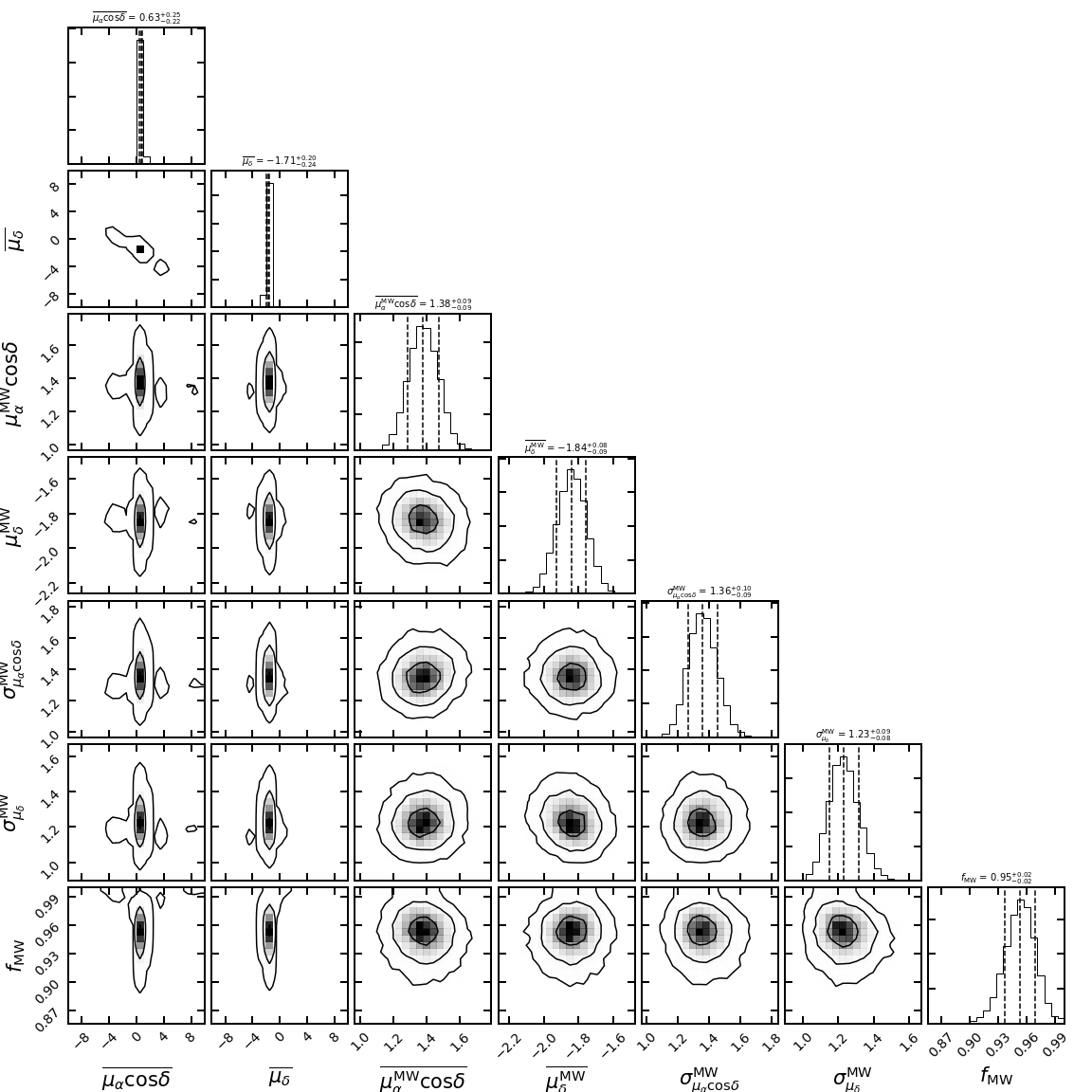}
\caption{Analogous to Figure~\ref{fig:ret2_corner} but for Tucana IV.}
\label{fig:tuc4_corner}
\end{figure*}

\begin{figure*}[h!]
\plotone{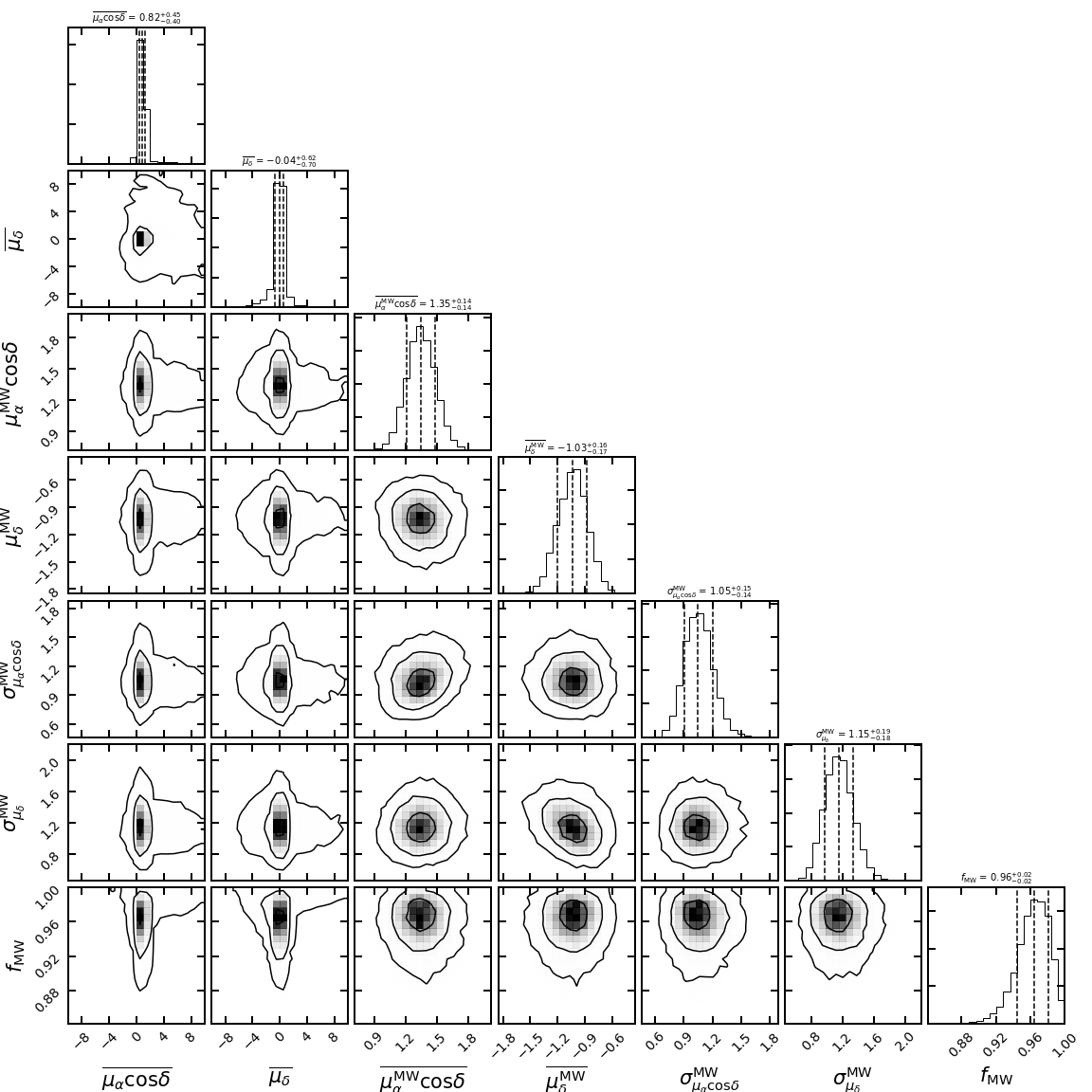}
\caption{Analogous to Figure~\ref{fig:ret2_corner} but for Horologium II.}
\label{fig:hor2_corner}
\end{figure*}

\begin{figure*}[h!]
\plotone{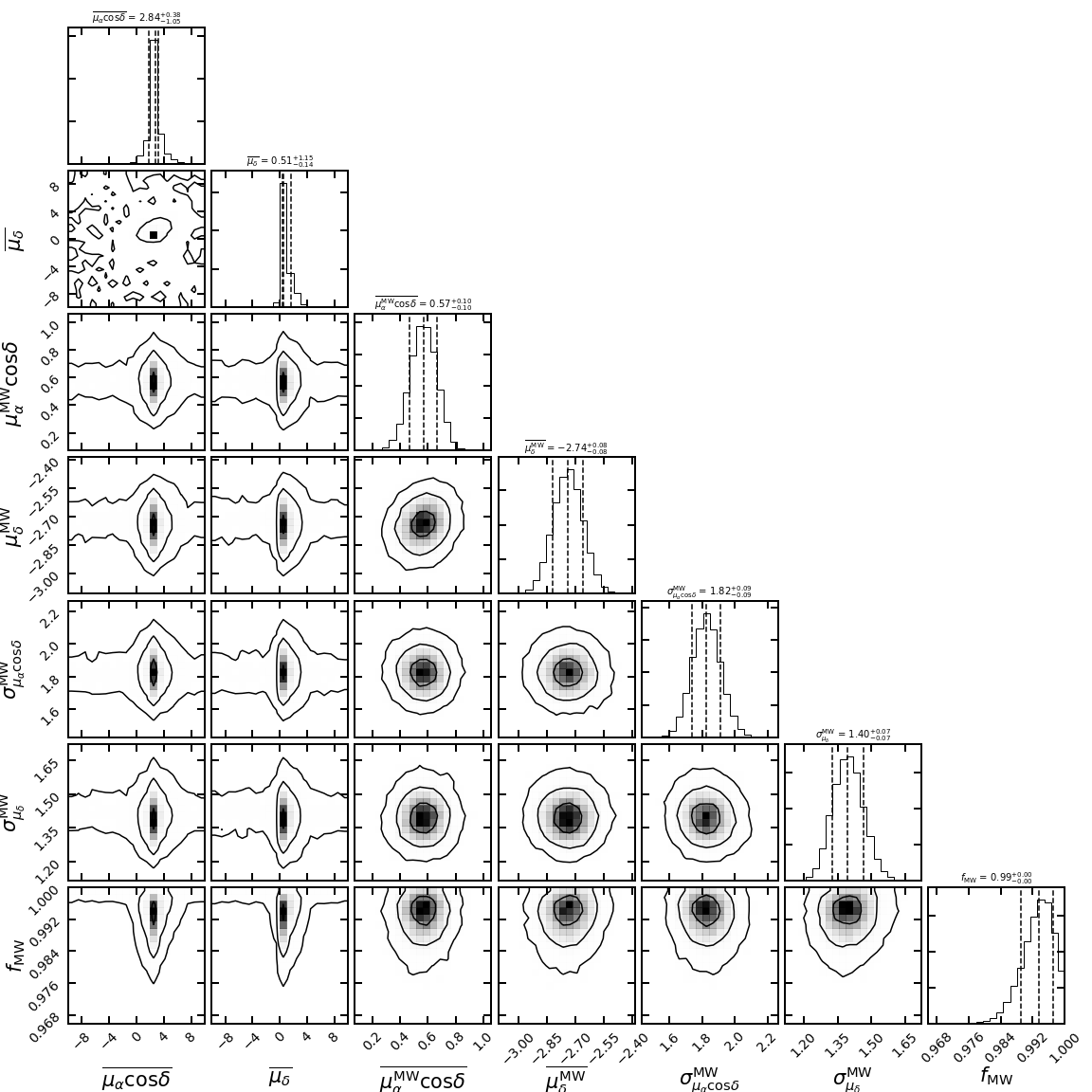}
\caption{Analogous to Figure~\ref{fig:ret2_corner} but for Cetus II.}
\label{fig:cet2_corner}
\end{figure*}

\begin{figure*}[h!]
\plotone{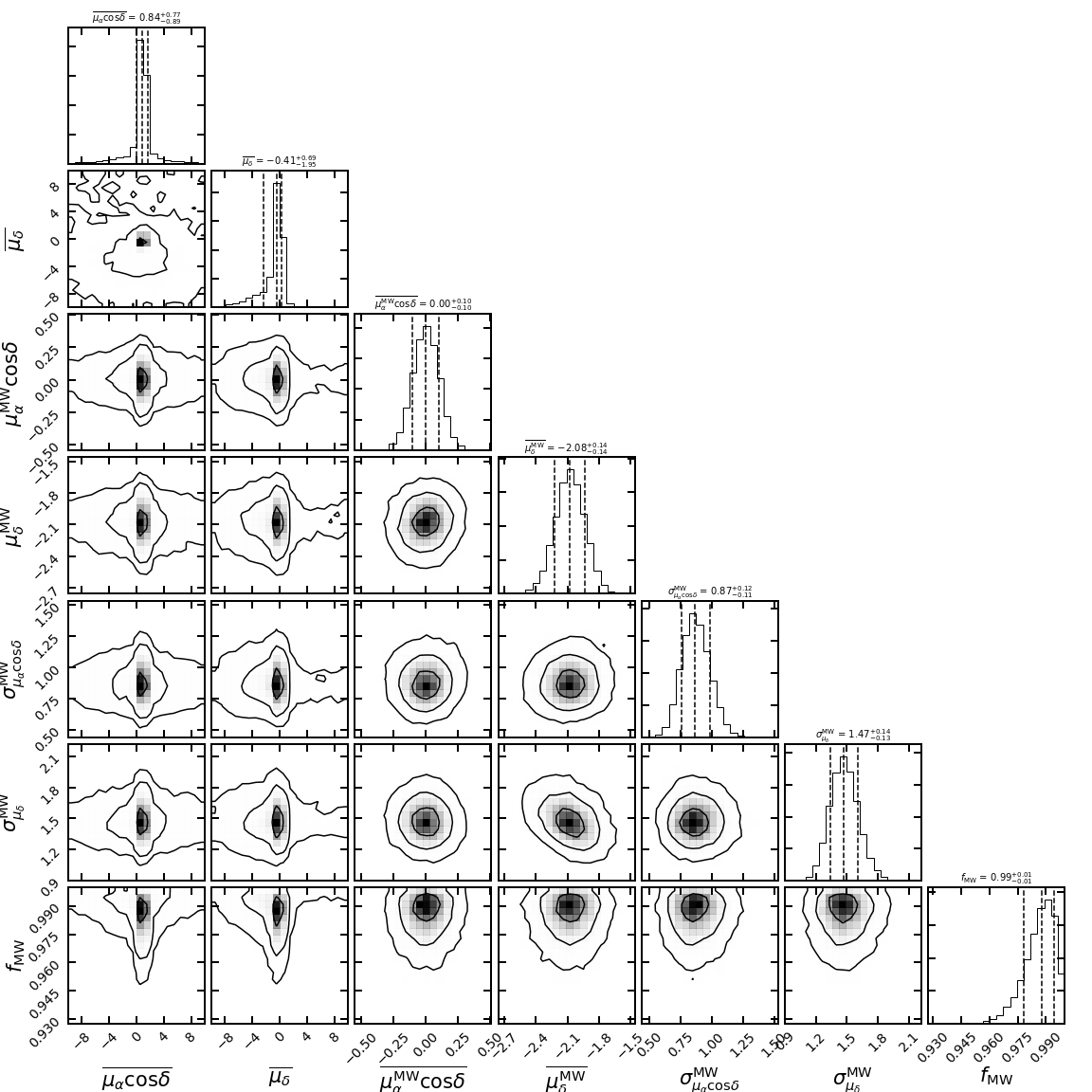}
\caption{Analogous to Figure~\ref{fig:ret2_corner} but for Kim 2.}
\label{fig:ind1_corner}
\end{figure*}

We present the diagnostic plots and  corner plots of posterior distributions in Figure~\ref{fig:col1}-~\ref{fig:ind1_corner} for the remaining satellites. The plots are  similar to those as shown in Figure~\ref{fig:ret2} and Figure~\ref{fig:ret2_corner}  for Reticulum II. 
In contrast to Figure~\ref{fig:ret2}, we show all stars with $p_i > 0.1$ except for Tucana II and Tucana III where we display $p_i>0.5$.

\newpage

\begin{figure*}[h!]
\includegraphics[scale=.34]{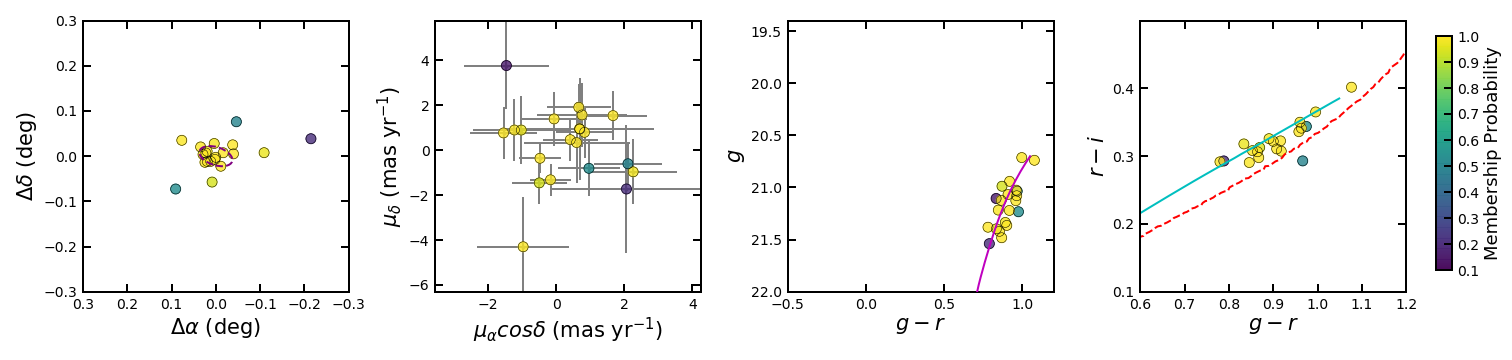}
\caption{Analogous to Figure~\ref{fig:ret2} but for Eridanus II.}
\label{fig:eri2}
\end{figure*}

\begin{figure*}[h!]
\includegraphics[scale=.34]{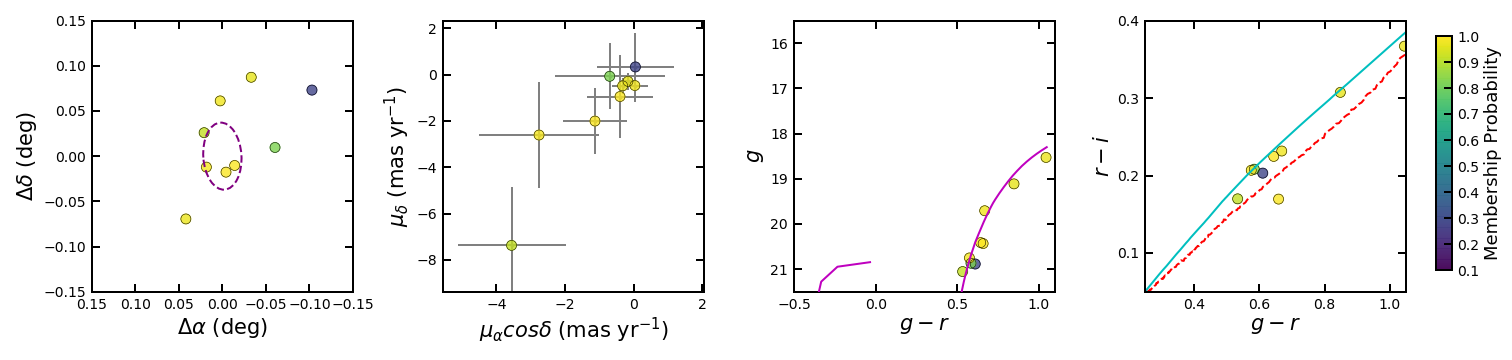}
\caption{Analogous to Figure~\ref{fig:ret2} but for Grus I.}
\label{fig:gru1}
\end{figure*}

\begin{figure*}[h!]
\includegraphics[scale=.34]{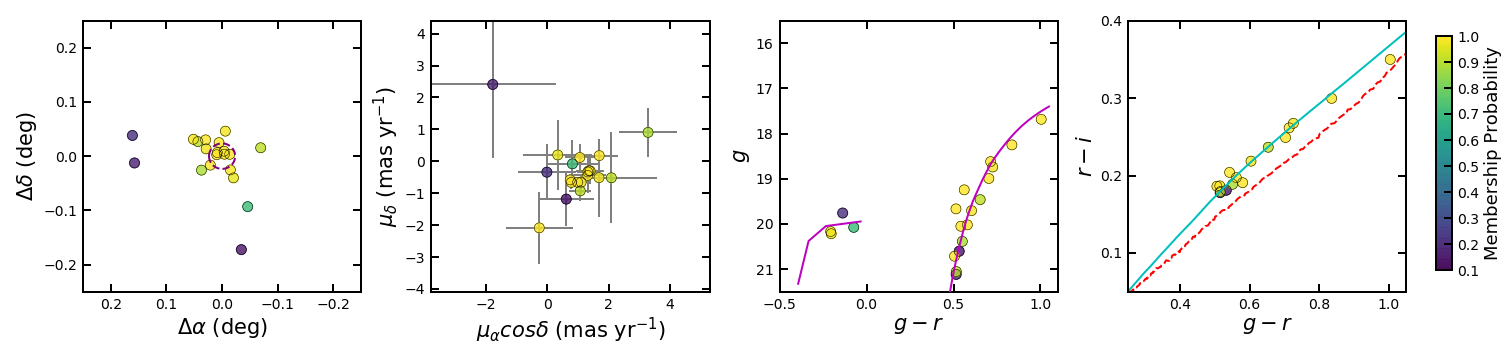}
\caption{Analogous to Figure~\ref{fig:ret2} but for Horologium I.}
\label{fig:hor1}
\end{figure*}

\begin{figure*}[h!]
\includegraphics[scale=.34]{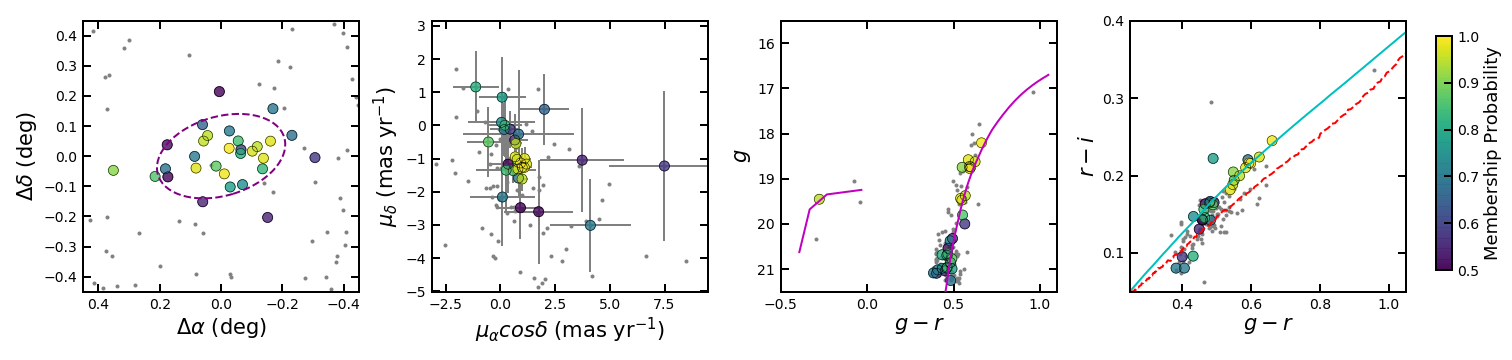}
\caption{Analogous to Figure~\ref{fig:ret2} but for Tucana II.}
\label{fig:tuc2}
\end{figure*}

\begin{figure*}[h!]
\includegraphics[scale=.34]{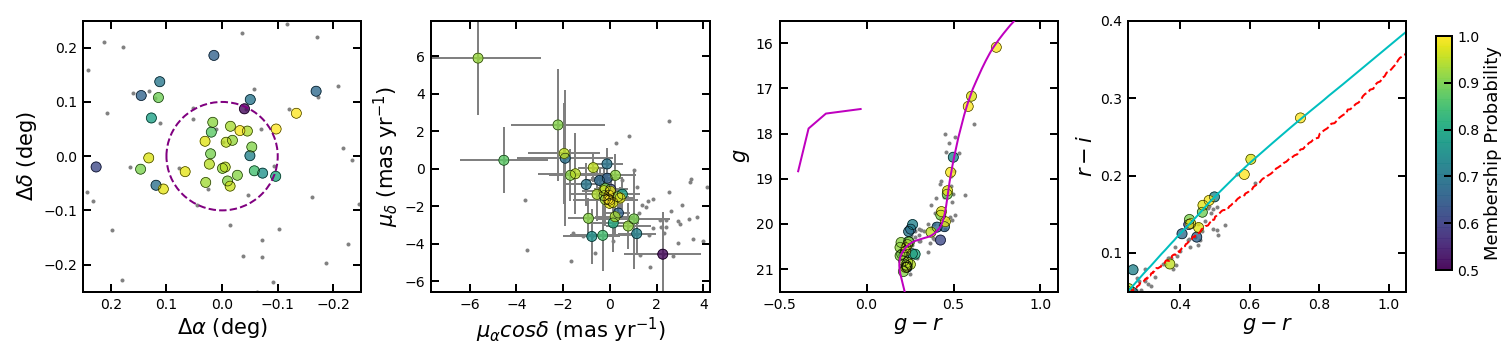}
\caption{Analogous to Figure~\ref{fig:ret2} but for Tucana III.}
\label{fig:tuc3}
\end{figure*}

\clearpage
\begin{turnpage}
\begin{deluxetable*}{r c c r r c c c c r r c r}

\tabletypesize{\scriptsize}
\tablecaption{Proper motion members from the likelihood fit with membership probability $p_i > 0.1$.
\label{tab:membership}
}
\tablehead{Dwarf & ID (DES)\tablenotemark{a} & ID (Gaia)\tablenotemark{a} &  R.A.\tablenotemark{b} & Dec.\tablenotemark{b}  & $g$\tablenotemark{c} & $r$\tablenotemark{c} & $i$\tablenotemark{c} & $G$\tablenotemark{d} & \multicolumn{1}{c}{$\mu_\alpha \cos{\delta}$}  &  \multicolumn{1}{c}{$\mu_\delta$} & $p_i$ & MP$i$\tablenotemark{e} \\ 
 &  & & (deg) & (deg) & (mag) & (mag) & (mag) & (mag) & \multicolumn{1}{c}{${\rm (mas \, yr^{-1}})$}  &  \multicolumn{1}{c}{${\rm (mas \, yr^{-1}})$} & & 
}
\startdata
 col1 & 435071974 & 2908280823235180672 & $ 82.874025$  & $-28.030109$ & 20.822 & 20.054 & 19.791 & 20.254  & $-0.321 \pm  0.874$ & $ 0.229 \pm  0.985$ & 0.98 & $1$  \\
 col1 & 435071863 & 2908279414485908480 & $ 82.861250$  & $-28.028666$ & 20.176 & 19.367 & 19.081 & 19.537  & $-0.467 \pm  0.430$ & $ 0.992 \pm  0.574$ & 0.97 & $1$  \\
 col1 & 435072141 & 2908279483205382784 & $ 82.829585$  & $-28.033424$ & 20.142 & 19.385 & 19.103 & 19.541  & $ 0.004 \pm  0.464$ & $-0.592 \pm  0.627$ & 0.97 & $1$  \\
 col1 & 435072438 & 2908232410363815552 & $ 82.879960$  & $-28.039009$ & 19.735 & 18.869 & 18.557 & 19.035  & $ 0.410 \pm  0.297$ & $-0.486 \pm  0.416$ & 0.97 & $1$  \\
 col1 & 435070706 & 2908281098113356672 & $ 82.855848$  & $-28.007312$ & 21.335 & 20.809 & 20.627 & 20.989  & $ 0.237 \pm  2.933$ & $-2.258 \pm  2.995$ & 0.92 & $1$  \\
 col1 & 435075684 & 2908231963687185536 & $ 82.871261$  & $-28.094459$ & 20.530 & 19.790 & 19.582 & 19.984  & $-1.483 \pm  0.604$ & $-0.195 \pm  0.763$ & 0.78 & $0$  \\
 col1 & 435070077 & 2908280445278290560 & $ 82.821741$  & $-27.995826$ & 21.035 & 20.333 & 20.134 & 20.569  & $-0.230 \pm  1.095$ & $ 0.797 \pm  1.336$ & 0.70 & $0$  \\
 col1 & 435075223 & 2908232071063767424 & $ 82.914629$  & $-28.085861$ & 20.160 & 19.461 & 19.261 & 19.648  & $ 0.629 \pm  0.455$ & $ 0.781 \pm  0.665$ & 0.24 & $0$  \\
 col1 & 435079410 & 2908225710214769152 & $ 82.792278$  & $-28.162965$ & 20.499 & 19.732 & 19.505 & 19.936  & $ 0.488 \pm  0.559$ & $ 0.783 \pm  0.795$ & 0.15 & $0$  \\
 col1 & 435068966 & 2908257806507777152 & $ 83.000291$  & $-27.976304$ & 20.693 & 19.925 & 19.683 & 20.123  & $-0.202 \pm  0.688$ & $ 0.829 \pm  0.727$ & 0.11 & $1$  \\
 col1 & 435068915 & 2908234540667629952 & $ 82.984642$  & $-27.975305$ & 20.904 & 20.149 & 19.938 & 20.354  & $ 0.850 \pm  0.920$ & $-1.619 \pm  0.942$ & 0.11 & $0$  \\
  eri3 & 115993672 & 4745740262792352128 & $ 35.702833$  & $-52.284756$ & 19.318 & 18.623 & 18.403 & 18.796  & $ 1.114 \pm  0.256$ & $-0.419 \pm  0.252$ & 1.00 & $1$  \\
 eri3 & 115993659 & 4745740262792353536 & $ 35.693772$  & $-52.282666$ & 20.340 & 20.586 & 20.817 & 20.588  & $ 0.645 \pm  1.205$ & $-1.141 \pm  1.290$ & 1.00 & $-1$  \\
 eri3 & 115993945 & 4745740262793792640 & $ 35.699386$  & $-52.286849$ & 21.389 & 20.793 & 20.589 & 20.969  & $ 2.351 \pm  4.156$ & $-0.514 \pm  3.576$ & 1.00 & $1$  \\
 eri3 & 115993325 & 4745740335808220800 & $ 35.688382$  & $-52.276405$ & 20.236 & 20.423 & 20.589 & 20.400  & $ 1.807 \pm  0.960$ & $-2.196 \pm  0.938$ & 0.98 & $-1$  \\
 eri3 & 115994829 & 4745739438158625408 & $ 35.734790$  & $-52.303653$ & 20.578 & 20.041 & 20.038 & 20.411  & $-0.623 \pm  0.891$ & $ 0.659 \pm  0.826$ & 0.73 & $0$  \\
 eri3 & 115992692 & 4745743423889682176 & $ 35.620614$  & $-52.264944$ & 21.224 & 20.800 & 20.680 & 20.981  & $ 4.340 \pm  3.237$ & $ 0.456 \pm  4.716$ & 0.43 & $1$  \\
 phe2 & 137808662 & 6497787062124139904 & $354.991914$  & $-54.406305$ & 18.880 & 18.127 & 17.858 & 18.250  & $ 0.595 \pm  0.209$ & $-1.169 \pm  0.227$ & 1.00 & $1$  \\
 phe2 & 137806191 & 6497793143797836032 & $354.982488$  & $-54.368989$ & 18.569 & 17.764 & 17.477 & 17.890  & $ 0.493 \pm  0.146$ & $-1.055 \pm  0.173$ & 1.00 & $1$  \\
 phe2 & 137809630 & 6497787062124293376 & $355.002579$  & $-54.415568$ & 21.138 & 20.595 & 20.415 & 20.749  & $ 3.384 \pm  1.919$ & $-1.767 \pm  1.748$ & 0.99 & $1$  \\
 phe2 & 617652809 & 6497792868919926016 & $355.032863$  & $-54.391801$ & 19.419 & 18.839 & 18.637 & 18.946  & $ 0.586 \pm  0.289$ & $-1.027 \pm  0.360$ & 0.99 & $1$  \\
 phe2 & 137807790 & 6497792971999300608 & $354.999279$  & $-54.390529$ & 20.286 & 20.431 & 20.608 & 20.404  & $ 1.995 \pm  0.908$ & $-0.031 \pm  1.199$ & 0.97 & $-1$  \\
 phe2 & 137811333 & 6497786581087794816 & $354.954566$  & $-54.440401$ & 19.410 & 18.816 & 18.607 & 18.941  & $ 0.052 \pm  0.311$ & $-0.812 \pm  0.344$ & 0.93 & $1$  \\
 phe2 & 137806663 & 6497793113733183872 & $355.008566$  & $-54.373974$ & 21.151 & 20.625 & 20.447 & 20.773  & $-1.992 \pm  1.433$ & $ 0.256 \pm  2.187$ & 0.92 & $1$  \\
 phe2 & 137807073 & 6497787886758025600 & $354.932206$  & $-54.380084$ & 20.299 & 20.464 & 20.644 & 20.424  & $-0.125 \pm  0.871$ & $ 0.384 \pm  1.070$ & 0.83 & $-1$  \\
 phe2 & 137809363 & 6497787749319062272 & $354.947494$  & $-54.412762$ & 21.324 & 20.755 & 20.580 & 20.934  & $ 0.632 \pm  2.257$ & $-1.962 \pm  2.657$ & 0.82 & $1$  \\
 phe2 & 617656251 & 6497789505960781312 & $355.155429$  & $-54.440715$ & 20.825 & 20.282 & 20.114 & 20.429  & $ 1.132 \pm  1.224$ & $-0.191 \pm  1.164$ & 0.11 & $1$  \\
 pic1 & 507874118 & 4784435444228398720 & $ 70.936636$  & $-50.283391$ & 19.646 & 18.930 & 18.656 & 19.092  & $-0.268 \pm  0.422$ & $ 0.458 \pm  0.540$ & 1.00 & $1$  \\
 pic1 & 507874680 & 4784435547307612160 & $ 70.925860$  & $-50.290798$ & 19.127 & 18.293 & 17.970 & 18.420  & $-0.090 \pm  0.320$ & $-0.229 \pm  0.446$ & 1.00 & $1$  \\
 pic1 & 507875217 & 4784435341149180160 & $ 70.938406$  & $-50.297109$ & 19.469 & 18.670 & 18.359 & 18.789  & $ 0.118 \pm  0.406$ & $ 0.491 \pm  0.506$ & 1.00 & $1$  \\
 pic1 & 507873863 & 4784435482884682368 & $ 70.953691$  & $-50.279079$ & 20.335 & 19.698 & 19.447 & 19.817  & $ 0.785 \pm  0.698$ & $-0.049 \pm  0.894$ & 1.00 & $1$  \\
 pic1 & 507874403 & 4784435547310313344 & $ 70.928587$  & $-50.285425$ & 20.207 & 19.565 & 19.306 & 19.687  & $ 0.215 \pm  0.641$ & $ 1.537 \pm  0.842$ & 1.00 & $1$  \\
 pic1 & 507874307 & 4784435650386828800 & $ 70.931597$  & $-50.283887$ & 20.742 & 20.213 & 20.010 & 20.339  & $ 0.317 \pm  1.052$ & $-3.142 \pm  1.363$ & 0.99 & $1$  \\
 pic1 & 507873043 & 4784799111995400704 & $ 71.009510$  & $-50.266991$ & 21.262 & 20.654 & 20.420 & 20.784  & $-6.615 \pm  3.534$ & $ 1.694 \pm  2.526$ & 0.96 & $1$  \\
  ret3 & 378640368 & 4680599524606215296 & $ 56.390209$  & $-60.449138$ & 20.456 & 19.844 & 19.658 & 20.054  & $-0.781 \pm  0.889$ & $-1.047 \pm  1.118$ & 0.99 & $1$  \\
 ret3 & 378640453 & 4680600246160717312 & $ 56.360442$  & $-60.452279$ & 20.059 & 19.399 & 19.205 & 19.622  & $-0.782 \pm  0.715$ & $ 0.299 \pm  0.830$ & 0.99 & $0$  \\
 ret3 & 378644536 & 4680596058569220224 & $ 56.431472$  & $-60.528601$ & 19.217 & 18.514 & 18.333 & 18.741  & $-1.192 \pm  0.445$ & $-2.027 \pm  0.514$ & 0.98 & $0$  \\
 ret3 & 378643163 & 4680599082224649216 & $ 56.403424$  & $-60.502436$ & 20.723 & 20.225 & 20.086 & 20.451  & $-1.255 \pm  1.452$ & $-1.333 \pm  1.942$ & 0.92 & $0$  \\
 ret3 & 378632896 & 4680790599111264768 & $ 56.315672$  & $-60.309597$ & 19.446 & 18.768 & 18.587 & 19.022  & $-1.301 \pm  0.507$ & $-1.167 \pm  0.567$ & 0.90 & $0$  \\
 ret3 & 378642244 & 4680599112289740032 & $ 56.414505$  & $-60.487047$ & 20.602 & 20.877 & 21.134 & 20.878  & $ 2.685 \pm  3.068$ & $ 6.494 \pm  4.286$ & 0.69 & $-1$  \\
 ret3 & 378637914 & 4680600628413758720 & $ 56.315797$  & $-60.403005$ & 20.983 & 20.540 & 20.462 & 20.763  & $ 5.734 \pm  2.779$ & $-0.103 \pm  3.801$ & 0.21 & $0$  \\
\enddata
\tablecomments{This table includes all stars with $p_i > 0.1$ for 17 satellites with a detection or potential detection of the systemic proper motion\footnote{Full member files are available at \url{https://github.com/apace7/gaia_cross_des_proper_motions}.}. This table is available in its entirety in the electronic edition of the journal. A portion is reproduced here to provide guidance on form and content.\\
(a) DES IDs are from \code{COADD\_OBJECT\_ID} column in DES DR1; Gaia IDs are from \code{SOURCE\_ID} column in \gaia DR2. \\
(b) R.A. and Dec. are from Gaia DR2 catalog (J2015.5 Epoch).\\
(c) $g$-, $r$- and $i$-band magnitudes are reddening corrected weighted average photometry (\code{WAVG\_MAG\_PSF\_DERED}) from DES DR1 catalog.\\
(d) $G$-band magnitudes are from Gaia DR2 catalog without reddening correction.\\
(e) MP is the column to assess the metallicity of the star based on its location in the color-color diagram (see \S\ref{sec:color} for details). Here we define ${\rm MP} = -1$ for blue stars with $g-r < 0.35$ without a metallicity assessment; ${\rm MP} = 1$ (${\rm MP} = 0$) for stars with $g-r > 0.35$ and a location on color-color diagram  above (below) the empirical stellar locus, indicating that they are possible metal-poor (metal-rich) stars and therefore likely to be members (non-members) of a ultra-faint satellite. 
}

\end{deluxetable*}

\end{turnpage}

\end{document}